\documentclass[10pt,a4paper,twocolumn]{article}
\newif\ifsubmission\submissionfalse
\newif\iftwocol\twocolfalse
\newif\ifrsc\rsctrue
\expandafter\ifx\csname ifrsc\endcsname\relax\expandafter\newif\csname ifrsc\endcsname\fi

\usepackage[T1]{fontenc}
\usepackage[utf8]{inputenc}
\iftwocol
\else
  \usepackage{lmodern}
  \ifsubmission
    \usepackage[margin=1in]{geometry}
  \else
    \usepackage[margin=0.75in]{geometry}
  \fi
\fi
\usepackage{graphicx}
\usepackage{booktabs}
\usepackage{array}
\usepackage{tabularx}
\usepackage{siunitx}
\usepackage{amsmath}
\usepackage{xcolor}
\usepackage{setspace}
\iftwocol
  \biboptions{sort&compress}
\else
  \ifrsc
    \usepackage[super,sort&compress]{natbib}
  \else
    \usepackage[numbers,sort&compress]{natbib}
  \fi
\fi
\usepackage{lineno}
\iftwocol
  \usepackage{placeins}
\else
  \usepackage{placeins}
\fi
\usepackage{xurl}
\usepackage[hidelinks]{hyperref}
\usepackage[font=small,labelfont=bf,skip=4pt]{caption}

\iftwocol
  \makeatletter
  \def\els@aparagraph[#1]#2{\elsparagraph[#1]{#2}}
  \def\els@bparagraph#1{\elsparagraph*{#1}}
  \makeatother
\fi

\DeclareRobustCommand{\_}{\textunderscore\allowbreak}
\newcolumntype{Y}{>{\raggedright\arraybackslash}X}
\newcolumntype{P}[1]{>{\raggedright\arraybackslash}p{#1}}

\newenvironment{widefigure}[1][tbp]{\iftwocol\begin{figure*}[#1]\else\begin{figure}[#1]\fi\centering}{\iftwocol\end{figure*}\else\end{figure}\fi}
\newenvironment{widetable}[1][tbp]{\iftwocol\begin{table*}[#1]\else\begin{table}[#1]\fi\centering}{\iftwocol\end{table*}\else\end{table}\fi}

\ifsubmission
  \linenumbers
  \usepackage[nolists,nomarkers,figuresonly]{endfloat}
  \DeclareDelayedFloatFlavor{widefigure}{figure}
\fi

\newcommand{\kjma}{kJ\,(mol of atoms)$^{-1}$}
\newcommand{\evat}{eV\,atom$^{-1}$}

\newcommand{\tbd}[1]{\textcolor{red}{[#1]}}
\newsavebox{\nfigbox}
\newcommand{\nfig}[1]{\sbox{\nfigbox}{\includegraphics{#1}}%
  \ifdim\wd\nfigbox>\linewidth\includegraphics[width=\linewidth]{#1}\else\usebox{\nfigbox}\fi}
\newcommand{\nfigOr}[2]{\IfFileExists{#1}{\nfig{#1}}{\tbd{#2}}}
\ifrsc
  \newcommand{\suppfig}[1]{Fig.~S#1 in the ESI$^\dagger$}
  \newcommand{\supptab}[1]{Table~S#1 in the ESI$^\dagger$}
  \newcommand{\suppnote}[1]{Supplementary Note~#1 in the ESI$^\dagger$}
  \newcommand{\suppinfo}{ESI}
\else
  \newcommand{\suppfig}[1]{Supplementary Fig.~S#1}
  \newcommand{\supptab}[1]{Supplementary Table~S#1}
  \newcommand{\suppnote}[1]{Supplementary Note~#1}
  \newcommand{\suppinfo}{Supplementary Information}
\fi

\newcommand{\dsName}[1]{\texorpdfstring{\textbf{#1}}{#1}}
\newcommand{\dsValues}{\dsName{Values}}
\newcommand{\dsDiagrams}{\dsName{Diagrams}}
\newcommand{\dsHulls}{\dsName{Hulls}}
\newcommand{\dsParity}{\dsName{Parity}}
\newcommand{\dsRegistry}{\dsName{Registry}}
\newcommand{\dsAssemblages}{\dsName{Assemblages}}
\newcommand{\dsBibliography}{\dsName{Bibliography}}
\newcommand{\dsBench}{\dsName{Bench}}
\newcommand{\dsValuesFormal}{ORCHID-Values}
\newcommand{\dsDiagramsFormal}{ORCHID-Diagrams}
\newcommand{\dsHullsFormal}{ORCHID-Hulls}
\newcommand{\dsParityFormal}{ORCHID-Parity}
\newcommand{\dsRegistryFormal}{ORCHID-Registry}
\newcommand{\dsAssemblagesFormal}{ORCHID-Assemblages}
\newcommand{\dsBibliographyFormal}{ORCHID-Bibliography}
\newcommand{\dsBenchFormal}{ORCHID-Bench}

\providecommand{\BlindExpertAgreeEquivPct}{}\renewcommand{\BlindExpertAgreeEquivPct}{97.9}
\providecommand{\BlindExpertAgreeRawPct}{}\renewcommand{\BlindExpertAgreeRawPct}{93.8}
\providecommand{\BlindExpertWilsonEquiv}{}\renewcommand{\BlindExpertWilsonEquiv}{0.89--1.00}
\providecommand{\BlindExpertWilsonRaw}{}\renewcommand{\BlindExpertWilsonRaw}{0.83--0.98}
\providecommand{\asmAuditFields}{}\renewcommand{\asmAuditFields}{185}
\providecommand{\asmAuditPrecHi}{}\renewcommand{\asmAuditPrecHi}{1.00}
\providecommand{\asmAuditPrecLo}{}\renewcommand{\asmAuditPrecLo}{0.87}
\providecommand{\asmAuditPrecWeighted}{}\renewcommand{\asmAuditPrecWeighted}{0.96}
\providecommand{\asmExpertAccepted}{}\renewcommand{\asmExpertAccepted}{92}
\providecommand{\asmExpertProposals}{}\renewcommand{\asmExpertProposals}{93}
\providecommand{\asmExpertReviewed}{}\renewcommand{\asmExpertReviewed}{126}

\providecommand{\auditAllRateUnits}{}\renewcommand{\auditAllRateUnits}{18.2}

\providecommand{\auditCoreRateHi}{}\renewcommand{\auditCoreRateHi}{4}
\providecommand{\auditCoreRateLo}{}\renewcommand{\auditCoreRateLo}{3}

\providecommand{\auditPostAllRateUnits}{}\renewcommand{\auditPostAllRateUnits}{23.64}

\providecommand{\auditPostRateClaims}{}\renewcommand{\auditPostRateClaims}{1.0}
\providecommand{\auditPostRateDft}{}\renewcommand{\auditPostRateDft}{2.73}
\providecommand{\auditPostRateElem}{}\renewcommand{\auditPostRateElem}{1.82}
\providecommand{\auditPostRateFig}{}\renewcommand{\auditPostRateFig}{0.0}
\providecommand{\auditPostRatePhase}{}\renewcommand{\auditPostRatePhase}{0.33}
\providecommand{\auditPostRateUnits}{}\renewcommand{\auditPostRateUnits}{9.09}

\providecommand{\auditValueCIUnits}{}\renewcommand{\auditValueCIUnits}{6.4--18.1}

\providecommand{\auditValueRateUnits}{}\renewcommand{\auditValueRateUnits}{10.9}
\providecommand{\benchClaudeRecordsDegraded}{}\renewcommand{\benchClaudeRecordsDegraded}{0.86}
\providecommand{\benchClaudeRecordsVector}{}\renewcommand{\benchClaudeRecordsVector}{1.00}
\providecommand{\benchOracleAllOne}{}\renewcommand{\benchOracleAllOne}{all}
\providecommand{\benchSpend}{}\renewcommand{\benchSpend}{8.46}

\providecommand{\biasMP}{}\renewcommand{\biasMP}{2.0}
\providecommand{\biasOQMD}{}\renewcommand{\biasOQMD}{2.0}
\providecommand{\ciHmCal}{}\renewcommand{\ciHmCal}{2.2--4.6}

\providecommand{\costAuthorHours}{}\renewcommand{\costAuthorHours}{100}
\providecommand{\costCentsPerAccepted}{}\renewcommand{\costCentsPerAccepted}{6}
\providecommand{\costPipelineUSD}{}\renewcommand{\costPipelineUSD}{4,000}
\providecommand{\costTotalUSD}{}\renewcommand{\costTotalUSD}{8,600}

\providecommand{\cuzrActHoldFinal}{}\renewcommand{\cuzrActHoldFinal}{1.7}
\providecommand{\cuzrActHoldMlip}{}\renewcommand{\cuzrActHoldMlip}{5.1}
\providecommand{\cuzrActHoldSgte}{}\renewcommand{\cuzrActHoldSgte}{2.6}
\providecommand{\cuzrCongruentHigh}{}\renewcommand{\cuzrCongruentHigh}{106}
\providecommand{\cuzrEutectoidLow}{}\renewcommand{\cuzrEutectoidLow}{16}
\providecommand{\cuzrFitRelease}{}\renewcommand{\cuzrFitRelease}{\texttt{778cc6bfa}}
\providecommand{\cuzrInvFinal}{}\renewcommand{\cuzrInvFinal}{9}
\providecommand{\cuzrInvMlip}{}\renewcommand{\cuzrInvMlip}{0}
\providecommand{\cuzrInvPrereg}{}\renewcommand{\cuzrInvPrereg}{4}
\providecommand{\cuzrInvSgte}{}\renewcommand{\cuzrInvSgte}{12}
\providecommand{\cuzrInvTotal}{}\renewcommand{\cuzrInvTotal}{13}
\providecommand{\cuzrKemsDiffHi}{}\renewcommand{\cuzrKemsDiffHi}{8.0}
\providecommand{\cuzrKemsDiffLo}{}\renewcommand{\cuzrKemsDiffLo}{3.7}
\providecommand{\cuzrKemsSigmaHi}{}\renewcommand{\cuzrKemsSigmaHi}{0.26}
\providecommand{\cuzrKemsSigmaLo}{}\renewcommand{\cuzrKemsSigmaLo}{0.08}
\providecommand{\cuzrZpfHoldFinal}{}\renewcommand{\cuzrZpfHoldFinal}{0.11}
\providecommand{\cuzrZpfHoldMlip}{}\renewcommand{\cuzrZpfHoldMlip}{1.7}
\providecommand{\cuzrZpfHoldSgte}{}\renewcommand{\cuzrZpfHoldSgte}{0.12}
\providecommand{\doiAssessed}{}\renewcommand{\doiAssessed}{\href{https://doi.org/10.5281/zenodo.23735996}{10.5281/zenodo.23735996}}

\providecommand{\doiBibliography}{}\renewcommand{\doiBibliography}{\href{https://doi.org/10.5281/zenodo.23736021}{10.5281/zenodo.23736021}}
\providecommand{\doiCodeMcp}{}\renewcommand{\doiCodeMcp}{\href{https://doi.org/10.5281/zenodo.23748397}{10.5281/zenodo.23748397}}

\providecommand{\doiDftBenchmark}{}\renewcommand{\doiDftBenchmark}{\href{https://doi.org/10.5281/zenodo.23735943}{10.5281/zenodo.23735943}}
\providecommand{\doiEspei}{}\renewcommand{\doiEspei}{\href{https://doi.org/10.5281/zenodo.23735788}{10.5281/zenodo.23735788}}
\providecommand{\doiExamples}{}\renewcommand{\doiExamples}{\href{https://doi.org/10.5281/zenodo.23736043}{10.5281/zenodo.23736043}}
\providecommand{\doiHulls}{}\renewcommand{\doiHulls}{\href{https://doi.org/10.5281/zenodo.23735891}{10.5281/zenodo.23735891}}
\providecommand{\doiLiterature}{}\renewcommand{\doiLiterature}{\href{https://doi.org/10.5281/zenodo.23735569}{10.5281/zenodo.23735569}}
\providecommand{\doiPhaseDiagrams}{}\renewcommand{\doiPhaseDiagrams}{\href{https://doi.org/10.5281/zenodo.23735859}{10.5281/zenodo.23735859}}
\providecommand{\doiPhaseRegistry}{}\renewcommand{\doiPhaseRegistry}{\href{https://doi.org/10.5281/zenodo.23735970}{10.5281/zenodo.23735970}}

\providecommand{\doiUmbrella}{}\renewcommand{\doiUmbrella}{\href{https://doi.org/10.5281/zenodo.23736065}{10.5281/zenodo.23736065}}
\providecommand{\espeiReadyOneIn}{}\renewcommand{\espeiReadyOneIn}{seven}
\providecommand{\espeiVersion}{}\renewcommand{\espeiVersion}{0.9.1}
\providecommand{\floorHmCal}{}\renewcommand{\floorHmCal}{3.2}
\providecommand{\floorHmEmf}{}\renewcommand{\floorHmEmf}{4.6}

\providecommand{\floorHmNotPrinted}{}\renewcommand{\floorHmNotPrinted}{3.5}

\providecommand{\floorModelHm}{}\renewcommand{\floorModelHm}{3.9}
\providecommand{\floorTinv}{}\renewcommand{\floorTinv}{11}
\providecommand{\floorXphase}{}\renewcommand{\floorXphase}{0.011}
\providecommand{\hfConfigList}{}\renewcommand{\hfConfigList}{\texttt{literature\_values}, \texttt{papers}, \texttt{experimental\_context}, \texttt{dft\_parity}, \texttt{phase\_diagram\_invariants}, \texttt{phase\_diagram\_index}, \texttt{hull\_vertices}, \texttt{phases}, \texttt{structure\_types}, \texttt{assessed\_phase\_assemblages}, \texttt{bibliography\_papers}, \texttt{bibliography\_edges}}
\providecommand{\invValExpInSamplePct}{}\renewcommand{\invValExpInSamplePct}{100}
\providecommand{\invValExpMAE}{}\renewcommand{\invValExpMAE}{12.9}
\providecommand{\invValExpMedian}{}\renewcommand{\invValExpMedian}{0.1}
\providecommand{\invValExpSigma}{}\renewcommand{\invValExpSigma}{5.9}
\providecommand{\invValExpWithinTen}{}\renewcommand{\invValExpWithinTen}{73}

\providecommand{\invValModelSigma}{}\renewcommand{\invValModelSigma}{2.5}

\providecommand{\maeKim}{}\renewcommand{\maeKim}{3.8}
\providecommand{\maeMP}{}\renewcommand{\maeMP}{9.6}
\providecommand{\maeMPCalorimetry}{}\renewcommand{\maeMPCalorimetry}{9.0}
\providecommand{\maeMPEmf}{}\renewcommand{\maeMPEmf}{12.3}
\providecommand{\maeMPIntermetallic}{}\renewcommand{\maeMPIntermetallic}{8.3}
\providecommand{\maeMPMedian}{}\renewcommand{\maeMPMedian}{9.8}
\providecommand{\maeMPNoOutliers}{}\renewcommand{\maeMPNoOutliers}{9.2}
\providecommand{\maeMPNonmetal}{}\renewcommand{\maeMPNonmetal}{14.9}
\providecommand{\maeMPWithModel}{}\renewcommand{\maeMPWithModel}{8.5}
\providecommand{\maeMPWithUnsigned}{}\renewcommand{\maeMPWithUnsigned}{9.6}
\providecommand{\maeOQMD}{}\renewcommand{\maeOQMD}{8.3}
\providecommand{\maeOQMDCalorimetry}{}\renewcommand{\maeOQMDCalorimetry}{7.6}
\providecommand{\maeOQMDEmf}{}\renewcommand{\maeOQMDEmf}{9.6}
\providecommand{\maeOQMDIntermetallic}{}\renewcommand{\maeOQMDIntermetallic}{7.6}
\providecommand{\maeOQMDIntermetallicEV}{}\renewcommand{\maeOQMDIntermetallicEV}{0.079}
\providecommand{\maeOQMDMedian}{}\renewcommand{\maeOQMDMedian}{8.3}
\providecommand{\maeOQMDNoOutliers}{}\renewcommand{\maeOQMDNoOutliers}{7.9}
\providecommand{\maeOQMDNonmetal}{}\renewcommand{\maeOQMDNonmetal}{11.5}
\providecommand{\maeOQMDWithModel}{}\renewcommand{\maeOQMDWithModel}{7.9}

\providecommand{\maeOQMDeV}{}\renewcommand{\maeOQMDeV}{0.086}
\providecommand{\mcpToolList}{}\renewcommand{\mcpToolList}{\texttt{search\_systems}, \texttt{get\_system\_data}, \texttt{get\_invariant\_reactions}, \texttt{get\_formation\_enthalpy}, \texttt{get\_phase\_diagram}, \texttt{list\_tdb\_files}, \texttt{get\_tdb}, \texttt{get\_paper}, \texttt{search\_papers}, \texttt{get\_paper\_relations}, \texttt{get\_assessment\_evidence}, \texttt{compare\_to\_dft}, \texttt{get\_phase}, \texttt{search\_phases}, \texttt{get\_structure\_type}}
\providecommand{\medKim}{}\renewcommand{\medKim}{1.2}
\providecommand{\nAccepted}{}\renewcommand{\nAccepted}{66,551}
\providecommand{\nAsmBoundaryDashed}{}\renewcommand{\nAsmBoundaryDashed}{1,028}
\providecommand{\nAsmBoundaryMixed}{}\renewcommand{\nAsmBoundaryMixed}{1,292}
\providecommand{\nAsmBoundarySolid}{}\renewcommand{\nAsmBoundarySolid}{20,782}
\providecommand{\nAsmCentroid}{}\renewcommand{\nAsmCentroid}{21,616}
\providecommand{\nAsmDiagrams}{}\renewcommand{\nAsmDiagrams}{2,178}
\providecommand{\nAsmDiagramsDashed}{}\renewcommand{\nAsmDiagramsDashed}{1,021}
\providecommand{\nAsmEspeiDatasets}{}\renewcommand{\nAsmEspeiDatasets}{1,068}
\providecommand{\nAsmEspeiExcluded}{}\renewcommand{\nAsmEspeiExcluded}{13,252}
\providecommand{\nAsmEspeiPoints}{}\renewcommand{\nAsmEspeiPoints}{9,877}

\providecommand{\nAsmPoints}{}\renewcommand{\nAsmPoints}{23,129}
\providecommand{\nAsmPole}{}\renewcommand{\nAsmPole}{1,513}
\providecommand{\nAsmSgteAgree}{}\renewcommand{\nAsmSgteAgree}{17,876}
\providecommand{\nAsmSgteComparable}{}\renewcommand{\nAsmSgteComparable}{21,488}
\providecommand{\nAsmSgteComparisons}{}\renewcommand{\nAsmSgteComparisons}{30,769}

\providecommand{\nAsmSgteNotComparable}{}\renewcommand{\nAsmSgteNotComparable}{9,281}
\providecommand{\nAsmSgtePointsAgreeAny}{}\renewcommand{\nAsmSgtePointsAgreeAny}{8,333}
\providecommand{\nAsmSgtePointsComparable}{}\renewcommand{\nAsmSgtePointsComparable}{9,425}

\providecommand{\nAsmTieLines}{}\renewcommand{\nAsmTieLines}{11,961}
\providecommand{\nAsmTwoPhase}{}\renewcommand{\nAsmTwoPhase}{18,490}

\providecommand{\nAuditsWord}{}\renewcommand{\nAuditsWord}{six}
\providecommand{\nBasisInferred}{}\renewcommand{\nBasisInferred}{2,117}
\providecommand{\nBasisUnstatedFormation}{}\renewcommand{\nBasisUnstatedFormation}{2,255}
\providecommand{\nBenchCalls}{}\renewcommand{\nBenchCalls}{525}
\providecommand{\nBenchDocs}{}\renewcommand{\nBenchDocs}{20}
\providecommand{\nBenchOracleMetrics}{}\renewcommand{\nBenchOracleMetrics}{1,528}
\providecommand{\nBenchPages}{}\renewcommand{\nBenchPages}{86}
\providecommand{\nBibCitedSources}{}\renewcommand{\nBibCitedSources}{3,578}
\providecommand{\nBibEdges}{}\renewcommand{\nBibEdges}{60,235}

\providecommand{\nBibHandbookSources}{}\renewcommand{\nBibHandbookSources}{2,283}
\providecommand{\nBibLeads}{}\renewcommand{\nBibLeads}{2,806}

\providecommand{\nBibPapers}{}\renewcommand{\nBibPapers}{9,570}

\providecommand{\nBibRead}{}\renewcommand{\nBibRead}{1,427}
\providecommand{\nBibTdbSources}{}\renewcommand{\nBibTdbSources}{1,886}
\providecommand{\nBibWithDOI}{}\renewcommand{\nBibWithDOI}{7,813}
\providecommand{\nBiblio}{}\renewcommand{\nBiblio}{4,882}
\providecommand{\nBinWithDiagram}{}\renewcommand{\nBinWithDiagram}{851}
\providecommand{\nBinWithDiagramValues}{}\renewcommand{\nBinWithDiagramValues}{50,038}

\providecommand{\nBlindExpertAgree}{}\renewcommand{\nBlindExpertAgree}{45}
\providecommand{\nBlindExpertAgreeEquiv}{}\renewcommand{\nBlindExpertAgreeEquiv}{47}

\providecommand{\nBlindExpertN}{}\renewcommand{\nBlindExpertN}{48}
\providecommand{\nBlindRuleAgree}{}\renewcommand{\nBlindRuleAgree}{28}

\providecommand{\nContextItems}{}\renewcommand{\nContextItems}{1,504}

\providecommand{\nCuZrAsmFit}{}\renewcommand{\nCuZrAsmFit}{7}
\providecommand{\nCuZrAsmPoints}{}\renewcommand{\nCuZrAsmPoints}{28}
\providecommand{\nCuZrAssessments}{}\renewcommand{\nCuZrAssessments}{11}
\providecommand{\nCuZrDemoDense}{}\renewcommand{\nCuZrDemoDense}{177}
\providecommand{\nDFT}{}\renewcommand{\nDFT}{6,150}

\providecommand{\nDatasetsWord}{}\renewcommand{\nDatasetsWord}{eight}
\providecommand{\nDftOutlier}{}\renewcommand{\nDftOutlier}{130}
\providecommand{\nDiagramSystems}{}\renewcommand{\nDiagramSystems}{1,234}
\providecommand{\nDiagrams}{}\renewcommand{\nDiagrams}{2,722}
\providecommand{\nDiagramsFailed}{}\renewcommand{\nDiagramsFailed}{52}
\providecommand{\nDroppedFields}{}\renewcommand{\nDroppedFields}{2,781}
\providecommand{\nDupGroups}{}\renewcommand{\nDupGroups}{3,330}
\providecommand{\nDupRemoved}{}\renewcommand{\nDupRemoved}{3,550}
\providecommand{\nElements}{}\renewcommand{\nElements}{79}

\providecommand{\nEspeiAssessedDatasets}{}\renewcommand{\nEspeiAssessedDatasets}{643}

\providecommand{\nEspeiDatasets}{}\renewcommand{\nEspeiDatasets}{6,453}
\providecommand{\nEspeiDatasetsAll}{}\renewcommand{\nEspeiDatasetsAll}{8,589}

\providecommand{\nEspeiEvalExcluded}{}\renewcommand{\nEspeiEvalExcluded}{387}
\providecommand{\nEspeiEvaluated}{}\renewcommand{\nEspeiEvaluated}{8,589}
\providecommand{\nEspeiExcluded}{}\renewcommand{\nEspeiExcluded}{24}
\providecommand{\nEspeiExpDatasets}{}\renewcommand{\nEspeiExpDatasets}{1,118}
\providecommand{\nEspeiInvDatasets}{}\renewcommand{\nEspeiInvDatasets}{1,546}
\providecommand{\nEspeiModelDatasets}{}\renewcommand{\nEspeiModelDatasets}{504}
\providecommand{\nEspeiNoPrintedSigma}{}\renewcommand{\nEspeiNoPrintedSigma}{558}

\providecommand{\nEspeiReadyApprox}{}\renewcommand{\nEspeiReadyApprox}{11,000}

\providecommand{\nEspeiSingleReader}{}\renewcommand{\nEspeiSingleReader}{9}
\providecommand{\nEspeiSystems}{}\renewcommand{\nEspeiSystems}{690}
\providecommand{\nEspeiValues}{}\renewcommand{\nEspeiValues}{12,513}

\providecommand{\nExclBasis}{}\renewcommand{\nExclBasis}{2,154}
\providecommand{\nExclCompConflictMP}{}\renewcommand{\nExclCompConflictMP}{3}
\providecommand{\nExclCompConflictOQMD}{}\renewcommand{\nExclCompConflictOQMD}{3}
\providecommand{\nExclDFT}{}\renewcommand{\nExclDFT}{1,545}
\providecommand{\nExclGateFlaggedMP}{}\renewcommand{\nExclGateFlaggedMP}{115}
\providecommand{\nExclGateFlaggedOQMD}{}\renewcommand{\nExclGateFlaggedOQMD}{118}
\providecommand{\nExclNoFormula}{}\renewcommand{\nExclNoFormula}{436}
\providecommand{\nExclOffStoichMP}{}\renewcommand{\nExclOffStoichMP}{7}
\providecommand{\nExclOffStoichOQMD}{}\renewcommand{\nExclOffStoichOQMD}{12}
\providecommand{\nExclRefStateMP}{}\renewcommand{\nExclRefStateMP}{45}
\providecommand{\nExclRefStateOQMD}{}\renewcommand{\nExclRefStateOQMD}{50}
\providecommand{\nExclStructUnmatchedMP}{}\renewcommand{\nExclStructUnmatchedMP}{1}
\providecommand{\nExclStructUnmatchedOQMD}{}\renewcommand{\nExclStructUnmatchedOQMD}{10}
\providecommand{\nExp}{}\renewcommand{\nExp}{36,288}

\providecommand{\nExprTermGibbs}{}\renewcommand{\nExprTermGibbs}{555}

\providecommand{\nExprTermRecords}{}\renewcommand{\nExprTermRecords}{838}
\providecommand{\nFloorCleanedClasses}{}\renewcommand{\nFloorCleanedClasses}{11}
\providecommand{\nFloorMinPairs}{}\renewcommand{\nFloorMinPairs}{20}
\providecommand{\nFloorMinPairsCleaned}{}\renewcommand{\nFloorMinPairsCleaned}{5}
\providecommand{\nFloorMinStatesCleaned}{}\renewcommand{\nFloorMinStatesCleaned}{3}
\providecommand{\nFloorPriorClasses}{}\renewcommand{\nFloorPriorClasses}{46}
\providecommand{\nFormUnitNotNormalized}{}\renewcommand{\nFormUnitNotNormalized}{572}

\providecommand{\nGateAcceptedAfter}{}\renewcommand{\nGateAcceptedAfter}{65,284}
\providecommand{\nGateAcceptedBefore}{}\renewcommand{\nGateAcceptedBefore}{76,600}

\providecommand{\nGateCompConflictsLabel}{}\renewcommand{\nGateCompConflictsLabel}{74}
\providecommand{\nGateCompFromFormula}{}\renewcommand{\nGateCompFromFormula}{419}

\providecommand{\nGateCorrected}{}\renewcommand{\nGateCorrected}{15,528}

\providecommand{\nGateDemoted}{}\renewcommand{\nGateDemoted}{9,378}

\providecommand{\nGateDupMerged}{}\renewcommand{\nGateDupMerged}{3,998}

\providecommand{\nGateInvClassConflict}{}\renewcommand{\nGateInvClassConflict}{67}
\providecommand{\nGateInvClassFilled}{}\renewcommand{\nGateInvClassFilled}{3,022}
\providecommand{\nGateInvDuplicate}{}\renewcommand{\nGateInvDuplicate}{179}
\providecommand{\nGateInvPhaseRule}{}\renewcommand{\nGateInvPhaseRule}{303}
\providecommand{\nGateInvPhasesFromReaction}{}\renewcommand{\nGateInvPhasesFromReaction}{219}
\providecommand{\nGateInvTempCheck}{}\renewcommand{\nGateInvTempCheck}{5}
\providecommand{\nGateMixToForm}{}\renewcommand{\nGateMixToForm}{286}

\providecommand{\nGatePageUnitFixes}{}\renewcommand{\nGatePageUnitFixes}{211}
\providecommand{\nGatePageValueFixes}{}\renewcommand{\nGatePageValueFixes}{108}
\providecommand{\nGatePhaseListFixes}{}\renewcommand{\nGatePhaseListFixes}{916}
\providecommand{\nGateRecordsIn}{}\renewcommand{\nGateRecordsIn}{77,540}
\providecommand{\nGateRecordsOut}{}\renewcommand{\nGateRecordsOut}{73,542}

\providecommand{\nGateSignFromPage}{}\renewcommand{\nGateSignFromPage}{10}
\providecommand{\nGateSmallFormation}{}\renewcommand{\nGateSmallFormation}{3}
\providecommand{\nGateStatePhasesMissing}{}\renewcommand{\nGateStatePhasesMissing}{8,944}
\providecommand{\nGateStatePhasesPage}{}\renewcommand{\nGateStatePhasesPage}{16}
\providecommand{\nGateStatePhasesQuote}{}\renewcommand{\nGateStatePhasesQuote}{264}
\providecommand{\nGateStatePhasesRecovered}{}\renewcommand{\nGateStatePhasesRecovered}{280}
\providecommand{\nGateSystemReassigned}{}\renewcommand{\nGateSystemReassigned}{327}
\providecommand{\nGateTableCelsius}{}\renewcommand{\nGateTableCelsius}{137}
\providecommand{\nGateTempCorrected}{}\renewcommand{\nGateTempCorrected}{2,933}
\providecommand{\nGateTempStdRef}{}\renewcommand{\nGateTempStdRef}{2,790}

\providecommand{\nGateUnitPrefix}{}\renewcommand{\nGateUnitPrefix}{11}

\providecommand{\nGateVisualChecks}{}\renewcommand{\nGateVisualChecks}{3,303}
\providecommand{\nGateVisualEntries}{}\renewcommand{\nGateVisualEntries}{3,262}
\providecommand{\nGenTable}{}\renewcommand{\nGenTable}{41,106}
\providecommand{\nGibbsCoefT}{}\renewcommand{\nGibbsCoefT}{0}

\providecommand{\nHfConfigs}{}\renewcommand{\nHfConfigs}{12}
\providecommand{\nHmFormRecords}{}\renewcommand{\nHmFormRecords}{9,137}
\providecommand{\nHullMPAgree}{}\renewcommand{\nHullMPAgree}{3,827}
\providecommand{\nHullMPElements}{}\renewcommand{\nHullMPElements}{88}
\providecommand{\nHullMPSystems}{}\renewcommand{\nHullMPSystems}{3,828}
\providecommand{\nHullMPVertices}{}\renewcommand{\nHullMPVertices}{12,829}
\providecommand{\nHullMPWithCompound}{}\renewcommand{\nHullMPWithCompound}{2,084}
\providecommand{\nHullOQMDElements}{}\renewcommand{\nHullOQMDElements}{89}
\providecommand{\nHullOQMDSystems}{}\renewcommand{\nHullOQMDSystems}{3,916}
\providecommand{\nHullOQMDVertices}{}\renewcommand{\nHullOQMDVertices}{15,314}
\providecommand{\nHullOQMDWithCompound}{}\renewcommand{\nHullOQMDWithCompound}{2,256}
\providecommand{\nHullSGTEAssessments}{}\renewcommand{\nHullSGTEAssessments}{2,695}
\providecommand{\nHullSGTESystems}{}\renewcommand{\nHullSGTESystems}{1,223}
\providecommand{\nHullVertices}{}\renewcommand{\nHullVertices}{28,143}
\providecommand{\nInvTable}{}\renewcommand{\nInvTable}{26,693}
\providecommand{\nInvValBeyond}{}\renewcommand{\nInvValBeyond}{86}

\providecommand{\nInvValCandidates}{}\renewcommand{\nInvValCandidates}{10,207}
\providecommand{\nInvValEutectic}{}\renewcommand{\nInvValEutectic}{2,014}
\providecommand{\nInvValEutectoid}{}\renewcommand{\nInvValEutectoid}{281}
\providecommand{\nInvValExp}{}\renewcommand{\nInvValExp}{3,616}
\providecommand{\nInvValExpInSample}{}\renewcommand{\nInvValExpInSample}{3,599}
\providecommand{\nInvValExpSystems}{}\renewcommand{\nInvValExpSystems}{401}
\providecommand{\nInvValMapped}{}\renewcommand{\nInvValMapped}{8,770}
\providecommand{\nInvValModel}{}\renewcommand{\nInvValModel}{3,380}

\providecommand{\nInvValPeritectic}{}\renewcommand{\nInvValPeritectic}{1,135}
\providecommand{\nInvValPeritectoid}{}\renewcommand{\nInvValPeritectoid}{186}
\providecommand{\nInvariantHits}{}\renewcommand{\nInvariantHits}{24,849}
\providecommand{\nInvariants}{}\renewcommand{\nInvariants}{15,336}
\providecommand{\nKimCandidates}{}\renewcommand{\nKimCandidates}{1,436}
\providecommand{\nKimDiffer}{}\renewcommand{\nKimDiffer}{19}
\providecommand{\nKimFormulas}{}\renewcommand{\nKimFormulas}{606}
\providecommand{\nKimOverlap}{}\renewcommand{\nKimOverlap}{160}
\providecommand{\nLatticeABC}{}\renewcommand{\nLatticeABC}{8,329}
\providecommand{\nLatticeAngstrom}{}\renewcommand{\nLatticeAngstrom}{7,690}
\providecommand{\nLatticeNoUnit}{}\renewcommand{\nLatticeNoUnit}{639}
\providecommand{\nMPDefault}{}\renewcommand{\nMPDefault}{1,070}
\providecommand{\nMPWithModel}{}\renewcommand{\nMPWithModel}{3,106}
\providecommand{\nMagnitudeSuspect}{}\renewcommand{\nMagnitudeSuspect}{499}
\providecommand{\nMcpTools}{}\renewcommand{\nMcpTools}{15}
\providecommand{\nModel}{}\renewcommand{\nModel}{26,153}

\providecommand{\nNiThreeTiPaper}{}\renewcommand{\nNiThreeTiPaper}{1}
\providecommand{\nNiThreeTiPaperWord}{}\renewcommand{\nNiThreeTiPaperWord}{paper}
\providecommand{\nNiThreeTiTdb}{}\renewcommand{\nNiThreeTiTdb}{13}
\providecommand{\nOQMDDefault}{}\renewcommand{\nOQMDDefault}{1,107}
\providecommand{\nOQMDWithModel}{}\renewcommand{\nOQMDWithModel}{3,352}
\providecommand{\nOutlierUnexplained}{}\renewcommand{\nOutlierUnexplained}{42}
\providecommand{\nOutlierUnexplainedDefault}{}\renewcommand{\nOutlierUnexplainedDefault}{25}
\providecommand{\nOutliersExamined}{}\renewcommand{\nOutliersExamined}{74}
\providecommand{\nPagesNulled}{}\renewcommand{\nPagesNulled}{2}
\providecommand{\nPairsHmCal}{}\renewcommand{\nPairsHmCal}{80}

\providecommand{\nPapers}{}\renewcommand{\nPapers}{1,321}
\providecommand{\nParityCompounds}{}\renewcommand{\nParityCompounds}{965}
\providecommand{\nParityExp}{}\renewcommand{\nParityExp}{1,364}
\providecommand{\nParityMP}{}\renewcommand{\nParityMP}{3,679}
\providecommand{\nParityModel}{}\renewcommand{\nParityModel}{2,543}
\providecommand{\nParityOQMD}{}\renewcommand{\nParityOQMD}{3,967}
\providecommand{\nParityPoints}{}\renewcommand{\nParityPoints}{7,646}
\providecommand{\nParityRecords}{}\renewcommand{\nParityRecords}{4,067}
\providecommand{\nParityUnspec}{}\renewcommand{\nParityUnspec}{160}
\providecommand{\nParquetColumns}{}\renewcommand{\nParquetColumns}{165}
\providecommand{\nPerFormulaFixed}{}\renewcommand{\nPerFormulaFixed}{26}
\providecommand{\nPlausChecked}{}\renewcommand{\nPlausChecked}{6,476}
\providecommand{\nPlausCheckedExp}{}\renewcommand{\nPlausCheckedExp}{3,006}
\providecommand{\nPlausFail}{}\renewcommand{\nPlausFail}{785}

\providecommand{\nPlausPass}{}\renewcommand{\nPlausPass}{5,691}
\providecommand{\nPlausPassExp}{}\renewcommand{\nPlausPassExp}{2,551}

\providecommand{\nPlausTdb}{}\renewcommand{\nPlausTdb}{438}
\providecommand{\nPrecisionGone}{}\renewcommand{\nPrecisionGone}{12}

\providecommand{\nPrecisionSample}{}\renewcommand{\nPrecisionSample}{300}

\providecommand{\nPrecisionSurvive}{}\renewcommand{\nPrecisionSurvive}{288}
\providecommand{\nPrecisionSurviveAccepted}{}\renewcommand{\nPrecisionSurviveAccepted}{260}
\providecommand{\nPressureAssumed}{}\renewcommand{\nPressureAssumed}{73,537}
\providecommand{\nPressureNotOneAtm}{}\renewcommand{\nPressureNotOneAtm}{5}
\providecommand{\nPrimarySources}{}\renewcommand{\nPrimarySources}{3,578}
\providecommand{\nPropCodesCleaned}{}\renewcommand{\nPropCodesCleaned}{528}
\providecommand{\nPropertyCodes}{}\renewcommand{\nPropertyCodes}{96}
\providecommand{\nPropertyGroups}{}\renewcommand{\nPropertyGroups}{11}
\providecommand{\nRAMulti}{}\renewcommand{\nRAMulti}{69,253}
\providecommand{\nRANotClassified}{}\renewcommand{\nRANotClassified}{400}
\providecommand{\nRASingle}{}\renewcommand{\nRASingle}{1,342}
\providecommand{\nRASingleText}{}\renewcommand{\nRASingleText}{2,547}
\providecommand{\nRThreeFixIntroduced}{}\renewcommand{\nRThreeFixIntroduced}{12}
\providecommand{\nRThreeResample}{}\renewcommand{\nRThreeResample}{155}
\providecommand{\nRThreeScopes}{}\renewcommand{\nRThreeScopes}{7}
\providecommand{\nRecoveryChecks}{}\renewcommand{\nRecoveryChecks}{888}

\providecommand{\nRecoveryFinalChecks}{}\renewcommand{\nRecoveryFinalChecks}{114}
\providecommand{\nRecoveryNumerical}{}\renewcommand{\nRecoveryNumerical}{23}
\providecommand{\nRecoveryOriginalFailed}{}\renewcommand{\nRecoveryOriginalFailed}{54}
\providecommand{\nRecoveryPromoted}{}\renewcommand{\nRecoveryPromoted}{0}
\providecommand{\nRecoveryRemaining}{}\renewcommand{\nRecoveryRemaining}{44}
\providecommand{\nRecoverySourceCases}{}\renewcommand{\nRecoverySourceCases}{21}
\providecommand{\nRecoveryVariants}{}\renewcommand{\nRecoveryVariants}{57}

\providecommand{\nRegDOIs}{}\renewcommand{\nRegDOIs}{3,070}
\providecommand{\nRegGas}{}\renewcommand{\nRegGas}{231}
\providecommand{\nRegLiquid}{}\renewcommand{\nRegLiquid}{1,304}

\providecommand{\nRegModels}{}\renewcommand{\nRegModels}{22,556}
\providecommand{\nRegNames}{}\renewcommand{\nRegNames}{43,269}
\providecommand{\nRegNamesCorpus}{}\renewcommand{\nRegNamesCorpus}{10,916}
\providecommand{\nRegNamesMP}{}\renewcommand{\nRegNamesMP}{677}
\providecommand{\nRegNamesOQMD}{}\renewcommand{\nRegNamesOQMD}{677}
\providecommand{\nRegNamesPaper}{}\renewcommand{\nRegNamesPaper}{8,368}
\providecommand{\nRegNamesTdb}{}\renewcommand{\nRegNamesTdb}{22,556}
\providecommand{\nRegPhases}{}\renewcommand{\nRegPhases}{10,430}
\providecommand{\nRegReview}{}\renewcommand{\nRegReview}{1,236}
\providecommand{\nRegStructureTypes}{}\renewcommand{\nRegStructureTypes}{1,563}
\providecommand{\nRegStructureTypesUsed}{}\renewcommand{\nRegStructureTypesUsed}{711}
\providecommand{\nRegSystems}{}\renewcommand{\nRegSystems}{1,241}
\providecommand{\nRegTierA}{}\renewcommand{\nRegTierA}{6,228}
\providecommand{\nRegTierAInterstitial}{}\renewcommand{\nRegTierAInterstitial}{23}
\providecommand{\nRegTierB}{}\renewcommand{\nRegTierB}{154}
\providecommand{\nRegTierC}{}\renewcommand{\nRegTierC}{1,248}

\providecommand{\nRegUnresolved}{}\renewcommand{\nRegUnresolved}{29}
\providecommand{\nRegWithRefs}{}\renewcommand{\nRegWithRefs}{9,903}
\providecommand{\nRegWithStability}{}\renewcommand{\nRegWithStability}{8,549}
\providecommand{\nRegWithValues}{}\renewcommand{\nRegWithValues}{4,974}
\providecommand{\nResBeyondMP}{}\renewcommand{\nResBeyondMP}{1}
\providecommand{\nResBeyondOQMD}{}\renewcommand{\nResBeyondOQMD}{1}

\providecommand{\nSigmaMethodCalibrated}{}\renewcommand{\nSigmaMethodCalibrated}{47,252}

\providecommand{\nSigmaMethodPrior}{}\renewcommand{\nSigmaMethodPrior}{12,904}
\providecommand{\nSigmaPrinted}{}\renewcommand{\nSigmaPrinted}{5,852}

\providecommand{\nSignVerified}{}\renewcommand{\nSignVerified}{11}
\providecommand{\nSourceDisputed}{}\renewcommand{\nSourceDisputed}{512}
\providecommand{\nSplitTest}{}\renewcommand{\nSplitTest}{8,473}
\providecommand{\nSplitTestPapers}{}\renewcommand{\nSplitTestPapers}{145}
\providecommand{\nSplitTrain}{}\renewcommand{\nSplitTrain}{58,315}
\providecommand{\nSplitTrainPapers}{}\renewcommand{\nSplitTrainPapers}{1,053}
\providecommand{\nSplitVal}{}\renewcommand{\nSplitVal}{6,754}
\providecommand{\nSplitValPapers}{}\renewcommand{\nSplitValPapers}{123}

\providecommand{\nSystemMismatch}{}\renewcommand{\nSystemMismatch}{424}
\providecommand{\nSystems}{}\renewcommand{\nSystems}{1,474}
\providecommand{\nSystemsBinary}{}\renewcommand{\nSystemsBinary}{977}
\providecommand{\nSystemsHigher}{}\renewcommand{\nSystemsHigher}{59}
\providecommand{\nSystemsTernary}{}\renewcommand{\nSystemsTernary}{365}
\providecommand{\nSystemsUnary}{}\renewcommand{\nSystemsUnary}{73}
\providecommand{\nTKRecords}{}\renewcommand{\nTKRecords}{21,703}
\providecommand{\nTKRelabelled}{}\renewcommand{\nTKRelabelled}{15,971}
\providecommand{\nTKTerms}{}\renewcommand{\nTKTerms}{16}
\providecommand{\nTKVariants}{}\renewcommand{\nTKVariants}{187}
\providecommand{\nTdbFiles}{}\renewcommand{\nTdbFiles}{2,774}
\providecommand{\nTdbLinked}{}\renewcommand{\nTdbLinked}{12,833}

\providecommand{\nTernFromDiagramsApprox}{}\renewcommand{\nTernFromDiagramsApprox}{9,700}

\providecommand{\nTernFromUnionApprox}{}\renewcommand{\nTernFromUnionApprox}{34,000}

\providecommand{\nText}{}\renewcommand{\nText}{5,743}

\providecommand{\nTierFlash}{}\renewcommand{\nTierFlash}{5,856}
\providecommand{\nTierFlashClaude}{}\renewcommand{\nTierFlashClaude}{59,435}
\providecommand{\nTierJudge}{}\renewcommand{\nTierJudge}{198}
\providecommand{\nTierPro}{}\renewcommand{\nTierPro}{8,053}

\providecommand{\nUnsigned}{}\renewcommand{\nUnsigned}{42}
\providecommand{\nUnsignedFixed}{}\renewcommand{\nUnsignedFixed}{6}
\providecommand{\nUnspec}{}\renewcommand{\nUnspec}{4,951}

\providecommand{\nValues}{}\renewcommand{\nValues}{73,542}
\providecommand{\nValuesBinary}{}\renewcommand{\nValuesBinary}{58,050}
\providecommand{\nValuesHigher}{}\renewcommand{\nValuesHigher}{869}
\providecommand{\nValuesTernary}{}\renewcommand{\nValuesTernary}{11,853}
\providecommand{\nValuesUnary}{}\renewcommand{\nValuesUnary}{2,770}
\providecommand{\nZenodoDataRecords}{}\renewcommand{\nZenodoDataRecords}{9}
\providecommand{\nZenodoRecords}{}\renewcommand{\nZenodoRecords}{10}
\providecommand{\pctRAMulti}{}\renewcommand{\pctRAMulti}{94}

\providecommand{\pearsonMPChange}{}\renewcommand{\pearsonMPChange}{stays at 0.94}

\providecommand{\pilotRequests}{}\renewcommand{\pilotRequests}{127}
\providecommand{\pilotRowsFlash}{}\renewcommand{\pilotRowsFlash}{388}
\providecommand{\pilotRowsOpus}{}\renewcommand{\pilotRowsOpus}{553}
\providecommand{\pilotVerbatimFlash}{}\renewcommand{\pilotVerbatimFlash}{306}
\providecommand{\pilotVerbatimOpus}{}\renewcommand{\pilotVerbatimOpus}{546}

\providecommand{\precisionExportDate}{}\renewcommand{\precisionExportDate}{22~October~2026}
\providecommand{\rThreeFinalWrongFive}{}\renewcommand{\rThreeFinalWrongFive}{2}
\providecommand{\rThreeFinalWrongFour}{}\renewcommand{\rThreeFinalWrongFour}{0}
\providecommand{\rThreeFinalWrongOne}{}\renewcommand{\rThreeFinalWrongOne}{1}
\providecommand{\rThreeFinalWrongSix}{}\renewcommand{\rThreeFinalWrongSix}{1}
\providecommand{\rThreeFinalWrongThree}{}\renewcommand{\rThreeFinalWrongThree}{3}
\providecommand{\rThreeFinalWrongTwo}{}\renewcommand{\rThreeFinalWrongTwo}{0}
\providecommand{\rThreeFirstWrongFive}{}\renewcommand{\rThreeFirstWrongFive}{11}

\providecommand{\rThreeNFive}{}\renewcommand{\rThreeNFive}{25}
\providecommand{\rThreeNFiveB}{}\renewcommand{\rThreeNFiveB}{15}
\providecommand{\rThreeNFour}{}\renewcommand{\rThreeNFour}{20}
\providecommand{\rThreeNOne}{}\renewcommand{\rThreeNOne}{25}
\providecommand{\rThreeNSix}{}\renewcommand{\rThreeNSix}{30}
\providecommand{\rThreeNThree}{}\renewcommand{\rThreeNThree}{25}
\providecommand{\rThreeNTwo}{}\renewcommand{\rThreeNTwo}{15}
\providecommand{\rThreePhasesOnPageFiveB}{}\renewcommand{\rThreePhasesOnPageFiveB}{12}

\providecommand{\rmseRatioMP}{}\renewcommand{\rmseRatioMP}{1.6}
\providecommand{\rmseRatioOQMD}{}\renewcommand{\rmseRatioOQMD}{1.7}

\providecommand{\sigLit}{}\renewcommand{\sigLit}{7.2}
\providecommand{\sigMP}{}\renewcommand{\sigMP}{8.6}
\providecommand{\sigOQMD}{}\renewcommand{\sigOQMD}{7.5}
\providecommand{\sigmaMlipFit}{}\renewcommand{\sigmaMlipFit}{9.7}
\providecommand{\sigmaMlipNow}{}\renewcommand{\sigmaMlipNow}{9.3}
\providecommand{\spearmanMP}{}\renewcommand{\spearmanMP}{0.86}
\providecommand{\spearmanOQMD}{}\renewcommand{\spearmanOQMD}{0.91}
\providecommand{\tabAsmBasisBody}{}\renewcommand{\tabAsmBasisBody}{deterministic adjacency & 8,035 \\
reader consensus & 6,793 \\
deterministic printed label & 3,205 \\
adjudicator & 2,803 \\
deterministic phase rules & 2,186 \\
expert reviewed & 87 \\
dft hull corroborated & 20 \\}
\providecommand{\tabAuditBody}{}\renewcommand{\tabAuditBody}{Elements \& compositions (A1) & accepted literature records; 110, seed 20261024 & 4/110 (3.6\%; 1.4--9.0) & 17/110 (15.5\%; 9.9--23.4) & 8: at.\% divided twice; header names wrong element; binary filed under ternary & 1.82\%\newline all 4.55\% \\
Phases \& mapping (A2) & registry, Assemblages and literature phase labels; 300, seed 20261024 & pooled 9/300 (3.0\%; 1.6--5.6)\newline registry: 5/100 (5.0\%)\newline Assemblages labels: 2/100 (2.0\%)\newline lit. model\_phase: 2/100 (2.0\%) & same (no metadata errors) & 8: $\varepsilon$$\to$BCC\_A2; $\beta$PtZr (B2)$\to$PTZR\_B33; first sublattice dropped & 0.33\%\newline all 0.33\% \\
Units, conditions \& uncertainties (A3) & accepted thermochemical records; 110, seed 20261027 & 12/110 (10.9\%; 6.4--18.1) & 20/110 (18.2\%; 12.1--26.4) & 10: H(T)$-$H(298) stored at 298 K; per-formula $\Delta$H as per-atom; unsigned $\Delta$H (Co17Y2) & 9.09\%\newline all 23.64\% \\
DFT pairing \& hulls (A4) & default-subset DFT--experiment pairs; 110, seed 424242 & 4/110 (3.6\%; 1.4--9.0) & same (no metadata errors) & 6: wrong polymorph (B2$\to$Cm); per-formula/per-atom slip; off-stoich. vs stoich. DFT & 2.73\%\newline all 4.55\% \\
Figures (A5) & parity pairs and invariant matches behind figures; 200, seed 20261024 & 0/200 (0.0\%; 0.0--1.9) & same (no metadata errors) & 10: duplicate invariant rows; page no. parsed as year; dataset numbering differs & 0.0\%\newline all 0.0\% \\
Paper claims (A6) & literature records vs record-level claims; 100, seed 20261024 & 3/100 (3.0\%; 1.0--8.5) & same (no metadata errors) & 8: superseded $\sigma$ rule in prose; hand-typed counts; units break Usage claim & 1.0\%\newline all 6.0\% \\}
\providecommand{\tabBenchBody}{}\renewcommand{\tabBenchBody}{table cells exact & pdfplumber (det.) & 0.76 & -- & -- & -- \\
table cells exact & OCR cells (det.) & 0.45 & 0.45 & 0.16 & 0.17 \\
invariant rows recall & Camelot (det.) & 0.62 & -- & -- & -- \\
invariant rows recall & AI reader & 0.99 & 0.99 & 0.85 & 0.85 \\
property records recall & Camelot (det.) & 0.94 & -- & -- & -- \\
property records recall & AI reader$^{a}$ & 0.01 & 0.00 & 0.01 & 0.00 \\
boundary recall & tracer (det.) & 0.58 & 0.54 & 0.74 & 0.74 \\
field recall & tracer (det.) & 0.26 & 0.17 & 0.07 & 0.07 \\
field recall & AI reader & 0.45 & 0.36 & 0.39 & 0.37 \\
structure labels & AI reader$^{a}$ & 0.00 & 0.00 & 0.00 & 0.00 \\
text values recall & regex (det.) & 0.97 & 0.86 & 0.74 & 0.74 \\
text values recall & AI reader & 0.53 & 0.34 & 0.31 & 0.27 \\
text values precision & AI reader & 0.99 & 1.00 & 0.98 & 0.96 \\}
\providecommand{\tabCalBody}{}\renewcommand{\tabCalBody}{MP & all-metal & on hull, single & 216 & 9.2 & 6.7--11.1 & 5.0 & 0.054 \\
MP & all-metal & polymorphs & 78 & 6.0 & 3.6--10.8 & 4.8 & 0.083 \\
MP & all-metal & above hull & 113 & 7.3 & 4.6--9.6 & 6.3 & 0.069 \\
MP & magnetic 3d & on hull, single & 68 & 5.2 & 3.5--7.6 & 1.2 & 0.096 \\
MP & magnetic 3d & polymorphs & 65 & 5.0 & 2.4--10.7 & 0.1 & 0.100 \\
MP & magnetic 3d & above hull & 32 & 6.0 & 1.8--8.8 & 5.3 & 0.083 \\
MP & p-block/anion & on hull, single & 115 & 6.5 & 3.7--16.3 & 1.6 & 0.077 \\
MP & p-block/anion & polymorphs & 317 & 10.9 & 6.7--17.0 & -2.1 & 0.046 \\
MP & p-block/anion & above hull & 66 & 15.1 & 8.3--23.3 & 7.8 & 0.033 \\
\midrule
OQMD & all-metal & on hull, single & 36 & 9.1 & 4.9--12.3 & 5.2 & 0.055 \\
OQMD & all-metal & polymorphs & 277 & 7.1 & 4.9--9.6 & 1.9 & 0.071 \\
OQMD & all-metal & above hull & 102 & 6.9 & 4.6--9.6 & 4.9 & 0.073 \\
OQMD & magnetic 3d & on hull, single & 15 & 3.0 & 0.8--7.1 & 1.0 & 0.165 \\
OQMD & magnetic 3d & polymorphs & 110 & 4.8 & 3.6--6.9 & 0.1 & 0.104 \\
OQMD & magnetic 3d & above hull & 60 & 7.5 & 2.6--17.2 & 2.5 & 0.067 \\
OQMD & p-block/anion & on hull, single & 26 & 3.3 & 1.7--17.0 & 2.6 & 0.151 \\
OQMD & p-block/anion & polymorphs & 410 & 8.1 & 6.3--10.9 & 0.7 & 0.062 \\
OQMD & p-block/anion & above hull & 71 & 13.7 & 5.7--15.6 & 9.5 & 0.037 \\
\midrule
literature DFT & all-metal & -- & 230 & 8.4 & 5.9--11.4 & 2.0 & 0.060 \\
literature DFT & magnetic 3d & -- & 105 & 3.9 & 2.0--5.2 & 0.6 & 0.127 \\
literature DFT & p-block/anion & -- & 123 & 10.0 & 6.0--14.8 & 1.1 & 0.050 \\}
\providecommand{\tabCountsBody}{}\renewcommand{\tabCountsBody}{invariant reaction & 12,066 & 12,822 & 27 & 1,496 & 26,411 \\
formation & 2,930 & 5,007 & 2,438 & 458 & 10,833 \\
phase transition & 7,294 & 2,844 & 6 & 310 & 10,454 \\
crystal structure & 3,301 & 1,330 & 3,065 & 2,161 & 9,857 \\
phase equilibrium & 4,785 & 706 & 0 & 289 & 5,780 \\
heat capacity / entropy / enthalpy & 2,599 & 2,014 & 139 & 161 & 4,913 \\
mixing / partial & 1,576 & 499 & 431 & 20 & 2,526 \\
activity / EMF / Gibbs energy & 1,062 & 614 & 7 & 21 & 1,704 \\
model expression & 604 & 227 & 4 & 19 & 854 \\
other property & 69 & 86 & 33 & 16 & 204 \\
transformation enthalpy & 2 & 4 & 0 & 0 & 6 \\
\midrule
total & 36,288 & 26,153 & 6,150 & 4,951 & 73,542 \\
accepted & 31,541 & 23,988 & 6,133 & 4,889 & 66,551 \\}
\providecommand{\tabEspeiBody}{}\renewcommand{\tabEspeiBody}{\texttt{experimental/} & experimental thermochemical and phase-composition (ZPF) data & 1,118 & 2,387 & 185 \\
\texttt{invariants/} & ZPF data from experimental invariant-reaction tables & 1,546 & 5,061 & 464 \\
\texttt{model/} & model-calculated values (TDB named where known) & 504 & 782 & 82 \\
\texttt{dft/literature/} & DFT formation enthalpies printed in papers & 783 & 1,781 & 91 \\
\texttt{dft/mp/} & Materials Project on-hull formation energies & 1,251 & 1,251 & 381 \\
\texttt{dft/oqmd/} & OQMD on-hull formation energies & 1,251 & 1,251 & 381 \\}
\providecommand{\tabFloorBody}{}\renewcommand{\tabFloorBody}{formation enthalpy & method not printed & 3.5 \kjma & 290 (98) & 2.41--4.3 \\
formation enthalpy & solution / reaction calorimetry & 3.2 \kjma & 80 (47) & 2.2--4.6 \\
formation enthalpy & experimental, method not named$^{\dagger}$ & 4.1 \kjma & 29 (9) & 1.99--9.33 \\
formation enthalpy & Knudsen effusion / KEMS & 4.2 \kjma & 20 (15) & 1.15--7.65 \\
formation enthalpy & EMF$^{\dagger}$ & 4.6 \kjma & 17 (8) & 1.89--14.8 \\
formation enthalpy & combustion calorimetry$^{\dagger}$ & 1 \kjma & 15 (4) & 0.0881--5.33 \\
Gibbs energy of formation & method not printed & 1.2 \kjma & 37 (17) & 0.66--2.1 \\
invariant temperature & method not printed & 11 K & 4,912 (1,680) & 9.44--10.5 \\
transition temperature & method not printed & 16 K & 274 (53) & 10.5--26.2 \\
transition temperature & thermal analysis$^{\dagger}$ & 5 K & 9 (4) & 2.1--58.2 \\
phase composition & method not printed & 0.011 mole fraction & 3,014 (1,426) & 0.0105--0.0115 \\
lattice parameter & method not printed & 0.22 \% of value & 586 (192) & 0.18--0.25 \\
lattice parameter & X-ray / neutron diffraction & 0.13 \% of value & 30 (25) & 0.09--0.39 \\
lattice parameter & experimental, method not named$^{\dagger}$ & 0.13 \% of value & 9 (9) & 0.05--0.48 \\
cell or molar volume & method not printed$^{\dagger}$ & 0.54 \% of value & 43 (5) & 0.37--1.97 \\
absolute entropy & method not printed & 0.7 J\,K$^{-1}$\,mol$^{-1}$ & 77 (20) & 0.285--1.17 \\
absolute entropy & drop / heat-content calorimetry$^{\dagger}$ & 1 J\,K$^{-1}$\,mol$^{-1}$ & 19 (9) & 0.0734--1.93 \\
partial Gibbs energy & method not printed$^{\dagger}$ & 1.4 kJ\,mol$^{-1}$ & 13 (13) & 1.09--1.7 \\
\midrule
formation enthalpy & model-calculated (against experiment) & 3.9 \kjma & 2,045 (361) & 3.11--4.55 \\
invariant temperature & model-calculated (against experiment) & 4 K & 14,146 (5,377) & 3.11--4.3 \\
phase composition & model-calculated (against experiment) & 0.005 mole fraction & 16,730 (8,821) & 0.00445--0.00445 \\}
\providecommand{\tabGateBody}{}\renewcommand{\tabGateBody}{G1 & temperatures & 2,933 & 17 & 26 & 0 \\
G2 & energies, entropies, heat capacities, activities & 477 & 42 & 93 & 0 \\
G3 & signs & 41 & 0 & 45 & 0 \\
G4 & system assignment & 327 & 3 & 67 & 0 \\
G5 & property semantics & 392 & 189 & 42 & 0 \\
G6 & invariant phase lists & 933 & 0 & 376 & 0 \\
G7 & duplicates & 0 & 0 & 509 & 7,981 \\
G8 & evidence class & 53 & 0 & 734 & 0 \\
G9 & compositions & 4 & 0 & 269 & 0 \\
G10 & reader tier / agreement & 0 & 0 & 0 & 0 \\
G11 & CALPHAD plausibility fields & 0 & 0 & 0 & 0 \\
G12 & pressure basis & 0 & 0 & 0 & 0 \\
page & page checks (visual) & 2,663 & 88 & 1,155 & 0 \\}
\providecommand{\tabParityBody}{}\renewcommand{\tabParityBody}{MP & default & 1,070 & 9.6 & 6.0 & 14.0 & 0.86 & 0.94 & 0.66 \\
MP & per-compound median & 455 & 9.8 & 6.1 & 14.1 & 0.87 & 0.93 & 0.65 \\
MP & without unexplained outliers & 1,053 & 9.2 & 5.9 & 13.2 & 0.86 & 0.95 & 0.67 \\
MP & calorimetry only & 295 & 9.0 & 5.7 & 13.3 & 0.89 & 0.96 & 0.66 \\
\midrule
OQMD & default & 1,107 & 8.3 & 5.3 & 12.4 & 0.91 & 0.95 & 0.71 \\
OQMD & per-compound median & 481 & 8.3 & 5.2 & 12.3 & 0.92 & 0.95 & 0.71 \\
OQMD & without unexplained outliers & 1,085 & 7.9 & 5.1 & 11.6 & 0.91 & 0.96 & 0.72 \\
OQMD & calorimetry only & 303 & 7.6 & 4.9 & 11.7 & 0.94 & 0.97 & 0.74 \\}
\providecommand{\tabPropBody}{}\renewcommand{\tabPropBody}{\texttt{INVARIANT\_TEMPERATURE} & invariant-reaction temperature & K & ZPF & 26,411 \\
\texttt{TRANSITION\_TEMPERATURE} & transition temperature & K & ZPF & 10,430 \\
\texttt{HM\_FORM} & enthalpy of formation & kJ/mol-atom & HM\_FORM, HMR & 9,137 \\
\texttt{PHASE\_COMPOSITION} & phase composition at equilibrium & -- & ZPF & 4,841 \\
\texttt{LATTICE\_A} & lattice parameter $a$ & Å & -- & 4,226 \\
\texttt{LATTICE\_C} & lattice parameter $c$ & Å & -- & 2,781 \\
\texttt{CPM} & molar heat capacity & J/mol (basis unstated)/K & CPM & 2,322 \\
\texttt{HM\_MIX} & enthalpy of mixing & kJ/mol (basis unstated) & HMR & 1,345 \\
\texttt{LATTICE\_B} & lattice parameter $b$ & Å & -- & 1,322 \\
\texttt{SM} & molar entropy & J/mol-atom/K & SM, SMR & 1,053 \\
\texttt{PHASE\_ASSEMBLAGE} & phases present at a state & -- & -- & 939 \\
\texttt{HEAT\_CONTENT} & heat content $H(T)-H(298.15)$ & J/g at. & -- & 889 \\
\texttt{SM\_FORM} & entropy of formation & J/mol-atom/K & SM\_FORM, SMR & 825 \\
\texttt{PARTIAL\_ENTHALPY} & partial molar enthalpy & kJ/mol (basis unstated) & -- & 820 \\
\texttt{GM\_FORM} & Gibbs energy of formation & kJ/mol-atom & GMR & 792 \\
\texttt{GIBBS\_ENERGY} & Gibbs energy & J/mol & -- & 595 \\
\texttt{GIBBS\_ENERGY\_EXPRESSION\_TERM} & model expression & J/mol & -- & 555 \\
\texttt{EMF} & electromotive force & mV & -- & 429 \\
\texttt{ENTROPY} & heat capacity / entropy / enthalpy & J/(mol K) & -- & 327 \\
\texttt{GM\_MIX} & Gibbs energy of mixing & kJ/mol (basis unstated) & GMR & 256 \\}
\providecommand{\yearMax}{}\renewcommand{\yearMax}{2025}
\providecommand{\yearMin}{}\renewcommand{\yearMin}{1975}

\usepackage{xr}
\newcommand{\mainSecRef}[2]{Section~\ref{#1} of the main paper}
\newcommand{\mainFigRef}[3]{Fig.~\ref{#1}#3 of the main paper}
\newcommand{\siThisDocument}{SI~1}
\newcommand{\suppnotesection}[2]{\section*{Supplementary Note #1: #2}}

\makeatletter
\renewcommand\section{\@startsection{section}{1}{\z@}{-2.0ex plus -.4ex minus -.2ex}{.8ex plus .2ex}{\normalfont\normalsize\bfseries}}
\renewcommand\subsection{\@startsection{subsection}{2}{\z@}{-1.5ex plus -.3ex minus -.2ex}{.6ex plus .2ex}{\normalfont\normalsize\bfseries\itshape}}
\renewcommand\subsubsection{\@startsection{subsubsection}{3}{\z@}{-1.2ex plus -.2ex}{.5ex}{\normalfont\normalsize\itshape}}
\makeatother

\twocoltrue
\newcommand{\orchidTitle}{ORCHID: connecting finite-temperature thermodynamic evidence to computational materials discovery}

\newcommand{\orchidAuthA}{Raymundo Arr\'oyave}
\newcommand{\orchidAuthB}{James E. Saal}
\newcommand{\orchidAuthC}{Dongwon Shin}
\newcommand{\orchidAffA}{Department of Materials Science and Engineering, Texas A\&M University, College Station, TX 77843, USA}
\newcommand{\orchidAffB}{Citrine Informatics, Inc., Redwood City, CA, USA}
\newcommand{\orchidAffC}{Oak Ridge National Laboratory, Oak Ridge, TN, USA}
\newcommand{\orchidEmail}{rarroyave@tamu.edu}

\newcommand{\orchidAbstract}{%
Finite-temperature phase stability is essential to materials discovery, but the evidence needed to describe it remains scattered across decades of literature.
ORCHID organizes this evidence around the thermodynamic states that the observations describe, using tiered agentic extraction guided by thermodynamic rules, expert supervision and explicit validation.
The collection contains \nValues{} values from \nPapers{} papers, represented with their thermodynamic states, phase identities, provenance, evidence classes and uncertainty, together with assessed phase assemblages and calculated references.
A qualified subset is released in ESPEI-compatible form for CALPHAD assessment, and machine-readable access supports both human and AI-assisted reuse.
Blind expert evaluation of \nBlindExpertN{} diagram fields gives \BlindExpertAgreeRawPct\% agreement, increasing to \BlindExpertAgreeEquivPct\% after accounting for notation conventions.
On a common formation-enthalpy benchmark, eight MLIPs show residual spreads close to their DFT reference, motivating evaluation of finite-temperature phase stability through PhaseForge.
One-shot MLIP--PhaseForge descriptions of Cu--Zr recover 0.35--0.56 of observed phase occurrences at assessed field points.
ORCHID-assisted refinement through PhaseForge+ recovers \cuzrInvFinal{} of \cuzrInvTotal{} invariant temperatures within $2\sigma$ and reduces held-out activity error from \cuzrActHoldMlip{} to \cuzrActHoldFinal~kJ\,mol$^{-1}$.
By supplying interpreted literature evidence directly to gradient-based refinement through PhaseForge+, ORCHID connects evidence recovery to programmatic CALPHAD assessment.
With the initial description and fitting datasets prepared, the final Cu--Zr refinement completes in approximately 27 minutes, providing a foundation for continuously updated thermodynamic models and their use in materials discovery.%
}

\usepackage{fancyvrb}
\begin{document}
\twocolumn[\begin{@twocolumnfalse}
\begin{center}
{\LARGE\bfseries\sffamily\orchidTitle\par}\vspace{1em}
{\large\orchidAuthA$^{a,*}$, \orchidAuthB$^{b}$ and \orchidAuthC$^{c}$\par}
\vspace{0.7em}
{\small $^{a}$\orchidAffA\par $^{b}$\orchidAffB\par $^{c}$\orchidAffC\par
$^{*}$\href{mailto:\orchidEmail}{\orchidEmail}\par}
\end{center}
\begin{abstract}\noindent\orchidAbstract\end{abstract}
\vspace{1em}\end{@twocolumnfalse}]
\section{Introduction}
\label{sec:intro}

Understanding phase stability---that is, which phases are stable and under which conditions---constitutes the starting point for the exploration of any materials space\cite{arroyave2022phase}, an idea captured by Hume-Rothery's dictum that phase diagrams are ``the beginning of wisdom, not the end of it''\cite{schmidfetzer2014}.
For materials discovery, this understanding connects a predicted compound to the conditions under which it might be synthesized and retained during operation.
While knowledge of the 0~K ground state and its neighborhood helps identify competing phases, evaluating synthesizability requires understanding their relative stability at finite temperature, together with the kinetics and processing pathways through which they can be obtained.
Thermodynamic descriptions also provide the driving forces for phase transformations, linking processing conditions to the evolution of microstructure and, through it, to material properties.
This connection places phase stability at the foundation of integrated computational materials engineering (ICME).

\begin{widefigure}[!t]
\nfig{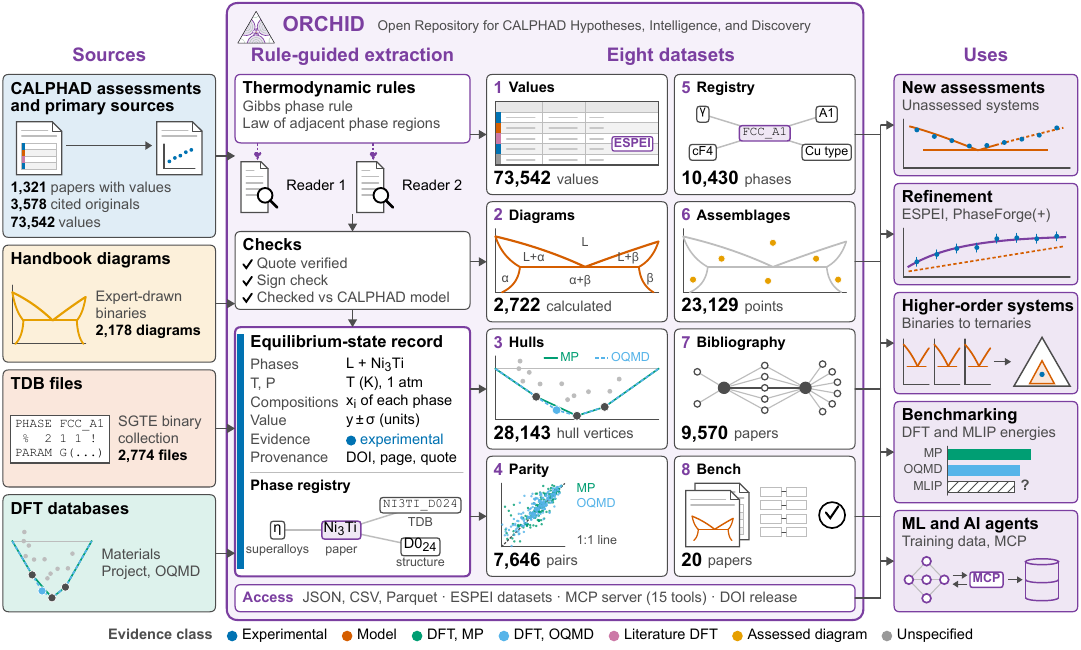}
\caption{\textbf{Connecting thermodynamic evidence to computational discovery.} Literature, assessed diagrams, CALPHAD models and first-principles databases are linked through extraction constrained by thermodynamic rules, source verification and phase identities. The resulting observations retain their states, evidence classes, uncertainties and provenance for model fitting, comparison and automated access. Glyphs are schematic except the downsampled DFT--experiment and citation-network tiles; no publisher page or handbook drawing is reproduced.}
\label{fig:hero}
\end{widefigure}

Open first-principles resources have made information on ground-state stability increasingly accessible.
The Materials Project\cite{jain2013}, the Open Quantum Materials Database (OQMD)\cite{saal2013,kirklin2015} and AFLOW\cite{curtarolo2012aflowlib}, for example, provide energies and structures from which competing compounds and their stability at 0~K can be examined, while the NOMAD repository\cite{draxl2019nomad} provides access to computational materials data contributed by the broader community.
The experimental evidence needed to establish phase stability at finite temperature, however, remains scattered across decades of literature, in reports of phase boundaries, invariant reactions, calorimetric measurements and activities.
Interpreting these measurements requires knowing the chemistry, temperature, pressure and phase constitution of the states they describe, information that is frequently distributed across a table, its caption, the surrounding text and earlier cited work.
Recovering a useful observation therefore requires reconstructing the thermodynamic state to which the reported value refers, together with its basis and source.

Once the states represented by these observations have been established, the observations can be brought together to constrain a thermodynamic description of the system.
The CALPHAD method provides a framework for constructing such descriptions by assessing parametric models of the Gibbs energies of the phases against experimental and computed evidence\cite{lukas2007,arroyave2022phase}.
Equilibrium is then obtained by minimizing the total Gibbs energy subject to conservation of mass at specified temperature and pressure.
By incorporating different observations into a common thermodynamic description, this approach also provides a basis for extrapolation to higher-order systems.
Open-source tools such as pycalphad\cite{otis2017} and ESPEI\cite{bocklund2019} support the calculation of phase equilibria and the estimation of model parameters and their uncertainties, respectively.
The SGTE unary descriptions\cite{dinsdale1991} and the binary assessments compiled by Hallstedt\cite{hallstedt2025} provide openly reusable models.
Thermodynamic database (TDB) files preserve these fitted descriptions, but the underlying evidence often has to be collected again when a model is assessed or refined.

The difficulty of recovering that evidence is partly linguistic and partly thermodynamic.
The same phase can appear under different names in papers, assessed diagrams, TDB files and first-principles databases, while interpreting a reported formation enthalpy requires knowing its composition, reference state and normalization basis.
Recent literature-extraction frameworks demonstrate that language models can recover structured materials information\cite{roy2026comproscanner}, and data resources coupled to predictive models illustrate the value of making such information reusable\cite{hodl2025surfpro}.
For thermodynamic evidence, the relations among phases and conditions provide additional constraints on the interpretation of a reported observation.
The Gibbs phase rule and the rules governing adjacent phase fields can be used to reject impossible readings and, where sufficient information is available, resolve ambiguous labels and identify phase assemblages.
These constraints provide a basis for guiding extraction with physical knowledge, while language models and expert review address cases that require further interpretation.

Commercial CALPHAD databases embody decades of expert assessment of published and, where available, proprietary measurements\cite{thermocalcAssessment,thermocalcDevelopment}.
The fitted models make this accumulated knowledge available for calculations, but the observations supporting them are not generally released together as an openly accessible, machine-readable dataset.
Constructing a new assessment or independently evaluating an existing one may therefore require collecting and interpreting evidence that has already been examined by others.
The open computational resources described above demonstrate the value of organizing materials data so that they can support further calculations and model development.
An analogous resource for experimental thermodynamic evidence would allow the observations underlying finite-temperature phase stability to be examined, combined and reused across assessments.

In this work, we present ORCHID (Open Repository for CALPHAD Hypotheses, Intelligence, and Discovery), which organizes thermodynamic evidence around the states that the observations describe (Fig.~\ref{fig:hero}).
Measurements of thermodynamic properties constrain the Gibbs energies of phases, while phase boundaries, invariant reactions and assessed assemblages constrain the conditions under which those phases coexist.
Describing these observations in terms of their chemistry, temperature, pressure and phase constitution allows different forms of evidence to be brought together while preserving their sources, evidence classes and uncertainties.
Observations that meet the requirements for thermodynamic assessment are released in ESPEI-compatible form\cite{bocklund2019}.
Through programmatic interfaces, researchers and AI agents can retrieve this evidence together with the information needed to interpret it, select observations for fitting, examine disagreements among sources and evaluate model predictions.

The evidence collected in ORCHID also provides a basis for evaluating whether computational models reproduce the competition among phases at finite temperature.
For machine-learned interatomic potentials (MLIPs), this extends evaluation beyond agreement with ground-state formation energies to the phase equilibria implied by their predictions.
PhaseForge connects these predictions to CALPHAD descriptions, allowing the resulting equilibria to be compared with the observations recorded in ORCHID\cite{zhu2025machinelearningpotentials,zhu2025accelerating,sariturk2025phaseforge}.
Where discrepancies are identified, the ESPEI-ready subset can support refinement through PhaseForge+, using selected observations for fitting and separate evidence for evaluation\cite{kunselman2025construction,sariturk2025phaseforgeplus}.
Making this process accessible to researchers and AI agents provides a basis for continuously refining thermodynamic descriptions as new evidence becomes available, supporting the connection between materials prediction, synthesis and processing needed for self-driving materials discovery.

\section{ORCHID: connecting thermodynamic evidence to models}
\label{sec:orchid-overview}\label{sec:release}

The use of a reported value in a thermodynamic assessment requires knowing the state to which that value refers. For a melting point, this requires identifying the participating phases, while a formation enthalpy must be associated with the structure of the compound and its reference state. Recovering this information from the literature is complicated by the different forms in which the evidence is reported, from invariant temperatures in tables to heat-capacity curves in figures. In ORCHID, this evidence is organized around a description of the thermodynamic state that includes the chemical system, coexisting phases, temperature, pressure and phase compositions, together with the reference state of the property (Fig.~\ref{fig:hero}).

Organizing evidence by thermodynamic state allows independent measurements, predictions from assessed models and results from first-principles calculations or machine-learned potentials to be compared while retaining the identity of the individual observations.
The present release connects these observations through their recorded state attributes; a planned extension will assign permanent state identifiers through an append-only registry.
This organization also requires tracking changes in phase constitution within a reported property series.
For heat-capacity, heat-content and lattice-parameter series that span a phase transition, the observations were attributed to the phases on the corresponding sides of the transition.
Cases that could not be resolved from the source were flagged and excluded from fitting.
The evidence class and source remain attached to each observation so that its origin can be examined alongside its thermodynamic interpretation.
The present release focuses primarily on binary systems, although the representation extends to higher-order systems.

Comparing observations from different sources requires establishing whether the reported phase names refer to the same phase.
Differences in naming among the experimental literature, thermodynamic models and first-principles databases make this correspondence difficult to establish from the names alone.
ORCHID therefore uses a phase registry that associates reported names with a common phase identity within a binary system, connecting the observations to the corresponding definitions in TDB models.
A structure prototype is assigned when the source information supports that identification unambiguously.

Establishing phase identity also allows evidence from assessed phase diagrams to be considered alongside numerical values from the literature.
These diagrams summarize an expert interpretation of the available experiments, identifying the phase equilibria considered most consistent with that evidence.
ORCHID captures this information from assessed binary diagrams in a phase-diagram handbook by recording the phases present at representative temperatures and compositions within the phase fields.
The resulting assemblages preserve information about phase coexistence in a form that can be used in thermodynamic calculations.
They are retained as a separate evidence class so that users can distinguish the assessment from the underlying measurements and decide whether to include it in a fit.

The assessed phase assemblages can be compared with equilibria calculated from thermodynamic models to identify differences in the predicted phase stability of the same system.
ORCHID supports this comparison through binary phase diagrams calculated with pycalphad\cite{otis2017} using the SGTE collection of binary assessments compiled by Hallstedt\cite{hallstedt2025}.
These finite-temperature descriptions are complemented by binary convex hulls derived from the Materials Project and OQMD, with the calculations linked to the model or database release used.
For compounds represented in these first-principles resources, calculated formation energies are paired with formation enthalpies from the experimental literature.
Comparison across \nParityCompounds{} compounds provides an empirical estimate of the uncertainty of DFT formation energies, which is used to determine their weight in the ESPEI datasets so that their influence in a fit reflects their agreement with experiment.

To support the construction and refinement of CALPHAD models, ORCHID releases observations as ESPEI datasets after checking that they meet the requirements for fitting.
This requires a fully specified equilibrium state, phases that can be identified in the CALPHAD model and consistency with the rules of phase equilibria.
Approximately \nEspeiReadyApprox{} values, or about one in \espeiReadyOneIn{}, meet these requirements.
For these observations, uncertainties are reported in physical units together with the corresponding weights, making explicit how uncertainty determines their influence in a fit.
The datasets also retain their evidence classes and quality flags so that users can select observations and reconsider their uncertainty budgets according to the purpose of the assessment (SI~1, Fig.~\ref{fig:espeirecord}).

These decisions can also be supported through programmatic access to the observations and their associated thermodynamic information.
An AI agent can query a chemical system, resolve phase identities across sources and inspect the evidence available for model development or evaluation.
Because the returned records include evidence classes, uncertainty budgets, provenance and fitting status, the agent can distinguish observations ready for fitting from those that require further interpretation.

\begin{widefigure}[!t]
\nfig{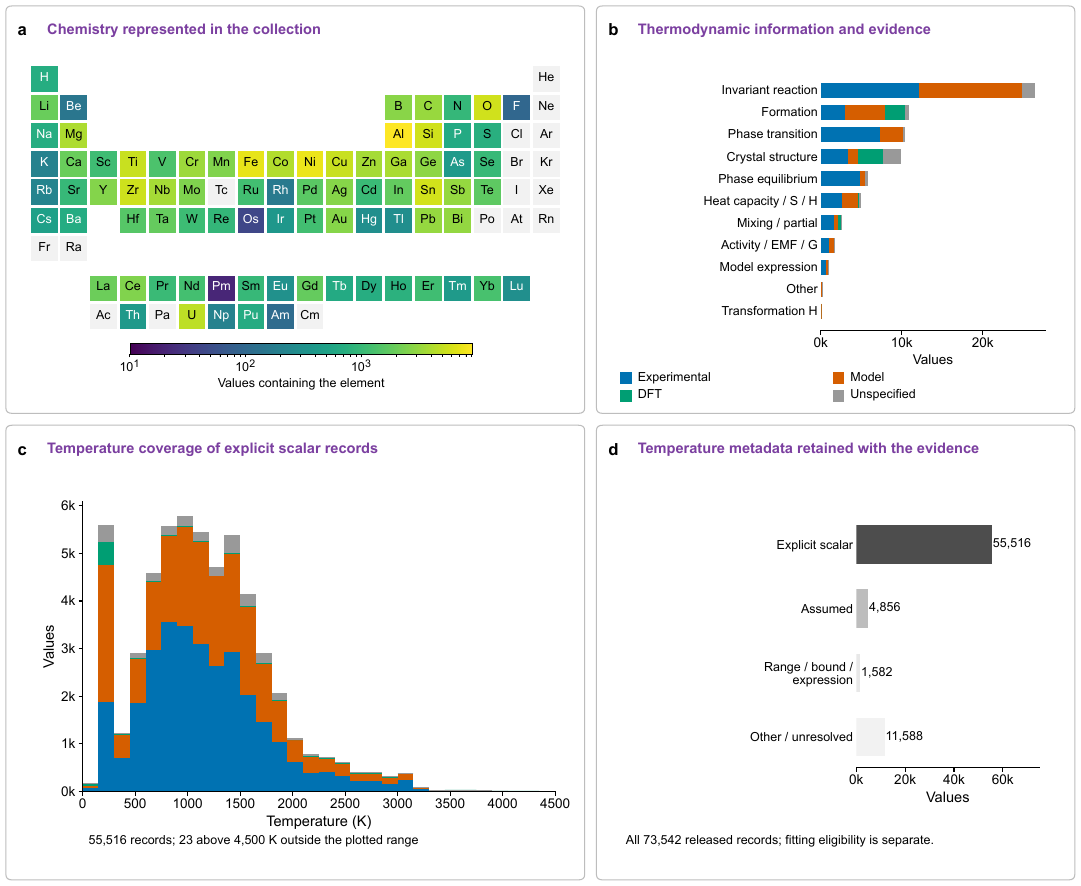}
\caption{\textbf{Chemistry, properties and temperatures represented in ORCHID.} \textbf{a}, Values by constituent element on a logarithmic scale; grey indicates no records. \textbf{b}, Property groups separated by evidence class. \textbf{c}, Temperatures of 55,516 records with an explicit scalar temperature, including invariant and transition temperatures as reported properties; 23 values above 4,500~K are outside the plotted range. \textbf{d}, A disjoint partition of the temperature metadata separates these scalar records from assumptions, ranges or bounds and expression terms, and other or unresolved cases. Counts describe released records, rather than independent measurements or fitting-eligible observations. The classification rules, binary-system coverage and source history are retained in SI~1.}
\label{fig:coverage}
\end{widefigure}

The range of tasks supported by these interfaces depends on the evidence available for the chemistry of interest.
The present release brings together approximately 74,000 literature values from 1,300 papers, spanning nearly 1,500 chemical systems and \nElements{} elements (Fig.~\ref{fig:coverage}).
Of these values, \nAccepted{} satisfy the extraction and consistency rules.
Coverage is predominantly binary, comprising approximately 58,000 values across nearly 1,000 systems, with additional observations for unary, ternary and higher-order systems.
This distribution reflects the literature and assessments used to construct the collection, so an unrepresented system or phase does not imply an absence of experimental evidence.
Approximately 850 binary systems in the literature collection also have calculated diagrams, allowing about 50,000 literature values to be examined alongside the corresponding model predictions.

Reusing this evidence requires preserving the distinctions among measurements, assessed interpretations and model predictions, even when they describe the same thermodynamic state.
Figure~\ref{fig:coverage} summarizes these evidence classes and distinguishes explicitly reported temperatures from assumptions, ranges and unresolved conditions.
Observations that are incomplete or do not meet the requirements for fitting remain accessible with their status recorded, allowing users to examine the broader evidence when selecting data for an assessment.
Figure~\ref{fig:anatomy} in SI~1 illustrates how records extracted from the same sentence can retain different evidence classes and how assumed conditions remain distinguishable from those reported by the authors.
The Appendix provides a visual guide to the eight connected data products.
SI~2 gives their inventory, field definitions and access instructions, while SI~1 describes the extraction, calibration and evaluation procedures used in the analyses.

\section{Physics-guided extraction and evaluation of the evidence}
\label{sec:methods}\label{sec:validation}

\subsection{Recovering thermodynamic states from the literature}
Recovering thermodynamic evidence requires establishing both what a source reports and the context needed to interpret it.
ORCHID combines independent readings of the source with constraints on the phases, compositions and conditions associated with an observation.
We assembled the literature by starting from the assessments in the SGTE binary collection\cite{hallstedt2025}, following their references to primary studies and including further assessments of the same systems.
The resulting corpus contains approximately 1,300 papers published between \yearMin{} and \yearMax.
Extraction focused on tables and running text, leaving numerical values available only in plots for later digitization.
Because a measurement may be reproduced in several papers, the records distinguish the reporting work from the original source wherever it could be identified, allowing reprinted evidence to be recognized.

\begin{widefigure}[!t]
\nfig{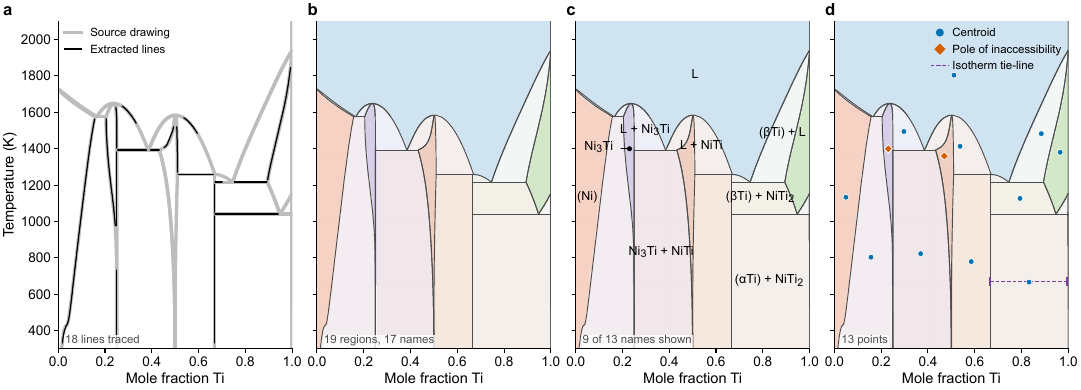}
\caption{\textbf{Thermodynamic constraints guide the interpretation of phase diagrams.} Synthetic Ni--Ti diagram generated from the SGTE assessment of Tokunaga et al. (2004), with known field identities. \textbf{a}, The source drawing (grey) and the boundary lines extracted from its vector paths (black). \textbf{b}, Enclosed fields. \textbf{c}, Assigned phase assemblages. \textbf{d}, Representative interior points and a drawing-based tie-line. Field names follow the benchmark truth key; narrow fields may not admit an interior point with the required margin. The diagram is synthetic, and no handbook image is reproduced. Geometry thresholds and label mappings are given in SI~1.}
\label{fig:annotation}
\end{widefigure}

For tables, two vision-language models independently interpreted the page image using instructions that incorporated the rules of phase equilibria.
Their readings were compared with the PDF text layer where available, and agreement combined with deterministic checks supported acceptance.
Disagreements were referred to a judge model or retained with their unresolved status recorded.
Running-text extraction also used paired readings, retaining a value only when the quoted sentence and numerical value were found together on the cited page.
Associated phases, methods and citations were retained when supported by that sentence or its immediate context.
Since agreement between readers can reproduce an error in the PDF text layer, suspect formation-enthalpy signs were checked against rendered pages.
These checks preserved the sign established by the source or recorded that the value was printed without one.

Interpreting a phase diagram requires combining the information printed in the source with the thermodynamic constraints on phase coexistence.
A deterministic pass first traces boundaries, calibrates axes and identifies phase fields.
Printed labels and the law of adjacent phase regions then constrain the assemblages in neighboring fields, while the phase rule constrains invariant reactions.
A field is assigned an assemblage directly when only one satisfies these constraints.
Otherwise, the remaining possibilities define the choices presented to the machine readers, with an adjudicator and, finally, a domain expert addressing unresolved cases.
Figure~\ref{fig:annotation} illustrates this progression on a synthetic Ni--Ti diagram with known field identities, showing how boundaries and neighboring phases provide information beyond that contained in an isolated label.

Once a phase field is identified, a temperature and composition within its interior define an observation of the phase assemblage.
The distance from the boundary provides a measure of how securely the selected point lies within that field.
For numerical records, checks similarly require a physically consistent set of phases and agreement between the assigned chemical system and the elements in the reported phase formulas.
The thermodynamic state associated with a value is then established by combining this information with the table headers, property units, reference state and conditions reported elsewhere in the source.
Corrections to the sign, unit, scale or chemical system require support from the rendered page, and the original representation is retained.
Observations with unresolved normalization or reference states remain available for inspection but can be excluded from fitting.
The reading rules and correction history are documented in SI~1, preserving the distinction between reported, assumed and inferred information.

\subsection{Human calibration and consistency across the collection}
We evaluated the extraction through controlled tests, expert readings and comparisons with thermodynamic assessments.
The controlled tests used synthetic documents generated from SGTE descriptions, providing known numerical values and phase identities that could be compared with the extracted results.
These documents were rendered as born-digital PDFs and degraded scans to examine the effect of document quality.
The AI readers recovered invariant-reaction tables almost completely from born-digital PDFs and substantially from degraded scans, whereas deterministic tools recovered fewer invariant rows and had difficulty interpreting degraded diagrams.
The benchmark also identified a prompt-dependent failure that produced empty tables, leading to a correction of the acceptance logic.
The test results and rendering variants are retained in SI~1, Fig.~S5 and Table~S6, allowing subsequent extraction methods to be evaluated under the same controlled conditions.

To evaluate the interpretation of assessed diagrams, a domain expert first reviewed \asmExpertReviewed{} difficult fields with access to the adjudicator's reasoning, accepting \asmExpertAccepted{} of the \asmExpertProposals{} proposals for which a phase assemblage had been named.
This review was followed by a blind evaluation of \nBlindExpertN{} fields, one from each of \nBlindExpertN{} diagrams that the expert had not previously reviewed.
Without seeing the machine's answer, the expert assigned phases that agreed with the extraction in \nBlindExpertAgree{} cases (\BlindExpertAgreeRawPct\%, Wilson 95\% interval \BlindExpertWilsonRaw).
Agreement increased to \nBlindExpertAgreeEquiv{} cases (\BlindExpertAgreeEquivPct\%, interval \BlindExpertWilsonEquiv) after accounting for two notation conventions and included all \nBlindRuleAgree{} fields identified through printed labels and thermodynamic rules.
The remaining discrepancy involved three closely spaced line compounds in an ambiguous drawing, for which the expert's interpretation was adopted.

\begin{widefigure}[!t]
\nfig{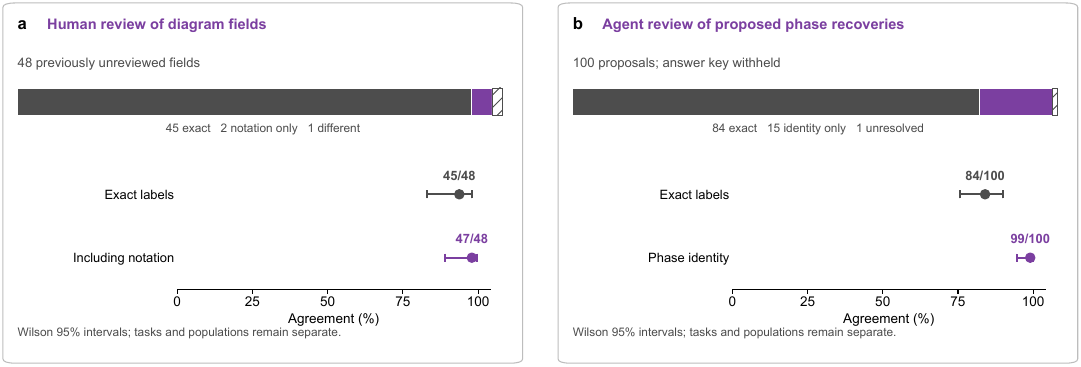}
\caption{\textbf{Source-based evaluation of phase interpretation.} \textbf{a}, A human expert assigned phases to 48 assessed-diagram fields without seeing the machine answer, agreeing on 45 fields as written and 47 after accounting for notation. \textbf{b}, An answer-key-blinded review in a separate assistant context examined 100 randomly selected proposals from a batch of 1,268 phase recoveries. Exact labels agreed in 84 cases and phase identity in 99, with one unresolved prototype retained as a disagreement. Wilson 95\% intervals describe field- or record-level agreement within these separate samples and omit within-source dependence. Panel b uses AI-assisted review, with shared-model capabilities and possible prior exposure to source pages. Sampling, adjudication and the separate evaluation of numerical records are documented in SI~1.}
\label{fig:reliability}
\end{widefigure}

\label{sec:precision}
For numerical extraction, evaluation addresses both the location of the reported value and the thermodynamic interpretation assigned to it.
All approximately 5,700 running-text records retain a passage found verbatim on the cited page that contains the printed number.
Source verification establishes where the value was reported, while phase-recovery review and the audits described below examine its assigned property, phase and reference state.
These checks evaluate specific components of the extracted records; estimating precision across the complete numerical collection requires an independent human evaluation of a representative sample.

A subsequent recovery pass assigned phase labels to numerical observations whose labels had remained unresolved during extraction.
To examine these assignments, a separate assistant context reread a seeded sample of 100 proposals from the 1,268-record batch, recording its interpretation of the source before viewing the proposed answers (Fig.~\ref{fig:reliability}b).
The readings agreed exactly in 84 cases and differed only in notation in another 15, giving agreement on phase identity in 99 of 100 cases (Wilson 95\% interval 94.55--99.82\%).
The remaining assignment was withheld because the source did not establish a unique phase identity.
This AI-assisted review measures agreement for the sampled recovery proposals, with record-level intervals that do not account for dependencies within a table.
The review used shared model capabilities and may have involved prior exposure to source pages, so it does not constitute an independent human evaluation.
Subsequent corrections and release snapshots are tracked separately.

\label{sec:audit}
To examine errors beyond phase assignment, separate language-model auditors combined targeted searches with seeded random samples, returning to source pages to investigate suspected discrepancies.
The audits distinguished errors in numerical values and thermodynamic states from errors in other metadata, identifying systematic problems involving units, phase mappings, compositions and the pairing of experimental observations with DFT entries.
Repeated sampling also identified errors introduced during correction, making verification of the revised records part of the audit procedure.
The evaluated snapshots and subsequent changes are documented in SI~1, together with sample sizes, error rates and correction details.

A final agent-assisted re-audit of the third-round release covered all six audit scopes, with its results and subsequent corrections reported in SI~1.
A source-based review of 200 records identified 189 that met all criteria for direct use in thermodynamic fitting (SI~1).
SI~1 also reports a preliminary human review of a frozen sample of 300 records, which provides additional feedback on the reviewed observations without estimating precision across the complete collection.

\begin{widefigure}[!t]
\nfig{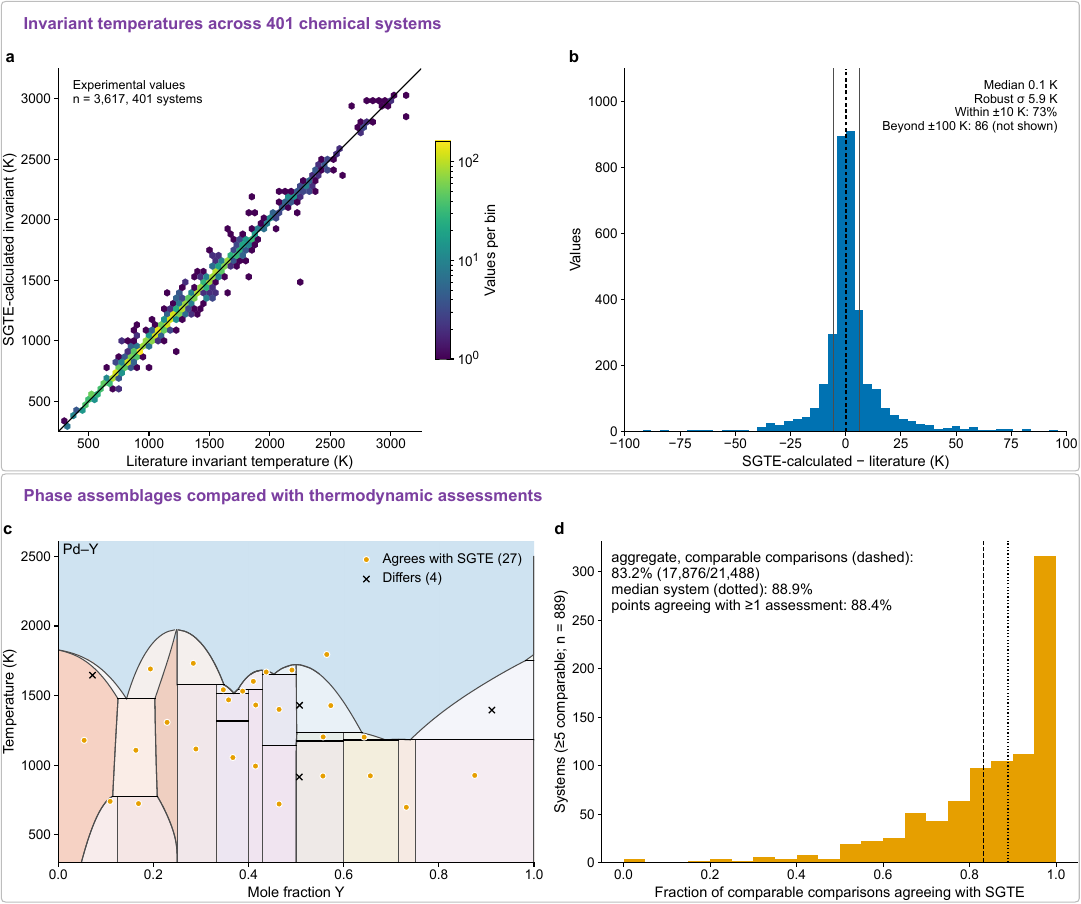}
\caption{\textbf{Finite-temperature evidence compared with CALPHAD assessments.} \textbf{a}, Experimental invariant temperatures against matching SGTE-calculated invariants for \nInvValExp{} observations in \nInvValExpSystems{} systems. \textbf{b}, Temperature residuals, with median \invValExpMedian~K and robust spread \invValExpSigma~K; \invValExpWithinTen\% lie within 10~K. The complete temperature comparison is retained in SI~1, Fig.~\ref{fig:invariants}. Of the matched observations, \nInvValExpInSample{} originate in a paper supplying an SGTE assessment, so agreement is a consistency check with substantial shared-source dependence. \textbf{c}, Assessed assemblage points over one calculated binary diagram, classified by agreement with the available assessments. \textbf{d}, Agreement fractions across systems with at least five comparable point--assessment pairs; dashed line, aggregate agreement; dotted line, median system. Of \nAsmSgteComparisons{} comparisons, \nAsmSgteNotComparable{} lack a comparable phase mapping; \nAsmSgteAgree{} of \nAsmSgteComparable{} comparable pairs agree. Separately, \nAsmSgtePointsAgreeAny{} of \nAsmSgtePointsComparable{} comparable points agree with at least one assessment. These two denominators answer different questions. All scientific panels retain the data and geometry of the source figures; no handbook drawing is reproduced.}
\label{fig:asm}
\end{widefigure}

The consistency of the recovered observations can also be examined by comparing them with equilibria calculated from assessed Gibbs-energy descriptions.
For invariant reactions, experimental temperatures are matched to SGTE predictions by chemical system, reaction type and phase identities, giving approximately 3,600 observations across 400 systems.
Where several assessments provide a match, their median prediction is used so that each observation enters the comparison once.
The temperature differences have a median of \invValExpMedian~K and a robust spread of \invValExpSigma~K, with \invValExpWithinTen\% lying within 10~K (Fig.~\ref{fig:asm}a,b).
Nearly all of these observations come from papers that also supply the corresponding SGTE assessments, making this comparison a check on the recovery of temperatures and reaction identities.
Disagreements identify observations or assessments for closer examination.

The comparison extends to phase assemblages recorded in the handbook, provided that the phase identities can be matched to those in the thermodynamic assessments.
Because several assessments of a system may incorporate different selections of evidence, agreement is evaluated both for individual point--assessment pairs and for points that match at least one assessment.
The first comparison gives agreement in \nAsmSgteAgree{} of \nAsmSgteComparable{} comparable pairs, while the second gives agreement for \nAsmSgtePointsAgreeAny{} of \nAsmSgtePointsComparable{} points (Fig.~\ref{fig:asm}c,d).
Examining disagreements revealed systematic annotation errors that were corrected before release, as well as unresolved differences between the handbook interpretation and the fitted Gibbs-energy descriptions.
Since the handbook diagram and the assessment are independent expert evaluations of the same system, often based on different selections of evidence and made at different times, a disagreement between them does not by itself indicate an error in the extraction, in the experiments or in the assessment, and each point therefore records the outcome of its comparison with every assessment of the system.
These differences remain accessible so that the assessed-diagram evidence retains its identity independently of agreement with a model.
An independent formation-enthalpy compilation\cite{kim2017} provides an additional check on extraction, units and signs, as documented in SI~1.

\subsection{Uncertainty and compatibility with CALPHAD assessment}
Assigning an observation a weight in a fit requires accounting for variation between independent measurements of the same thermodynamic state.
These measurements can differ by more than their reported uncertainties, indicating that the precision stated within a study does not capture all sources of variation between studies.
ORCHID estimates this additional variation by grouping measurements of the same property and state obtained using the same class of experimental method in different primary studies, excluding repeated reports of the same measurement.
Differences between pairs of measurements provide an estimate of the scatter associated with the method.
Giving equal weight to each state prevents frequently studied compounds from dominating the calibration and establishes a minimum uncertainty, or floor, for observations obtained by that method.

For formation enthalpies measured by solution and reaction calorimetry, \nPairsHmCal{} pairs spanning 47 states give an estimated scatter of approximately 3.1~\kjma, rounded upward to a floor of \floorHmCal~\kjma.
This estimate includes the variation present in the collected evidence, arising from experimental scatter and any remaining extraction or reference-state errors.
Where comparisons are insufficient to estimate a floor, documented literature estimates provide the starting uncertainty\cite{kubaschewski1993,drowart2005}.
The effective uncertainty used for fitting takes the larger of the reported uncertainty and the method floor, then incorporates reading resolution and applicable contributions from composition, reference state and inconsistency.
The reported uncertainty is preserved separately, and SI~1 documents the estimator, sampling requirements and combination rules.

Using these observations in ESPEI requires representing their thermodynamic states in the expected form and mapping their phases to the selected CALPHAD model.
We tested this compatibility by evaluating likelihoods against the named TDB files, obtaining finite values for all \nEspeiEvaluated{} released datasets tested.
Another \nEspeiEvalExcluded{} datasets failed the check and were excluded, with the reasons retained.
These checks establish compatibility with the fitting workflow, while the adequacy and uniqueness of the thermodynamic model are evaluated during assessment.
Physical uncertainties and the basis of the exported weights remain available so that assessors can select evidence classes and revise the weighting policy for a particular fit.
Preserving these choices allows the same observations to support both evaluation of computational predictions and refinement of thermodynamic descriptions.

\begin{widefigure}[!t]
\nfig{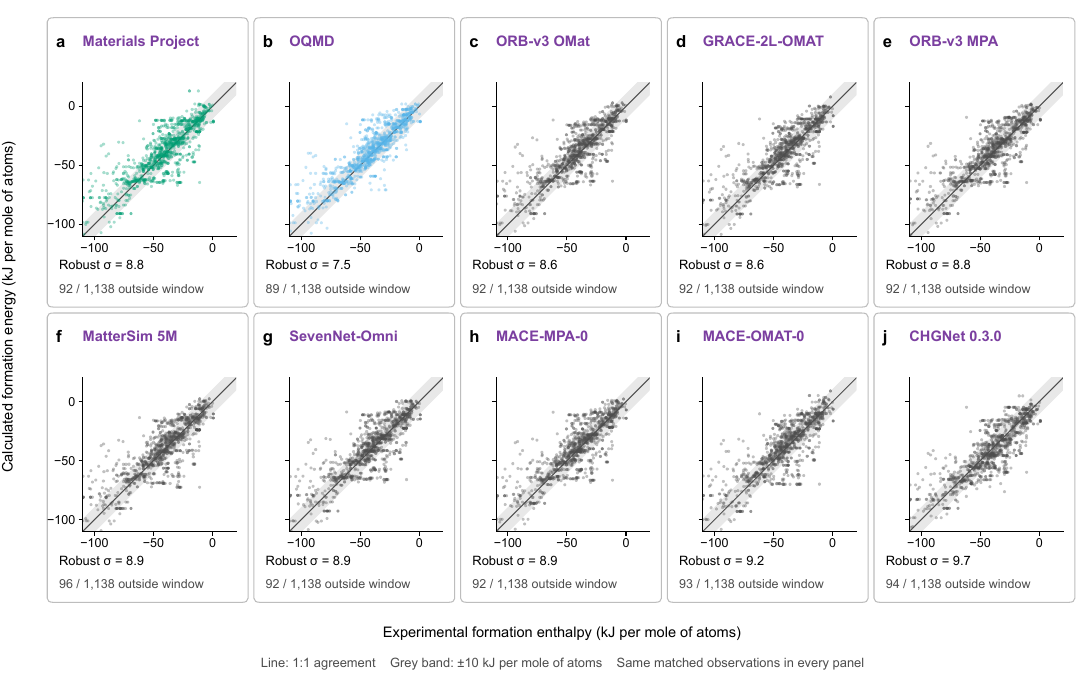}
\caption{\textbf{Formation-energy comparisons against the same experimental observations.} Calculated formation energies from the Materials Project (\textbf{a}), OQMD (\textbf{b}) and eight MLIPs (\textbf{c--j}) are paired with 1,138 experimental formation-enthalpy records representing 455 matched structures in 185 binary systems. All panels use identical axes; the solid line denotes 1:1 agreement and the grey band spans $\pm10$~\kjma. Panel annotations give the robust spread of the calculated-minus-experimental residuals, evaluated using all records, and the number outside the displayed window. The window is $-110$ to $20$~\kjma\ on both axes; no observations are excluded from the statistics on this basis. The MLIP energies use the relaxed structures and MP2020 correction convention of the matched benchmark described in SI~1. Similar deviations are visible across models; their correlations with Materials Project residuals are quantified in Fig.~\ref{fig:mlip}b.}
\label{fig:mlip-parity}
\end{widefigure}

\begin{widefigure}[!t]
\nfig{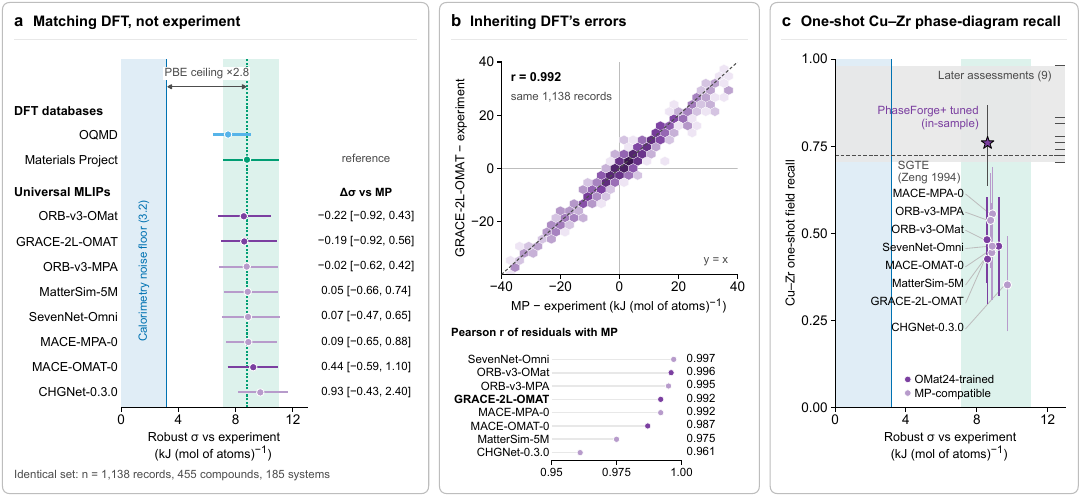}
\caption{\textbf{Similar formation-energy errors can accompany different finite-temperature failures.} \textbf{a}, Robust formation-enthalpy residual spread for eight MLIPs and two DFT references on the same 1,138 records (455 compounds, 185 binary systems), with 95\% intervals from 1,000 system-bootstrap resamples. All intervals for the MLIP-minus-MP spread include zero. The experimental band is the effective calorimetry floor used in this analysis. \textbf{b}, GRACE residuals against MP residuals; 25 pairs outside $\pm40$~\kjma\ are omitted from the plotted window. The strip gives residual correlations for all models. \textbf{c}, Weighted phase recall at 28 assessed Cu--Zr field points for one-shot PhaseForge descriptions, with point-bootstrap intervals. MLIP point estimates span 0.35--0.56 and their intervals overlap. The grey band represents nine later assessments; the dashed line is the SGTE assessment. The PhaseForge+ star is an in-sample result after refinement with ORCHID and is not a one-shot prediction. The benchmark inputs and construction assumptions are documented in SI~1.}
\label{fig:mlip}

\end{widefigure}

\begin{widefigure}[!t]
\nfig{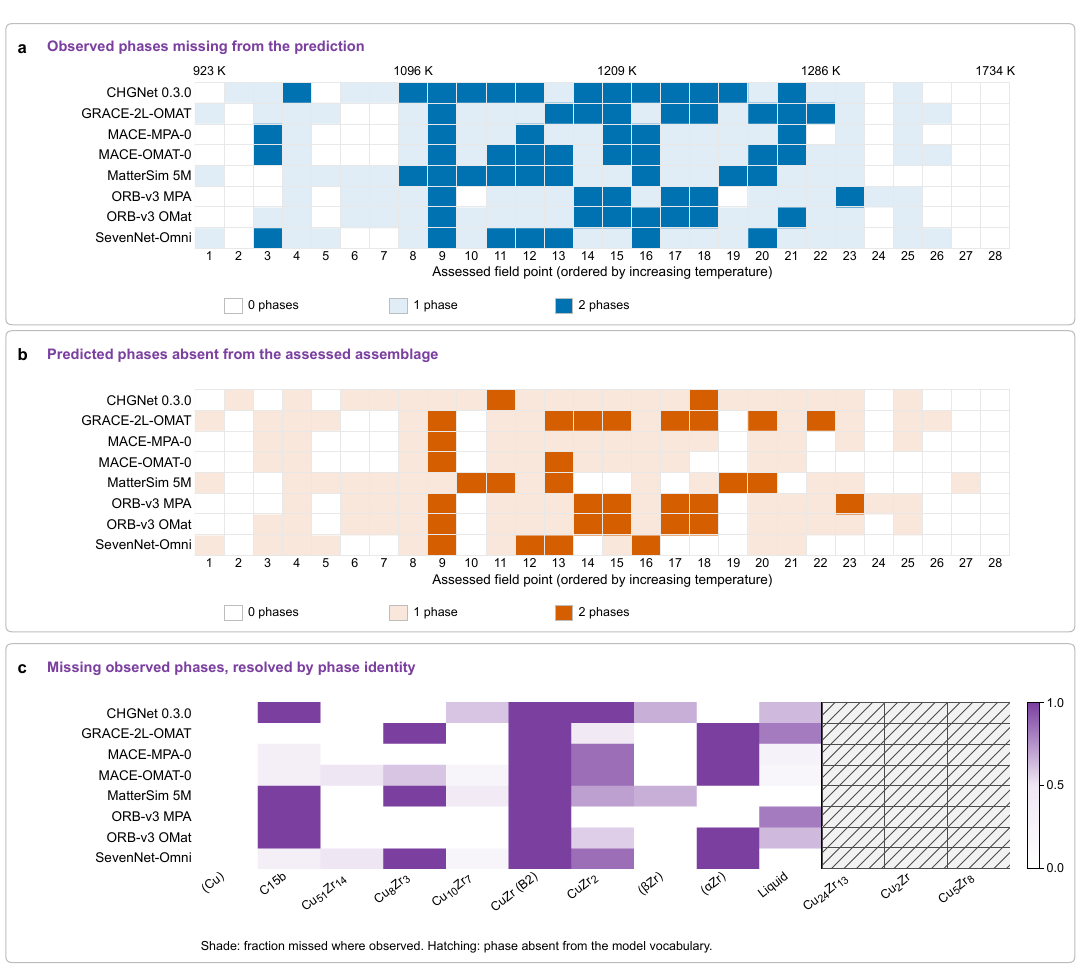}
\caption{\textbf{Finite-temperature errors depend on the state and on the phases represented by the model.} Eight one-shot MLIP--PhaseForge descriptions are evaluated at the same 28 assessed Cu--Zr field points, ordered by temperature and then composition. \textbf{a}, Number of observed phases missing from the prediction. \textbf{b}, Number of predicted phases absent from the assessed assemblage. \textbf{c}, Fraction of observations of a phase at which that phase is missed. Hatched cells identify phases absent from the model vocabulary; they are retained in the field scores but distinguished from represented phases that are predicted incorrectly. C15b and CuZr$_2$ use the physical-phase grouping of the benchmark. The panels compare state-specific errors within Cu--Zr; the overlapping aggregate-score intervals do not resolve a ranking of the MLIPs. Point coordinates, phase supports and grouping rules are given in SI~1.}
\label{fig:mlip-failures}
\end{widefigure}

\section{From ground-state energies to finite-temperature MLIP benchmarks}
\label{sec:benchmark}

Experimental formation enthalpies provide a common reference for comparing compound energies predicted by first-principles calculations and machine-learned interatomic potentials (MLIPs) (Fig.~\ref{fig:mlip-parity}).
Determining which phases coexist at finite temperature, however, requires the relative Gibbs energies of competing compounds, solid solutions and liquids.
Their balance determines the phase boundaries and invariant reactions, providing a test of how the computational description represents competition among phases.
ORCHID supplies both formation enthalpies and observations of finite-temperature equilibria, allowing us to examine whether agreement with the former carries over to predictions of the latter (Fig.~\ref{fig:mlip}).

\subsection{What formation-energy agreement can distinguish}
Comparing formation-energy errors among models requires accounting for differences in chemistry, structure matching and database reference conventions.
We evaluated eight potentials on a common subset of 1,138 experimental records representing 455 compounds in 185 binary systems, using relaxed Materials Project structures and the specified MP2020 correction convention.
Evaluating the same compounds removes differences in chemical coverage as a source of variation among models.
Uncertainty intervals were obtained by resampling chemical systems together with all their records, preserving the dependence among observations within a system.
These statistics describe the matched benchmark, while SI~1 reports the full DFT calibration, including the effects of chemical stratum, structure matching and reference conventions on the associated uncertainty weights.

The parity plots show similar deviations from experiment for DFT and the MLIPs (Fig.~\ref{fig:mlip-parity}).
Across the common benchmark, the robust spreads of MLIP formation-enthalpy residuals range from 8.6 to 9.7~\kjma, compared with 8.8~\kjma\ for the Materials Project (Fig.~\ref{fig:mlip}a).
All system-bootstrap intervals for the difference in spread between an MLIP and the Materials Project include zero.
The residuals are also strongly correlated with those of the Materials Project, with Pearson correlations of 0.961--0.997 (Fig.~\ref{fig:mlip}b).
These spreads exceed the effective calorimetry floor of 3.2~\kjma\ used in this comparison.
Within this benchmark, the potentials thus approach the agreement of their DFT reference with experiment, while their similar formation-energy errors provide little basis for distinguishing their ability to predict phase stability.

Evaluating these potentials for materials discovery therefore requires testing the phase stability that follows from their predicted energies.
Frameworks such as Matbench Discovery\cite{riebesell2025matbench} compare formation-energy and 0~K crystal-stability predictions, while observed phase boundaries, invariant reactions and phase constitutions provide additional tests over temperature and composition.
ORCHID supplies the experimental states needed for these comparisons, and PhaseForge connects MLIP predictions to the CALPHAD calculations through which their finite-temperature equilibria can be evaluated.

\subsection{One-shot phase-stability prediction through PhaseForge}
We demonstrate this connection for Cu--Zr by using PhaseForge to construct CALPHAD descriptions from eight MLIPs through the same procedure (Fig.~\ref{fig:espei}a)\cite{zhu2025machinelearningpotentials,zhu2025accelerating,sariturk2025phaseforge}.
The descriptions combine compound and solid-solution energies, liquid molecular dynamics and SGTE unary references, without fitting to the ORCHID observations used for evaluation.
We use ``one-shot'' to denote this absence of fitting to the benchmark evidence, rather than the number of calculations or the absence of assumptions in the free-energy model.
Applying a common construction procedure allows the resulting workflows to be evaluated by comparing their predicted equilibria with the phases recorded at specified temperatures and compositions (Fig.~\ref{fig:mlip}c).

For each phase $p$, the recall $R_p$ is the fraction of evidence points containing that phase at which the model also predicts its presence.
The prevalence-weighted score is $S=\sum_p n_pR_p/\sum_p n_p$, where $n_p$ is the number of evidence points containing the phase.
This score therefore measures the fraction of observed phase occurrences recovered by the model\cite{zhu2025machinelearningpotentials}.
Precision and F1 are reported alongside recall to account for predictions of additional phases absent from the observed assemblage.
Points within assessed phase fields and points on boundaries are evaluated separately, with the tolerance rules and bootstrap procedures documented in SI~1.
The benchmark thus evaluates the phases predicted at specified thermodynamic states, complementing the comparison of formation energies.

\begin{widefigure}[!t]
\nfig{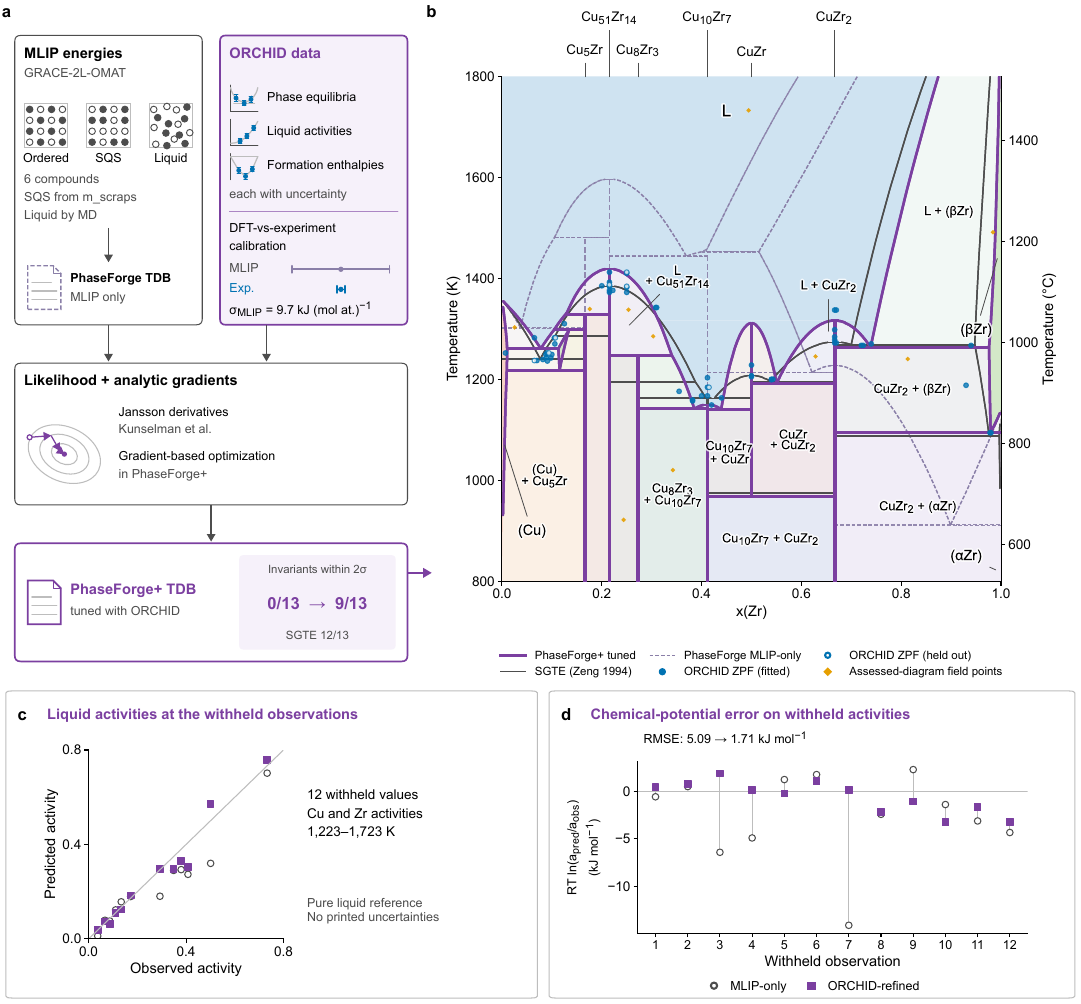}
\caption{\textbf{Refinement with ORCHID, illustrated for Cu--Zr.} \textbf{a}, PhaseForge constructs the starting description from MLIP compound, solid-solution and liquid energies on SGTE unary references. PhaseForge+ combines these with uncertainty-weighted ORCHID observations using the gradient-capable ESPEI development branch. \textbf{b}, Refined diagram (violet), initial MLIP description (dashed) and SGTE assessment (grey), with fitted and held-out literature points and assessed assemblages. Refinement recovers \cuzrInvFinal{} of \cuzrInvTotal{} invariant temperatures within $2\sigma$ and reduces held-out activity error from \cuzrActHoldMlip{} to \cuzrActHoldFinal~kJ\,mol$^{-1}$. \textbf{c}, Predicted and observed activities at all 12 withheld Cu and Zr observations, evaluated at their recorded temperatures and compositions with pure liquid reference states. \textbf{d}, Corresponding chemical-potential residuals, $RT\ln(a_\mathrm{pred}/a_\mathrm{obs})$; lines connect the initial and refined predictions for an observation. The sources provide no measurement uncertainties for these points. Holdouts are individual observations within Cu--Zr and may share sources with fitted evidence. The final fit adds assessed-field points after the preregistered evaluation, so its assessed-field recall is in sample. Remaining phase-stability discrepancies are discussed in the text; SI~1 documents construction, fitting and data splits.}
\label{fig:espei}

\end{widefigure}

At the 28 assessed Cu--Zr field points, the eight one-shot descriptions obtain weighted recalls of 0.35--0.56, compared with 0.72 for the SGTE assessment and 0.70--0.98 for nine later assessments (Fig.~\ref{fig:mlip}c).
The overlapping intervals do not resolve an ordering among the eight MLIPs, but examination of the predicted equilibria reveals distinct errors in phase stability.
Some descriptions underestimate liquid stability, while others overestimate it.
All eight descriptions leave B2 CuZr unstable at the four tested field points where it is observed, despite including this phase in their models.
Three other assessed phase classes are absent from the common starting model vocabulary, making their omission a consequence of the construction procedure.
Figure~\ref{fig:mlip-failures} separates these errors by state and phase identity, distinguishing missing observed phases from additional predicted phases.
These comparisons show how finite-temperature equilibria can reveal differences among workflows with similar aggregate formation-energy errors.

PhaseForge+ established the refinement of MLIP-derived CALPHAD descriptions using thermodynamic evidence\cite{sariturk2025phaseforgeplus}.
ORCHID extends this capability to the recovery and preparation of that evidence, connecting observations dispersed across the literature to the states, phase identities and uncertainty estimates required for fitting.
To our knowledge, this is the first integration of literature-derived thermodynamic evidence at this scale with the evaluation and refinement of MLIP-derived descriptions.
The Cu--Zr demonstration shows how this prepared evidence can enter gradient-based refinement directly.
This integration addresses both the preparation of the evidence and the optimization of the model, tasks that can otherwise require months of manual work.
It points toward a mode of CALPHAD assessment in which descriptions can be assembled and updated as new experimental and computational evidence becomes available.

\section{Using the evidence to refine thermodynamic descriptions}
\label{sec:refinement}

The comparison of predicted and observed phase stability also provides information that can be used to refine a thermodynamic description. This requires a choice of the observations to be included in the optimization and of the uncertainty assigned to them, while preserving separate evidence for evaluation. Since the ORCHID observations are already compatible with ESPEI, their thermodynamic states do not need to be reconstructed from the literature to perform this refinement. The retained uncertainty budgets determine the relative influence of experimental and computed information in the fit. The approach can be used to revise an existing assessment or refine a description initialized from first-principles or MLIP information. Here, we consider the latter case for Cu--Zr using PhaseForge+, with application to other systems requiring the corresponding state and phase mappings.

For this demonstration, PhaseForge+\cite{sariturk2025phaseforgeplus} starts from the GRACE-based description used in the benchmark.
ORCHID phase equilibria, activities and formation enthalpies constrain the refinement of liquid and solid-solution interaction parameters, compound energies and a CuZr entropy term\cite{kunselman2025construction}.
The optimization uses the analytical-gradient ESPEI implementation introduced by Kunselman et al.\cite{kunselman2025gradient}, through a development branch\cite{kunselman2024espeibranch} distinct from the released version used for the dataset compatibility checks.
Experimental formation enthalpies govern compound energies when both experimental and MLIP values are available, while calibration against experiment determines the uncertainties assigned to computed energies.
The fit uses a fixed data and calibration snapshot \cuzrFitRelease{}, preserving the connection between the reported results and the evidence used to obtain them as the collection is updated.

Refinement improves agreement with both the invariant temperatures and the observations withheld from fitting (Fig.~\ref{fig:espei}).
\cuzrInvFinal{} of the \cuzrInvTotal{} invariant temperatures fall within twice their uncertainty after refinement, compared with none for the starting MLIP description.
The held-out activity error decreases from \cuzrActHoldMlip{} to \cuzrActHoldFinal~kJ\,mol$^{-1}$, while the held-out phase-equilibrium error decreases from \cuzrZpfHoldMlip{} to \cuzrZpfHoldFinal~kJ\,mol$^{-1}$.
The corresponding SGTE errors are \cuzrActHoldSgte{} and \cuzrZpfHoldSgte~kJ\,mol$^{-1}$, respectively, and SGTE recovers \cuzrInvSgte{} invariant temperatures within $2\sigma$.
These results demonstrate that uncertainty-weighted observations from ORCHID can bring an MLIP-derived description into substantially closer agreement with experimental thermodynamics.

The evaluation distinguishes agreement with fitted evidence from predictions of withheld observations.
The weighted recall of 0.76 at the assessed field points measures agreement within the fitting set, while the activity and phase-equilibrium holdouts evaluate observations excluded from optimization within Cu--Zr.
The final model extends the preregistered fit, which recovered \cuzrInvPrereg{} of the \cuzrInvTotal{} invariant temperatures, through additional refinement performed after that initial evaluation; both stages remain in the demonstration record.
The remaining discrepancies identify specific targets for further improvement: the CuZr congruent melting temperature is \cuzrCongruentHigh~K too high, the eutectoid decomposition temperature is \cuzrEutectoidLow~K too low, and a narrow spurious ($\beta$Zr) field persists.
These results establish the improvement achieved for Cu--Zr and provide a basis for extending the evaluation to other chemical systems and phase models.

Traditional CALPHAD assessment often proceeds through staged parameter optimization, in which the assessor fixes some parameters while adjusting others, progressively releasing or revisiting parameter groups as the description develops\cite{lukas2007}.
This intervention reflects the difficulty of minimizing an error function whose dependence on model parameters is mediated by phase-equilibrium calculations.
Changes in the predicted phase assemblage can complicate the optimization, making convergence sensitive to the initial parameters and the sequence in which they are adjusted.
Much of the assessor's thermodynamic intuition therefore enters through the management of the fitting procedure itself.
The analytical-gradient ESPEI implementation\cite{kunselman2025gradient} provides derivative information to guide parameter updates, enabling the programmatic refinement used here, while ORCHID supplies the interpreted evidence and uncertainty estimates that define the fitting problem.

The complete workflow proceeds programmatically from MLIP predictions to a CALPHAD description constructed by PhaseForge and refined by PhaseForge+ using ORCHID observations.
With the initial description and fitting datasets prepared, the final Cu--Zr optimization completed in approximately 27 minutes on an Apple M1 Max Mac Studio (SI~1).
In our experience, conventional assessment from literature collection through final fitting can require months of expert work.
Bringing evidence preparation and gradient-based optimization into the same workflow makes rapid model assembly and refinement reproducible, with the source snapshot and fitting weights preserved and observations reserved for evaluation.

This integration provides a route toward continuous thermodynamic assessment.
When a source is added or corrected, its observations can be connected to the affected phases and models, checked for compatibility and incorporated into an updated fitting and evaluation problem.
Decisions about model form, contradictory evidence and missing phases remain guided by scientific criteria and expert judgment, while the workflow executes the corresponding calculations.
Retaining the evidence snapshot, software version, model form and fitting choices for each release makes these updates reproducible.
New or reinterpreted ORCHID observations can then enter PhaseForge+, and the refined description can be evaluated against its predecessor to establish what the new evidence improves.

\section{Reuse and prospects for materials discovery}
\label{sec:usage}

\begin{widefigure}[!t]
\nfig{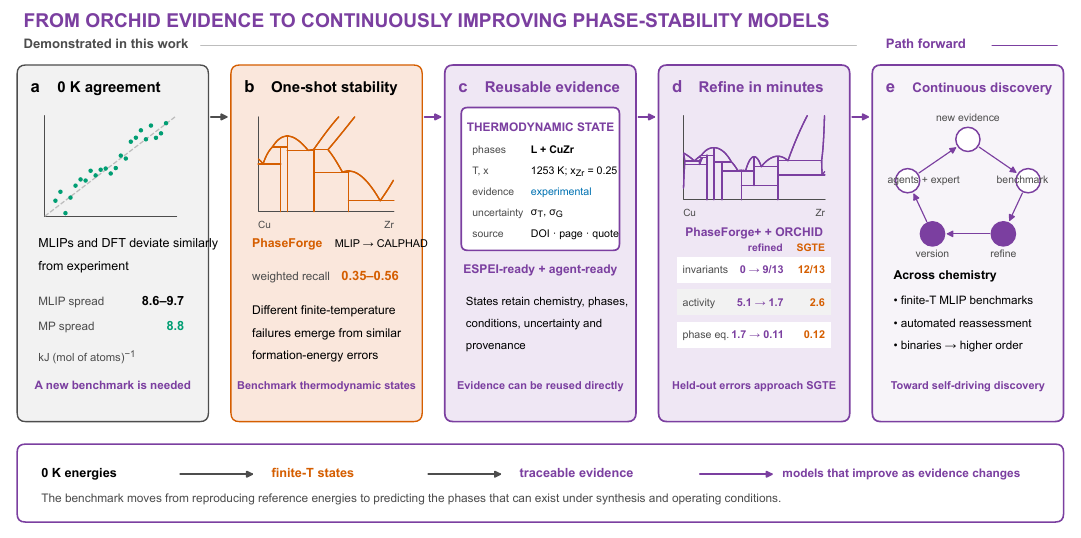}
\caption{\textbf{From ground-state agreement to continuously improving phase-stability models.} \textbf{a}, On the matched formation-enthalpy set, the residual spreads of eight MLIPs are similar to that of their Materials Project reference, limiting the ability of a 0~K energy benchmark to distinguish the models. \textbf{b}, One-shot PhaseForge descriptions expose different finite-temperature failures when evaluated against observed thermodynamic states. \textbf{c}, ORCHID makes these states reusable by retaining their chemistry, phases, conditions, uncertainty and provenance in forms that can be consumed by ESPEI and by agents. \textbf{d}, In the Cu--Zr demonstration, PhaseForge+ refines the prepared MLIP-derived description against this evidence in approximately 27 minutes. The refined model recovers 9 of 13 invariant temperatures within $2\sigma$, and its held-out activity and phase-equilibrium errors approach those of the SGTE assessment; the final fit is post hoc and its assessed-field score is in sample. \textbf{e}, The prospective loop extends this connection across chemistry by combining agents and expert judgment with new evidence, finite-temperature benchmarks, refinement and versioned releases. Panels \textbf{a--d} summarize results demonstrated here; panel \textbf{e} describes the path forward, for which cross-system performance remains to be established.}
\label{fig:synthesis}
\end{widefigure}

\begin{figure}[!t]
\centering
\nfig{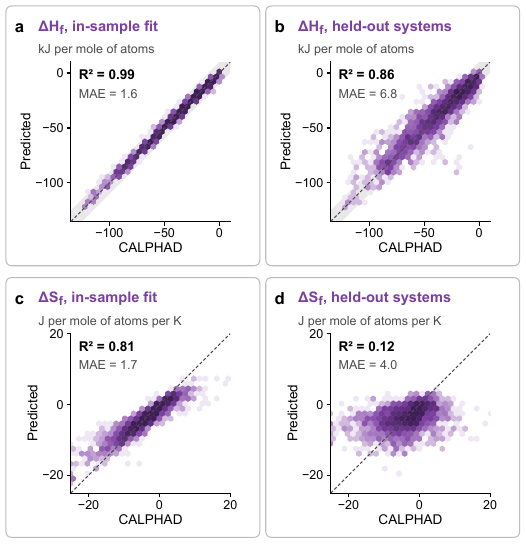}
\caption{\textbf{Binary compound thermochemistry as training data.} Formation enthalpy ($\Delta H_f$; \textbf{a}, \textbf{b}) and formation entropy ($\Delta S_f$; \textbf{c}, \textbf{d}) at 298.15~K relative to SER, predicted by models trained on 2,533 ($\Delta H_f$) and 2,490 ($\Delta S_f$) intermetallic compounds from 636 binary systems in ORCHID-Hulls and compared with the CALPHAD values. \textbf{a}, \textbf{c}, Predictions for the compounds used in training. \textbf{b}, \textbf{d}, Out-of-fold predictions with entire binary systems withheld (five folds grouped by system); the metrics are means over three random fold assignments, and the mean absolute error (MAE) is in the units of each row. Shading shows the density of points on a logarithmic scale, the dashed line marks 1:1 agreement and the grey band $\pm$10~kJ per mole of atoms. Entropy labels fixed by the Neumann--Kopp convention are excluded, and 18 entropy values outside the plotted window are drawn at its edge.}
\label{fig:imc}
\end{figure}

Putting continuous refinement into practice requires access to observations together with the information needed to interpret them (Fig.~\ref{fig:synthesis}).
ORCHID provides downloadable records and machine-readable metadata, complemented by a Model Context Protocol server through which programs and AI agents can query chemical systems, phase identities and associated observations.
Provenance, evidence classes and uncertainty budgets accompany the returned values, allowing an agent to investigate discrepancies and select evidence for fitting or evaluation.
Researchers can access the same information directly, so both forms of use draw on a common thermodynamic record.
These interfaces support the objectives of FAIR and AI-ready data\cite{wilkinson2016,esip2022}, with formats and usage documented in SI~2.

Provenance also guides the design of machine-learning evaluations because thermodynamic observations can share experimental methods, samples or an underlying measurement.
These dependencies extend across papers when later authors reproduce previously reported values.
The released paper and system folds support tests on sources or chemical systems absent from training, while grouping by the original experimental work and canonical evidence identity provides stricter control of shared observations.
Random record-level splits can distribute related observations across training and test sets, giving an apparent performance that does not establish transfer to unseen literature or chemistry.
Preserving these relationships allows researchers to construct evaluations that match the intended use of the model.
In addition to supporting the assessment of individual systems, the binary records in ORCHID can be used to train models that estimate the thermochemistry of compounds in systems that have not yet been assessed (Fig.~\ref{fig:imc}).
To illustrate this, we trained gradient-boosting and random-forest models on the formation enthalpies and entropies at 298.15~K of 2,533 intermetallic compounds in 636 binary systems, using elemental properties and the Miedema chemical enthalpy as descriptors.
For the compounds included in the training set, the predictions of these models agree closely with the CALPHAD values of the enthalpy and appear to describe the entropy satisfactorily.
In their intended application, however, these models are applied to systems that are not part of the training set, and when entire binary systems are withheld from training the behaviour of the two properties becomes quite different.
The enthalpy model retains much of its accuracy, with a mean absolute error of 6.8~kJ per mole of atoms that improves on the 9.9~kJ per mole of atoms of the Miedema estimate, whereas the error of the entropy model is only about 10\% below that of a prediction equal to the mean of the training data.
This behaviour of the entropy model is consistent with the way in which formation entropies are obtained in many assessments, where they are adjusted to reproduce the phase boundaries and are not constrained by thermal measurements.
The estimation of formation entropies for ternary and higher-order compounds will therefore require information beyond that contained in the binary assessments.

The predominantly binary coverage of ORCHID provides a foundation for extending the collection to higher-order systems.
Its \nDiagramSystems{} binaries with calculated diagrams define approximately \nTernFromDiagramsApprox{} ternary combinations for which all three constituent binaries have assessments.
Including literature and handbook coverage increases the number of ternary combinations supported by their constituent binaries to approximately \nTernFromUnionApprox{}.
Developing thermodynamic descriptions of these systems also requires evidence for genuinely ternary phases and the corresponding model parameters\cite{dshemuchadse2015ternary,omprakash2026beyond}.
Once binary and ternary descriptions are established, the lower frequency of genuinely higher-order phases may make further extrapolation more tractable\cite{steurer2016intermetallics}.
Extending coverage to higher-order observations, measurements available only in plots and kinetic properties would broaden the range of scientific questions the resource can address.

Self-driving materials discovery requires connecting proposed structures to the phase assemblages that can be obtained and retained under synthesis and processing conditions.
Finite-temperature thermodynamics establishes the competition among these phases, while kinetics and processing history determine the pathways through which stable or metastable structures can be reached.
ORCHID provides the thermodynamic evidence needed to develop this connection and a foundation for linking it to kinetic and processing data within ICME workflows.
By preserving the observations and interpretations underlying model predictions, it allows successive assessments to build on accumulated knowledge and supports the use of that knowledge in automated decisions about materials synthesis and processing.

\section{Conclusions}
\label{sec:conclusions}

ORCHID organizes thermodynamic evidence dispersed across the literature around the states described by the observations.
Thermodynamic constraints guide its recovery and interpretation, while source verification, expert review and controlled evaluations establish the basis for its use.
Retaining phase identities, provenance, evidence classes and uncertainty allows observations from different sources to be brought together without losing their individual meaning.
The ESPEI-compatible datasets and programmatic interfaces connect this evidence to CALPHAD assessment, making the same thermodynamic information available to researchers and AI agents.

The computational comparisons demonstrate why finite-temperature phase stability provides an essential complement to formation-energy benchmarks.
Eight MLIPs exhibit strongly correlated formation-enthalpy errors with spreads close to their DFT reference, yet their PhaseForge descriptions reveal distinct errors in the competition among phases in Cu--Zr.
Using ORCHID evidence, refinement through PhaseForge+ recovers \cuzrInvFinal{} of \cuzrInvTotal{} invariant temperatures within $2\sigma$ and reduces the held-out activity error from \cuzrActHoldMlip{} to \cuzrActHoldFinal~kJ\,mol$^{-1}$.
The final gradient-based refinement completed in approximately 27 minutes with the initial description and fitting datasets prepared.
This demonstration connects literature-derived evidence to rapid thermodynamic assessment and provides a reproducible procedure for extending the evaluation across chemical systems.

Connecting reusable thermodynamic evidence to programmatic model refinement opens a path toward continuously updated CALPHAD descriptions.
New measurements, corrected interpretations and improved computational predictions can enter successive assessments while preserving the evidence and modeling choices behind each revision.
ORCHID provides the foundation for this accumulation of knowledge, allowing the work invested in recovering and interpreting a source to support many subsequent assessments.
This capability brings thermodynamic model development closer to the pace of computational materials discovery and strengthens the connection between predicted materials, their phase stability and the conditions needed for synthesis and processing.

\section*{Data availability}
The datasets are released under CC~BY~4.0 and described in SI~2, with the copyright status of quoted source passages retained separately.
Versioned data are identified through the umbrella record (\doiUmbrella) and the individual dataset records listed in SI~2.
The data site is \url{https://data.orchid-project.org}, and repositories are under \url{https://github.com/orchid-calphad}.
The analyses and demonstrations retain their data snapshots so that later corrections do not silently change reported benchmark or fitting results.

\section*{Code availability}
The code is MIT-licensed under \url{https://github.com/orchid-calphad}.
Extraction, data construction, agent access, synthetic benchmarks and the runnable scientific examples are documented in SI~2.
SI~1 records the versions and development branches used for the calculations, including the gradient-capable ESPEI branch used for refinement.

\section*{Author contributions}
R.A., J.S. and D.S. conceived ORCHID and have worked jointly on all aspects of the project. All authors reviewed and approved the manuscript.

\paragraph{Use of AI tools.} AI tools were used extensively in this work, as described in the Methods: language models read the papers and the diagrams under the constraints and checks described there, and coding assistants (Claude Code and Codex) wrote and tested much of the software and drafted text that the authors edited. The authors take responsibility for all content, and the numbers in this paper are generated from the released files by scripts and were checked in the audits described above.

\section*{Conflicts of interest}
There are no conflicts to declare.

\section*{Acknowledgements}
This work was supported by the U.S. National Science Foundation under awards OAC-2608825 (CSSI Elements: PhaseForge) and DMR-2534344 (Autonomous Robotic Metallurgist: An NSF Materials Innovation Platform, ARM-MIP).
We thank B.~Hallstedt and the SGTE for releasing the binary collection under an open licence, and the Materials Project and OQMD teams for their open data.
We thank R.~Otis and B.~Bocklund for pycalphad and ESPEI, on which the calculated diagrams and the ESPEI datasets of ORCHID are built. We also thank C.~Kunselman, whose gradient-based optimization for ESPEI (pull request~\#268, with contributions from B.~Bocklund) makes likelihood gradients nearly as inexpensive as the likelihood itself and thereby allows assessments to be refined at a scale that MCMC sampling alone does not reach.

\clearpage
\onecolumn
\section*{Appendix: ORCHID data products and access}
\addcontentsline{toc}{section}{Appendix: ORCHID data products and access}

\begin{figure}[h!]
\centering
\nfig{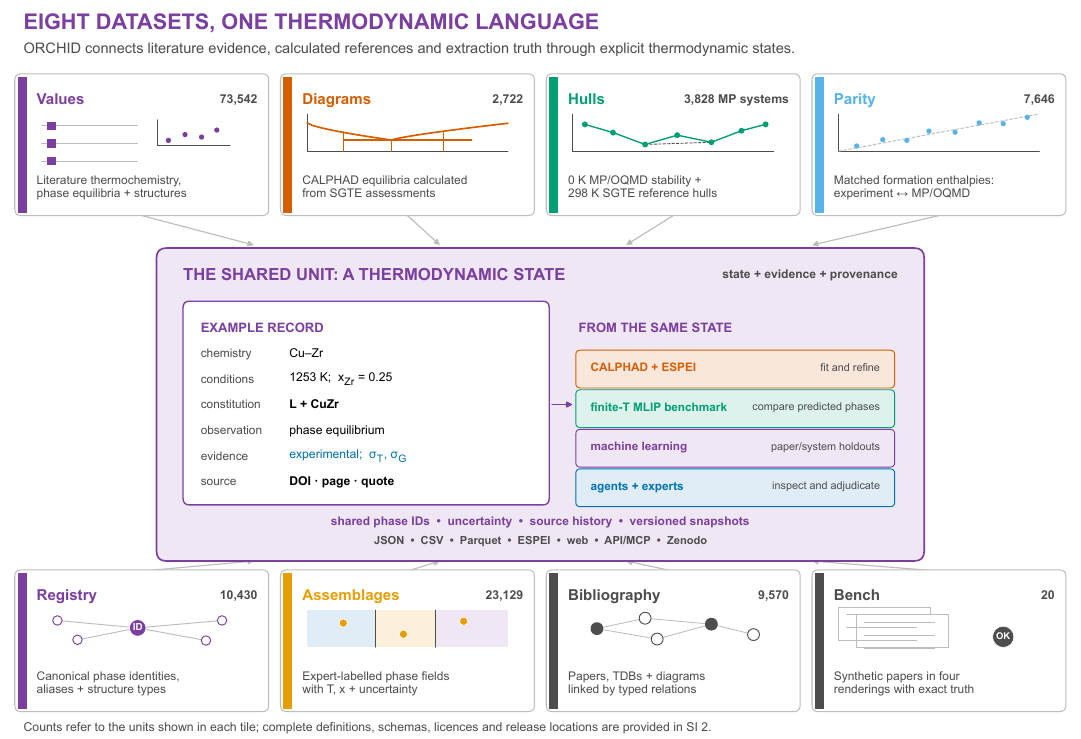}
\caption{\textbf{Eight datasets connected through a common thermodynamic language.} Each tile represents one released ORCHID data product and includes a visual summary of its scientific object. These products meet in an explicit thermodynamic state that retains the chemical system, conditions, phase constitution, observation, evidence class, uncertainty and source. Shared phase identities and provenance allow the same state to enter CALPHAD fitting, finite-temperature MLIP benchmarking, machine learning and joint analysis by agents and experts. Counts describe values, calculated diagrams, Materials Project systems, parity pairs, phase identities, assemblage points, papers or synthetic benchmark documents, as indicated in the corresponding tile.}
\label{fig:dataset-guide}
\end{figure}

The eight ORCHID datasets serve different scientific roles, but their connection is established by the thermodynamic state rather than by the folders in which they are distributed. As summarized in Fig.~\ref{fig:dataset-guide}, \dsValues{} and \dsAssemblages{} preserve reported observations, while \dsDiagrams{}, \dsHulls{} and \dsParity{} provide calculated references for their interpretation. \dsRegistry{} and \dsBibliography{} connect these observations to phase identities and sources, and \dsBench{} supplies documents with known answers for evaluating extraction.

Across these products, record, phase and source identifiers connect observations to their chemistry, conditions, phase constitution, evidence class, uncertainty and provenance. The present release establishes these connections through the recorded state attributes; permanent state identifiers issued by an append-only registry are planned for a subsequent release. This representation supports inspection against the source, export of eligible observations to ESPEI and evaluation of finite-temperature predictions. The file-level inventory, schemas, licences, release locations and programmatic access routes are given in SI~2.

\clearpage
\paragraph{What the datasets contain.}
The units counted in these products reflect different scientific objects. A literature value is an observation, a parity pair links an observation to a calculated entry, and a diagram or hull describes a chemical system under specified modeling assumptions. Their counts therefore cannot be added to obtain a number of independent measurements.

\begingroup
\small
\renewcommand{\arraystretch}{1.15}
\noindent\begin{tabularx}{\linewidth}{P{2.1cm}YY}
\toprule
Dataset & Scientific object & Principal use \\
\midrule
\dsValues{} & A reported property value with its state, source and evidence class & Select observations for assessment; export the eligible subset to ESPEI \\
\dsAssemblages{} & A phase assemblage at a temperature and composition in an assessed diagram & Test finite-temperature phase predictions or constrain a fit \\
\dsDiagrams{} & A phase diagram calculated from an identified TDB assessment & Compare reported equilibria with model predictions \\
\dsHulls{} & A convex hull and its entries for a chemical system and reference convention & Examine competing phases and calculated energetic stability \\
\dsParity{} & A literature formation enthalpy paired with a matched DFT entry & Compare computed and reported energies; calibrate DFT uncertainty \\
\dsRegistry{} & A phase identity linked to reported names and TDB models & Resolve names across sources and map observations to models \\
\dsBibliography{} & A source work and typed relations among works & Trace provenance and recognize shared or reprinted evidence \\
\dsBench{} & A synthetic document with known numerical and phase information & Evaluate extraction against an answer key \\
\bottomrule
\end{tabularx}
\endgroup

\paragraph{Following an observation into model evaluation.}
The archived Cu--Zr demonstration provides a concrete example. At 1573~K and $x_{\mathrm{Zr}}=0.21$, the activity record attributed to Zaitsev (2003) reports $a_{\mathrm{Cu}}=0.733$ and retains the observation identifier \texttt{ORC-5OZGTSNZVA.3a2b7e}. In the ESPEI activity dataset, the component, temperature and composition are accompanied by the pure-liquid reference state. This observation belongs to the held-out set shown in Fig.~\ref{fig:espei}c,d, where the initial and refined models predict activities of approximately 0.702 and 0.758, respectively. No printed measurement uncertainty is available for this observation, and none is inferred from its agreement with either model. The retained source and observation identifier connect the numerical comparison to the evidence, while the held-out assignment distinguishes its use in evaluation from the observations used to determine the model parameters. The archived inputs and uncertainty treatment for the demonstration are documented in SI~1.

\paragraph{Choosing evidence for a task.}
For CALPHAD fitting, begin with the ESPEI-compatible export of \dsValues{}, check the phase mapping to the intended TDB model and select evidence classes and uncertainty weights appropriate to the assessment. Assessed-diagram constraints from \dsAssemblages{} can be included as a distinct evidence class. For MLIP evaluation, use matched formation-enthalpy observations from \dsParity{} and finite-temperature observations from \dsValues{} or \dsAssemblages{}, reserving evaluation evidence before fitting. For extraction-method evaluation, use \dsBench{} and its known answers. SI~1 gives the validation and calibration procedures, and SI~2 describes retrieval and evidence selection, including source- and system-based splits for tests of transfer to new literature or chemistry.

\clearpage
\twocolumn

\FloatBarrier
\clearpage
\renewcommand{\mainSecRef}[2]{Section~\ref{#1}}
\renewcommand{\mainFigRef}[3]{Fig.~\ref{#1}#3}
\renewcommand{\suppnotesection}[2]{\section*{Supplementary Note #1: #2}\addcontentsline{toc}{section}{Supplementary Note #1: #2}}
\setcounter{figure}{0}\setcounter{table}{0}
\renewcommand{\thefigure}{S\arabic{figure}}
\renewcommand{\thetable}{S\arabic{table}}
\twocolumn[\begin{center}{\Large\bfseries\sffamily Supplementary Information 1: Methods and additional validation\par}\end{center}\vspace{1em}]
\section*{Scope and organization}
\addcontentsline{toc}{section}{Scope and organization}
\noindent \siThisDocument{} documents the extraction, uncertainty calibration and validation underlying the main paper.
Supplementary Notes~1--4 retain the extraction history, annotation procedures, audit findings and calculation details.
The original Supplementary Figs.~S1--S7 and Supplementary Tables~S1--S8 are followed by the detailed Methods and Validation sections, their additional figures and tables, the matched MLIP benchmark and the full construction and refinement figure notes.
SI~2 describes the repository organization and access conventions.

Counts use the shared manuscript snapshot in \texttt{numbers\_DD.tex}, derived from \texttt{numbers.tex} generated by \texttt{scripts/make\_numbers.py} from the released files.
The exceptions are the window-rule counts in Supplementary Table~S3 and the reader-pilot and sign-check counts in Supplementary Note~1, which come from reports of the extraction record store that are not released, and the recovery-trial counts in Supplementary Note~4, which come from a separate calculation audit.
The scientific figures are retained from those generated by \texttt{scripts/make\_figures.py}, with style definitions in \texttt{scripts/style.py}.
The provenance and release status of these supporting records are stated where the results are discussed.

\suppnotesection{1}{reading the values}
\label{snote:reading}
This note documents the empty-answer failure detected in one reader, the comparison of the two text readers and the verification of signs against printed pages.

\paragraph{Empty reads under one prompt version.} Reader agreement can support acceptance of a value only if both readers return data.
Under one version of the extraction prompt, Gemini returned a single placeholder row for tables that Claude transcribed in full.
The synthetic benchmark first exposed this failure, and an audit of the production records identified 98 affected tables in 79 papers.
Because the sanity filter counted rows rather than values, it had treated these empty reads as votes.

To correct this failure, two subsequent pipeline releases stopped sending the affected prompt to Gemini and changed the filter to count values.
Under this revised filter, an answer without values was treated as a failed read requiring a tie-break.
All table rows supported by only one visual reader were also sent to an independent second reader using the same crop, prompt and schema.
To document the outcome of these second readings, the added field \texttt{reader\_agreement} records the resulting support in release v2026.10, with \texttt{multi\_reader} for \nRAMulti{} values, \texttt{single\_reader} for \nRASingle{} and \texttt{single\_reader\_text\_layer} for \nRASingleText{}.
The \nRANotClassified{} values in unsettled invariant-table rows have no agreement classification and none is accepted.
The affected prompt was \texttt{thermo-table-v7.1}, and the corrections were introduced in pipeline releases v2.1.19q and v2.1.19r.
The record-store audit is documented in \path{build/sgte-corpus/audit/v71-empty-rows-20261022/REPORT.md}, which is not a released file.

\paragraph{Reader comparison pilot.} To evaluate replacing Gemini with Claude as the second text reader, we compared the readers on the same 127 requests.
On 94 passages of running text, they agreed in 89 cases on whether the passage contained data.
For these passages, Claude returned 553 rows and Gemini 388, with 491 distinct values found by both, 280 found only by Claude and 28 found only by Gemini.
About 17 of the values found only by Gemini were errors.

Verification of the returned quotes provided a further comparison of the readers.
The quotes were verbatim in 546 of Claude's 553 rows and 306 of Gemini's 388, with failures arising mostly from control characters in the PDF text layer.
In 32 passages, these characters also led to nine sign errors for Gemini and none for Claude.
Gemini, however, itemized samples somewhat better in passages describing experimental conditions.
The pilot compares the readers with one another and does not establish an independent human precision estimate for numerical extraction.
The comparison used Claude Opus and Gemini Flash on 22 October 2026 and is documented in \texttt{claude-reads/pilot-20261022/PILOT\_RESULTS.md}, which is not a released file.

\paragraph{Printed-page sign checks.} Loss of a minus sign in the PDF text layer changes the interpretation of a thermochemical value.
To check for this error, candidates identified by either of two independent criteria were examined individually on page images.
The set contained 144 candidates from 76 papers, with 67 identified by the first criterion, 122 by the second and 45 by both.
Of these candidates, 39 had a printed minus sign that the text layer had lost, either by dropping the glyph (22 cases), splitting it across a line break (9), or reading it as the digit 2 (8).
The other 105 were printed without a sign, mostly as magnitudes under the convention of the paper.

A sign verified on the page is recorded with \texttt{flag\_sign\_pdf\_page\_verified}.
Values printed without a sign are retained as printed and flagged \texttt{unsigned\_in\_print}.
In release v2026.10, \nSignVerified{} values have a page-verified sign, including \nGateSignFromPage{} set by the literature-level gates in Supplementary Note~3, and \nUnsigned{} are flagged as unsigned in print.
The source is the sign-check record \texttt{sign\_check.json}, which is not a released file.

\suppnotesection{2}{annotating assessed diagrams}
\label{snote:annotation}
Most fields of a handbook diagram can be named using the rules of phase diagrams alone.
For the fields left unresolved by these rules, pairs of vision models select among the permitted assemblages.
For these unresolved fields, disagreement between readers leads to successive readings by Gemini~3.1~Pro with Claude~Haiku~5.5, two Claude~Opus~5.5 readers, and then Claude~Fable~5.1 with GPT-6~Astra.
Agreement at this last stage is retained only as a proposal.
The rules then propagate names into thin fields, after which an adjudicator (Claude~Opus~5.5) examines the whole diagram, its phase table and all reader answers.
Based on this information, the adjudicator names the field, identifies a tracing fault or an undetermined region, or passes the field on, recording a reason for the decision.

For low-temperature fields below the range of the drawing, the adjudicator consulted Materials Project and OQMD hulls to identify the next stable phase.
Names obtained in this way carry the basis \texttt{dft\_hull\_corroborated}.
Their confidence is high when the two hulls agree and low when they disagree.

These names were evaluated through expert verification, an agent audit and a separate blind expert sample.
In the verification step, a domain expert reviewed the 126 hardest fields with the adjudicator's reason shown and accepted 92 of the 93 named proposals (a median of 41~s per field over the first 20 fields).
The accepted proposals included 19 of the 22 fields named with the help of the hulls and all 13 for which the two hulls agreed.

To examine agreement across naming routes, a blind stratified audit of 185 fields used independent adjudicator instances that saw the proposed name but not its basis.
This agent audit gave name precisions of 0.87--1.00 across the naming routes and about 0.96 when weighted by the number of fields.
The independent expert sample used a different form of blinding, in which the expert named each of \nBlindExpertN{} fields without seeing the machine's answer.
This sample gave \nBlindExpertAgree{} agreements (\BlindExpertWilsonRaw), increasing to \nBlindExpertAgreeEquiv{} when two conventions of notation were treated as equivalent (\BlindExpertWilsonEquiv).
These evaluations distinguish verification of a proposed answer from agreement with an independently assigned phase assemblage.

(Sources: \path{open-tdb/docs/reports/HANDBOOK_PHASE_DIAGRAM_PILOT.md}, \path{HUMAN_ASSISTED_ANNOTATION_STUDY.md}, \path{HULL_HEURISTIC_UNCERTAIN_LOW_T_FIELDS.md}; blind sample \texttt{tier5/blind-sample/score.json}.)

\suppnotesection{3}{literature-level gates, page checks and audits}
\label{snote:gates}
The literature-level gates described in Section~\ref{sdetail:methods} reduced the \nGateRecordsIn{} records remaining after clean-up to \nGateRecordsOut{} records, of which \nGateAcceptedAfter{} are accepted.
Supplementary Table~S8 records the actions of the individual gates.
To verify these gate actions, an AI agent confirmed changes of unit, sign, scale or chemical system on the rendered PDF page.
The ledger of \nGateVisualEntries{} checks retains the page, image hash, verdict and a short note of what is printed, without retaining the image or page text.
A check of a whole table applies to its rows, giving \nGateVisualChecks{} lookup keys for the gates.

After these page checks, adversarial audits searched the release, code and manuscript for classes of error.
Their native error rates and the post-correction re-audit rates are reported in the main text (\mainSecRef{sec:audit}{section on the adversarial audit}).
To retain the findings of these audits, the release change log and test suite record the classes found, corrections applied and permanent tests introduced to guard against recurrence.
Among the \rThreeNFiveB{} values sampled from the empty-phase demotions of the third audit round, \rThreePhasesOnPageFiveB{} had their phases printed in prose that the deterministic rule could not associate with the value.
A later page-checked recovery restored the printed phases of 1,267 records that had been demoted by the placeholder rule; these records were accepted again, although they remain excluded from ESPEI output.
Among the 1,687 placeholder records left unchanged by that recovery, a blind sample found a printed phase for 3 of 60 records (5.0\%, Wilson 95\% interval 1.7--13.7\%).
The records that still lack a real phase retain the flag \texttt{state\_phases\_missing}, and none is accepted.

In the final build of the third audit round (24 October 2026), the re-sample of records changed by that round found no wrong value among the split-row compositions (\rThreeFinalWrongTwo{} of \rThreeNTwo) or the phases recovered from quotes (\rThreeFinalWrongFour{} of \rThreeNFour), one among the parity pairs (\rThreeFinalWrongSix{} of \rThreeNSix), and residuals that predate the round in three other pools (\rThreeFinalWrongOne{} of \rThreeNOne, \rThreeFinalWrongThree{} of \rThreeNThree{} and \rThreeFinalWrongFive{} of \rThreeNFive). The later corrections are listed in the change log of the release.

(Sources: \texttt{tools/literature\_gates.py}; \texttt{literature/QUALITY\_FIXES.md} and \texttt{quality\_fixes.json}, which list every rule and count; ledger \texttt{tools/literature\_visual\_checks.json}.)

\suppnotesection{4}{details of the methods, comparisons and uses}
\label{snote:details}
\paragraph{One-shot phase-diagram score.} The one-shot score asks how much of the recorded phase equilibrium a description recovers when it is used without fitting to the benchmark evidence, and it treats each equilibrium recorded by ORCHID as a small classification problem in which every phase is either present or absent.

Each evidence point $i$ therefore has a reference assemblage $A_i$, the phases recorded at its temperature and composition, and a predicted assemblage $\hat A_i$, the phases with non-zero amount in the pycalphad equilibrium of the description at the same conditions ($P=101325$~Pa), with the names of both mapped to registry identities so that order--disorder partners and composition sets are compared correctly.

For each phase $p$, $n_p=|\{i:p\in A_i\}|$, $\mathrm{TP}_p=|\{i:p\in A_i\cap\hat A_i\}|$, $\mathrm{FP}_p=|\{i:p\in\hat A_i\setminus A_i\}|$, recall $R_p=\mathrm{TP}_p/n_p$ and precision $P_p=\mathrm{TP}_p/(\mathrm{TP}_p+\mathrm{FP}_p)$, and the prevalence-weighted recall $S=\sum_p n_pR_p/\sum_p n_p=\sum_p\mathrm{TP}_p/\sum_p n_p$ is then the fraction of the recorded occurrences of phases that the description recovers, so that a phase seen at many points counts in proportion to the evidence for it.

Recall alone does not penalize a phase that is predicted where it is not observed, as Cu$_{10}$Zr$_7$ is at many points of the Cu--Zr description from GRACE, and this is the reason that precision and F1, with the same weights, are reported with it.

Points placed inside fields of the assessed diagram are scored at their coordinates; points that lie on a boundary (tie-line ends and invariant vertices from the literature) are scored as the union of phases predicted within $\pm10$~K and $\pm0.01$ in mole fraction and reported separately. Confidence intervals resample whole equilibria or whole source studies. Phases recorded by the handbook but absent from every model's phase list are reported, and a score restricted to representable points is given alongside.

\paragraph{Machine-learning split.} Values reported in a paper share authors, samples and methods, which can make a test on that paper appear more accurate if some of its values were used for training.
To avoid this overlap, the \texttt{split} column of the Parquet tables assigns all values of a paper to the same set.
The assignment uses the first 8 bytes of \texttt{sha256("orchid-split-v1:" + paper\_id)}, read as an integer and taken modulo 100.
The resulting integers below 80 enter training, those below 90 enter validation, and the remainder enter the test set.
This deterministic rule requires no stored list of assignments.

To complement this fixed split with cross-validation, \texttt{fold\_lpo} leaves out whole papers, while \texttt{fold\_lso} leaves out whole chemical systems.
The corresponding sets of five folds are obtained from the SHA-256 of the paper identifier or chemical system, modulo 5.
The latter grouping tests prediction for systems absent from the training data.

\paragraph{Clean-up of Version~1.} The first release of the literature values (Version~1) contained four kinds of defect, addressed in order of increasing difficulty.
Duplicate records were identified by agreement in paper, page, property, value, unit, temperature, phases, quote and all fields distinguishing one state from another.
Merging \nDupGroups{} groups removed \nDupRemoved{} records, while retaining the original identifiers in \texttt{merged\_ids}.
After duplicate removal, reaction and transition labels were mapped to controlled vocabularies.
This mapping reduced \nTKVariants{} printed variants in \nTKRecords{} records to \nTKTerms{} terms, relabelling \nTKRelabelled{} records while preserving their printed labels.
Malformed property codes were treated similarly in \nPropCodesCleaned{} records.

After correcting the labels, the DFT comparison identified candidates for the next class of correction, involving the basis or sign of a value.
For this comparison, we examined \nOutliersExamined{} accepted experimental formation enthalpies that had a Materials Project or OQMD match and no flag of suspect magnitude.
These candidates differed from every matched DFT value by more than 50~\kjma, or had the opposite sign of a DFT value with $|\Delta| > 10$~\kjma.
A per-atom value was converted from a per-formula-unit basis only when the printed page established the column basis, giving \nPerFormulaFixed{} corrections documented by page checks.
For these basis conversions, an earlier rule had accepted changes when they improved agreement with DFT.
This rule was withdrawn because it used DFT to correct the experimental values subsequently used to calibrate DFT.
Unsigned magnitudes were flagged (\nUnsignedFixed), and the remaining \nOutlierUnexplained{} values were retained unchanged with \texttt{dft\_outlier\_unexplained}.
Of these, \nOutlierUnexplainedDefault{} remain in the default subset, while the others are excluded by its point rules.

After the basis and sign checks, the final corrections addressed garbled fields.
The \nPagesNulled{} entries that did not represent valid page numbers were set to null and read again on the rendered page.
In addition to these page-number corrections, one free-text reference state unsupported by the quote was removed.
The evidence for these and all other changes is retained in \texttt{literature/QUALITY\_FIXES.md} and \texttt{quality\_fixes.json}.

\paragraph{Assessed diagrams: annotation layers and points.} To extract assemblages from an assessed diagram, the fields must first be identified and named.
To identify these fields, a deterministic pass reads the vector drawing, reduces outlines to centrelines and calibrates both axes using the tick marks and printed labels.
It then finds the enclosed fields and assigns printed phase labels to the fields containing them.
The rules of phase diagrams allow further assignments because adjacent fields differ by exactly one phase, except across an invariant line or the line of a compound, and a phase of a two-phase field vanishes into a neighbouring field.
Using these constraints, the solver accepts a name only if a single assemblage is consistent with the whole diagram.
For fields unresolved by these constraints, vision-model readers and an adjudicator provide assignments, followed by expert review as described in Supplementary Note~2.

Once the fields have been named, a point deep inside a field provides the assemblage supported by the drawing without relying on the precise position of a drawn boundary.
The area centroid is used when it lies inside the field at a safe distance from all drawn lines.
Otherwise, the point is placed farthest from all field boundaries, at the pole of inaccessibility, i.e.\ the centre of the largest inscribed circle.
This alternative is needed because the centroid of a curved or crescent-shaped field can lie outside the field or on a boundary.
Fields too thin to satisfy the margin are listed as exclusions and provide no point.

Dashed boundaries represent estimates by the evaluator, but they are traced and used as field boundaries in the same way as solid lines.
To retain this distinction, a point records whether the boundaries along its isotherm and isopleth are solid, dashed or both (\nAsmBoundarySolid, \nAsmBoundaryDashed{} and \nAsmBoundaryMixed{} points, in \nAsmDiagramsDashed{} diagrams with dashed lines).
This boundary information is used to enlarge the reading uncertainty when a bounding line is dashed.
The annotation figure in the main text illustrates these steps using a synthetic diagram from \dsBench{}, without reproducing a handbook drawing.

For the points placed in these fields, the record contains temperature, mole fraction, pressure (1~atm, assumed) and assemblage, with phase-registry identities where the mapping is unambiguous.
To document the assignment, it also gives the naming basis (deterministic, reader consensus, adjudicator, DFT-hull corroborated, expert reviewed), confidence and review state.

The uncertainty budget of a point combines the limitations of reading the drawing and of the handbook method.
The reading component is propagated from axis calibration as the maximum calibration residual plus half a raster cell.
The method component uses a prior of 10~K and 0.02 in mole fraction, enlarged for dashed bounding lines.
Combining the reading and method contributions in this way, the larger component is used as the effective uncertainty.
A two-phase point may additionally contain a \texttt{tie\_line} block obtained by intersecting its isotherm with the drawn boundaries.
Because these endpoints are based on the drawing alone, they are marked as a weak constraint.
This tie-line block is omitted, with a recorded reason, when the isotherm does not meet two clean boundaries.

To use these point records in fitting, diagrams whose phases can be named in a TDB file also produce \nAsmEspeiDatasets{} ESPEI ZPF datasets containing \nAsmEspeiPoints{} points.
The \nAsmEspeiExcluded{} points without such names are excluded.
For these datasets, phase compositions are left unknown by default, using null vertices and a hyperplane vertex at the overall composition of the point.
An alternative to these unknown compositions supplies the tie-line endpoints.
The points are also compared with the equilibria of the SGTE assessments at the nearest grid cell, but the comparisons never alter the records.

In distributing the point records and their comparisons, the released content is restricted to derived phase names, assemblages, presence facts, coordinates, phase metadata and citations.
The builder enforces this restriction by failing on output resembling page images, crops or handbook text.
The records cite the original evaluation first and the handbook second, identifying the latter as the compilation.

\paragraph{Extraction benchmark.} To test reading against exact truth, \nBenchDocs{} synthetic documents containing \nBenchPages{} pages were generated from SGTE TDB files using pycalphad 0.11.2, one document per binary system (\dsBench).
The phase diagram in a document is drawn using a seeded choice of line width, colour, labelled or unlabelled fields, K or $^\circ$C, at.\% or mole fraction, insets, experimental-style markers and a shifted second assessment.
The document also contains phase-fraction and property plots, tables of invariant reactions, solid phases, formation properties and liquid mixing properties, and running text stating values in varied phrasing.
Layouts and units are varied across the tables.

To test the effect of document quality, the documents are rendered as born-digital vector PDFs, 300-dpi rasters, degraded 150-dpi scans and image-only PDFs.
Scan degradation includes contrast loss, blur, small rotations, noise and JPEG compression.
Before testing extraction, an oracle wrote the known truth in the extractor format to validate the scorer.
For this oracle, \benchOracleAllOne{} \nBenchOracleMetrics{} rate metrics (precision, recall, F1, accuracy, exact match) equal 1.0.

Supplementary Table~S6 compares deterministic extraction routes with the production AI reader, Gemini~3.8~Flash using the production prompts as a single reader.
On clean renderings, the AI reader recovers almost all invariant-table rows, with lower recall on degraded renderings.
Its running-text values are precise, but recall is limited by the paragraph filter of the production pipeline.
In contrast, Tesseract drops the reaction arrow and therefore never recovers invariant rows, while automatic axis detection fails on many degraded diagrams.

This benchmark exposed the empty reads discussed in Supplementary Note~1.
In that run, the AI reader returned placeholder rows, giving scores near zero for property records and structure labels.
In contrast to these placeholder outputs, Claude recovered, on the same requests, all property records from vector PDFs (\benchClaudeRecordsVector) and \benchClaudeRecordsDegraded{} from degraded scans.
Together, these benchmark runs involved \nBenchCalls{} calls at a list-price cost of US\$\benchSpend{}.
Supplementary Fig.~S5 reports the results across tasks, metrics and renderings.

\paragraph{Sign checks and the DFT comparison.} Values that are printed without a sign could bias the DFT comparison if they were taken as positive, but after the clean-up only \nUnsigned{} of them remain, and including them changes the Materials Project MAE of the experimental set from \maeMP{} to \maeMPWithUnsigned~\kjma, while the Pearson coefficient \pearsonMPChange, so that the choice of including or excluding them does not affect the comparison.

\paragraph{DFT parity in detail.} The experimental formation enthalpies are individual extractions from the papers of the corpus and not values taken from a compilation, and values reprinted from earlier work are linked to their original source where it could be resolved. The DFT parity compares each experimental formation enthalpy in the default subset with the structure-matched entries of the Materials Project and OQMD, and the mean absolute error (MAE) is \maeMP~\kjma\ for the Materials Project (\nMPDefault{} pairs) and \maeOQMD~\kjma\ for OQMD (\nOQMDDefault{} pairs). The point rules left out, for the Materials Project and OQMD respectively, \nExclRefStateMP{} and \nExclRefStateOQMD{} pairs whose reference state is not the solid standard reference, \nExclGateFlaggedMP{} and \nExclGateFlaggedOQMD{} whose literature value has a gate failure, \nExclOffStoichMP{} and \nExclOffStoichOQMD{} measured away from the stoichiometry of the compound, \nExclCompConflictMP{} and \nExclCompConflictOQMD{} whose composition conflicts with the label, and \nExclStructUnmatchedMP{} and \nExclStructUnmatchedOQMD{} whose named structure is absent from the database.

The OQMD value corresponds to \maeOQMDeV~\evat\ over pairs that are mostly intermetallic (\maeOQMDIntermetallicEV~\evat\ for the intermetallics alone), which may be compared with the 0.069~\evat\ that Kirklin \emph{et al.}\ reported for 278 binary intermetallics and the 0.096~\evat\ for 1,670 compounds dominated by oxides and halides, obtained with the same fitted elemental references\cite{kirklin2015}. The closeness of the overall values is therefore a coincidence of populations, and some of our measurements are among those to which those references were fitted.

When the median of the values of a compound is taken first, the MAEs become \maeMPMedian{} (MP) and \maeOQMDMedian~\kjma\ (OQMD). By chemistry, intermetallics agree better (MAE \maeMPIntermetallic{} and \maeOQMDIntermetallic) than borides, carbides, nitrides and chalcogenides (\maeMPNonmetal{} and \maeOQMDNonmetal). By method, calorimetric values agree better than EMF and vapour-pressure values (MP \maeMPCalorimetry{} versus \maeMPEmf, OQMD \maeOQMDCalorimetry{} versus \maeOQMDEmf). Adding the model values to the experimental set (\nMPWithModel{} MP and \nOQMDWithModel{} OQMD pairs) gives MAEs of \maeMPWithModel{} and \maeOQMDWithModel~\kjma.

Behind these averages lie the \nOutliersExamined{} formation enthalpies with large residuals, which the clean-up described above sorted into three classes, namely \nPerFormulaFixed{} per-formula values that had been labelled per atom (such as CuO, Cu$_2$O, CaO and Li$_2$O, which are printed in kJ\,mol$^{-1}$ of compound), \nUnsignedFixed{} magnitudes printed without a sign, and \nOutlierUnexplained{} values with no deterministic explanation, which are kept and flagged (\nOutlierUnexplainedDefault{} of them are in the default subset, and the others are excluded by its point rules). They include printed values that disagree with both DFT sources (e.g.\ Ca--Si compounds from one effusion study, with residuals above 150~\kjma), compounds for which the two DFT sources disagree with each other (e.g.\ Re$_2$Si), and values whose sign is printed opposite to that of DFT (e.g.\ AlB$_{12}$). Removing these values would lower the MAE to \maeMPNoOutliers{} (MP) and \maeOQMDNoOutliers~\kjma\ (OQMD), but they are kept because no rule justifies a change.

\paragraph{DFT calibration.} As for the experimental values, DFT uncertainty is described by a robust $\sigma$, here obtained from residuals against experiment.
For the Materials Project, the resulting $\sigma$ is smallest for magnetic 3d intermetallics and largest for compounds containing p-block or anion elements.
By comparison, the OQMD strata differ less and do not follow the same ordering (Supplementary Table~S5).
This difference between the two databases reflects their different treatment of energies, through the MP2020 corrections for anions and for the mixing of GGA and GGA+U calculations in the Materials Project and through elemental reference energies fitted to experimental formation enthalpies in OQMD, and part of the present experimental set is among the data to which the OQMD references were fitted.
The literature DFT values have a lower overall $\sigma$ (\sigLit~\kjma, compared with \sigMP{} and \sigOQMD~\kjma), but within each chemistry stratum they are not systematically closer to experiment (Supplementary Table~S5), so that their lower $\sigma$ reflects mainly a larger share of magnetic 3d intermetallics and a smaller share of p-block and anion compounds rather than more accurate calculations.

The residual distributions have heavy tails, with RMSE values \rmseRatioMP{} and \rmseRatioOQMD{} times the robust $\sigma$ for the Materials Project and OQMD, respectively.
A Gaussian likelihood based on this $\sigma$ assigns too little probability to large residuals, allowing outliers to influence a fit too strongly.
To limit this influence, a wider $\sigma$ or a heavy-tailed likelihood may therefore be preferable when fitting DFT data.
The estimate treats pairs as individual observations, while the confidence interval resamples whole chemical systems.
It also inherits errors in the machine-extracted experimental values used as the reference.

\paragraph{Cross-check with Kim \emph{et al.}} The compilation of experimental formation enthalpies by Kim \emph{et al.}\ includes intermetallic and inorganic compounds\cite{kim2017}.
Because several ORCHID values reproduce the same calorimetric measurements, agreement with this compilation primarily tests extraction, units and signs, rather than the measurements themselves.
There are \nKimCandidates{} accepted, signed experimental formation enthalpies in \kjma\ for \nKimFormulas{} formulas available for comparison.
The unit conversion requires care because the file of Kim \emph{et al.}\ does not state the unit, but the magnitudes identify the values as eV/atom.
After conversion, compounds were matched by reduced formula, using the median when either source contained multiple values.

The sources overlap in \nKimOverlap{} compounds, with a mean absolute difference of \maeKim{} and a median of \medKim~\kjma.
All \nKimOverlap{} compounds agree in sign.
The \nKimDiffer{} compounds differing by more than 10~\kjma, including Ge$_2$Hf$_3$, TaB$_2$ and AlIr, are listed with their record identifiers in \texttt{kim2017\_crosscheck.json} for review.

\paragraph{Invariant temperatures against the SGTE assessments.} Literature invariant temperatures can also be compared with the calculated invariants in \dsDiagrams{} using only released files, as shown in the invariant figure of the main text.
Because the assessments were fitted to many of these measurements, this comparison mainly tests extraction and phase identification.
For this comparison, a match requires agreement in chemical system, reaction class and the set of three phase-registry identities.
The reaction class distinguishes a eutectic, in which the liquid lies between the other phases, from a peritectic, in which it lies at one end, and from reactions without liquid.
Phase identities are obtained from the paper and TDB names in \dsRegistry{}.
When this procedure matches several assessments, their median invariant temperature is used to avoid favouring an individual assessment.

Of \nInvValCandidates{} accepted binary eutectic, peritectic, eutectoid and peritectoid temperatures, \nInvValMapped{} map to three identities, and \nInvValExp{} experimental values in \nInvValExpSystems{} systems match a calculated invariant.
The difference between the calculated and the literature temperature has a median of \invValExpMedian~K and a robust $\sigma$ of \invValExpSigma~K, and \invValExpWithinTen\% of the values lie within 10~K (mean absolute difference \invValExpMAE~K).
Model values agree more closely (\nInvValModel{} values, robust $\sigma$ \invValModelSigma~K), as is to be expected when a paper reports the assessment that the TDB file encodes, and \nInvValExpInSample{} of the \nInvValExp{} experimental values (\invValExpInSamplePct\%) come from a paper that is itself a source of one of the SGTE assessments of its system.

\paragraph{Using the ESPEI datasets.} The use for which ORCHID was built is the refinement of CALPHAD models, and it supplies the data needed at each step of an ESPEI fit\cite{bocklund2019}.
In the parameter-generation step, ESPEI fits end-member and interaction parameters to non-equilibrium thermochemical data. Such data are held in the folders \texttt{experimental/}, \texttt{dft/mp/}, \texttt{dft/oqmd/} and \texttt{dft/literature/}, the DFT folders with a calibrated $\sigma$.
Bayesian refinement by Markov chain Monte Carlo then fits all parameters to phase-equilibrium (ZPF), thermochemical and activity data.
For this step, the folders \texttt{experimental/} and \texttt{invariants/} supply measured thermochemical and phase-equilibrium data, while \dsAssemblages{} supplies ZPF points from assessed diagrams. Thermochemical weights follow the effective uncertainty of the observation; ZPF data retain the driving-force tolerance of the equilibrium likelihood, without converting their temperature uncertainty into a thermochemical weight.

In practice, a fit for one binary system can proceed as follows.
The SGTE TDB of the system, taken from the ORCHID catalog, is used as the starting model, with a phase-models file built from its phases, for which the phase registry gives the sublattices and constituents.
The ORCHID datasets of the system are then loaded with \texttt{load\_datasets}, after which \texttt{apply\_tags} is used to choose the evidence, for example to disable \texttt{model}, to use one DFT source and to down-weight \texttt{single\_reader}.
For these fits, the regenerated bundle should be loaded from a clean directory, retaining the exported \texttt{weight} of each record; the split manifest \texttt{UNCERTAINTY\_SPLITS.json} identifies the parent files that must be replaced rather than loaded alongside their children.
These weights already include the conversion from effective uncertainty, so any additional tag or property weight is a separate modeling choice that should be recorded with the release and ESPEI version.
The parent and source-study links remain available for sensitivity analyses of shared systematic effects, although covariance between those observations is not quantified by the exported weights.
The previously reported Cu--Zr fitting demonstrations retain their archived input manifests and weighting choices; their fitted parameters and scores have not been recomputed under this export revision.

At the last step, ESPEI is run with an \texttt{mcmc} block whose \texttt{input\_db} is the starting TDB.
This route was tested on Cu--Zr by refining six parameters of the 1994 assessment and, in a pre-registered fallback, seven parameters including a further liquid term.
Although the fit to ORCHID activities improved by about half, the invariant temperatures could not be kept within the pre-registered gate.
The main-text Cu--Zr example therefore used the gradient route described below, while the unsuccessful attempt is documented in \texttt{orchid-examples}.

The gradient route used for this refinement connects the data to the MLIP-based workflows of our group.
One of these, PhaseForge, computes end-member and special-quasirandom-structure energies with machine-learning interatomic potentials in the Alloy Theoretic Automated Toolkit and builds CALPHAD descriptions from them for high-throughput phase-diagram prediction\cite{zhu2025machinelearningpotentials,zhu2025accelerating,sariturk2025phaseforge}.
PhaseForge+ constructs physically informed CALPHAD models from MLIPs and tunes them against experimental observations by gradient-informed optimization\cite{kunselman2025construction,sariturk2025phaseforgeplus}. The experimental phase-equilibrium and thermochemical data of ORCHID are the observations that this kind of tuning needs.
The Pt--W example of PhaseForge+ encodes phase-equilibrium points read from a handbook diagram as ESPEI ZPF data with one phase composition for each point, and it also uses \texttt{\_\_HYPERPLANE\_\_} vertices, the encoding that \dsAssemblages{} uses for points inside two-phase fields, whose phase compositions are left unspecified\cite{sariturk2025phaseforgeplus}.
\dsAssemblages{} supplies \nAsmEspeiDatasets{} ZPF datasets in this encoding (\nAsmEspeiPoints{} points), so that they can be used without conversion.
To fit these data by gradients, PhaseForge+ relies on a branch of ESPEI, not yet merged, that adds analytic derivatives of the ZPF driving forces with respect to the model parameters\cite{kunselman2024espeibranch}.
The release is validated with ESPEI~0.9.1, whose dataset format that branch also reads, and the Cu--Zr example of the main text loaded the released datasets with the branch.

Alongside these fitting workflows, the labelled phase-stability observations, including assessed assemblages, invariants and calculated diagrams, also support training of phase-stability classifiers and Bayesian active learning of phase diagrams\cite{fan2026balpi}.
Beyond the ESPEI files, the DFT calibration offers a template for weighting computed (DFT or MLIP) energies against experiment.
In turn, the phase registry gives the structure types, prototypes and TDB definitions that are needed to choose which phases and end-members to compute.
Handbook phases that have no counterpart in the SGTE assessments of a system, which are found by the cross-check of \dsAssemblages, are concrete targets for such calculations.

\paragraph{Recovery of failed diagram calculations.}
The recovery review began with \nRecoveryOriginalFailed{} failed calculations.
An initial review identified nine recovery candidates and one entry unsuitable for the binary calculation, leaving \nRecoveryRemaining{} cases for further investigation.
Of these remaining cases, \nRecoveryNumerical{} underwent numerical recovery trials, while \nRecoverySourceCases{} required clarification of the source model or additional support in the calculation engine.
Those trials examined \nRecoveryVariants{} candidate variants through \nRecoveryChecks{} separately solved equilibrium checks retained in the review reports, and \nRecoveryPromoted{} candidates were admitted at that stage because several failed denser checks or represented only fragments of a diagram.

The final review extended the checks of the initial candidates and resolved a constituent-declaration inconsistency in the Mg--Y description against the COST 507 source and database implementations.
For that exact source file, the calculation reader removes an extra vacancy constituent group that conflicts with the two declared sublattices, while retaining the original file and every numerical parameter.
Four descriptions, Cu--O (1992), Al--Ti (1992), Cu--Nb (1990) and Mg--Y (1998), passed \nRecoveryFinalChecks{} checks across their sampled two-phase regions and were included in the release.
A subsequent audit of the entries marked successful identified two empty outputs, Ga--Se (2008) and Fe--Ti (2023), with neither mapped boundaries nor an equilibrium grid.
Repeating the calculations with the original initialization and with interior starting points did not recover a usable diagram, so both entries were retained as failed calculations, with the evidence recorded in \texttt{phase-diagrams/EMPTY\_OUTPUT\_REVIEW.json}.
The release therefore contains \nDiagrams{} calculated diagrams and \nDiagramsFailed{} entries retained explicitly as failed calculations, with the decisions and verification records in \texttt{phase-diagrams/RECOVERY\_REVIEW.json}.
These checks test consistency between mapped boundaries and separate equilibrium solutions from the same thermodynamic engine; they do not independently validate the source assessments or every boundary point.
The calculation tool rejects empty boundary outputs and provides explicit options for interior starting points and the phase exclusions specified by the source TDB, and the recovery work does not replace literature values.

\clearpage
\addcontentsline{toc}{section}{Supplementary Figures}

\begin{widefigure}[htbp]
\raggedright{\normalfont\normalsize\bfseries Supplementary Figures\par}\vspace{0.8ex}\centering
\nfig{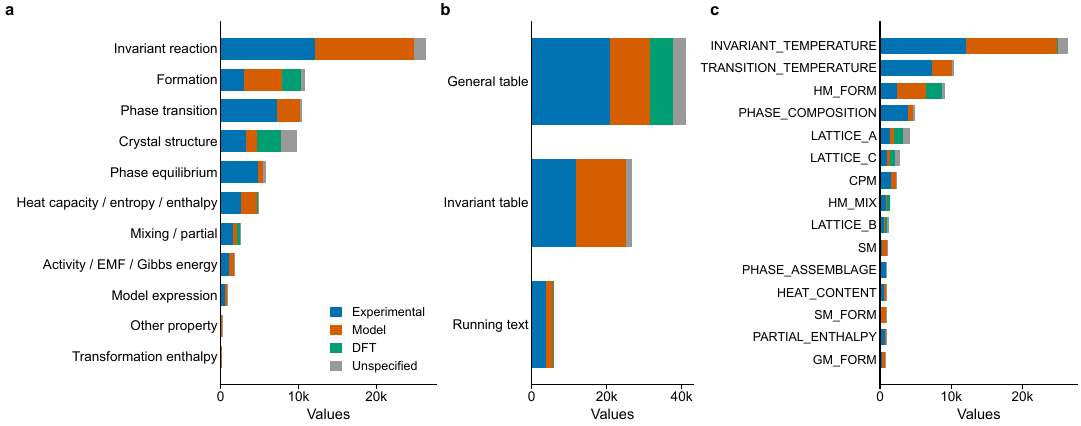}
\caption{\textbf{Half of the literature values are temperatures of invariant reactions and phase transitions, and nine in ten were read from tables.} \textbf{a}, Values by property group, stacked by evidence class. \textbf{b}, Values by record kind (general tables, invariant-reaction tables, running text). \textbf{c}, The 15 most frequent property codes. Colours: evidence class, as in every figure (blue experimental, vermilion model, green DFT, grey unspecified). n = \nValues{} values.}
\label{sfig:composition}
\end{widefigure}

\begin{widefigure}[tbp]
\nfig{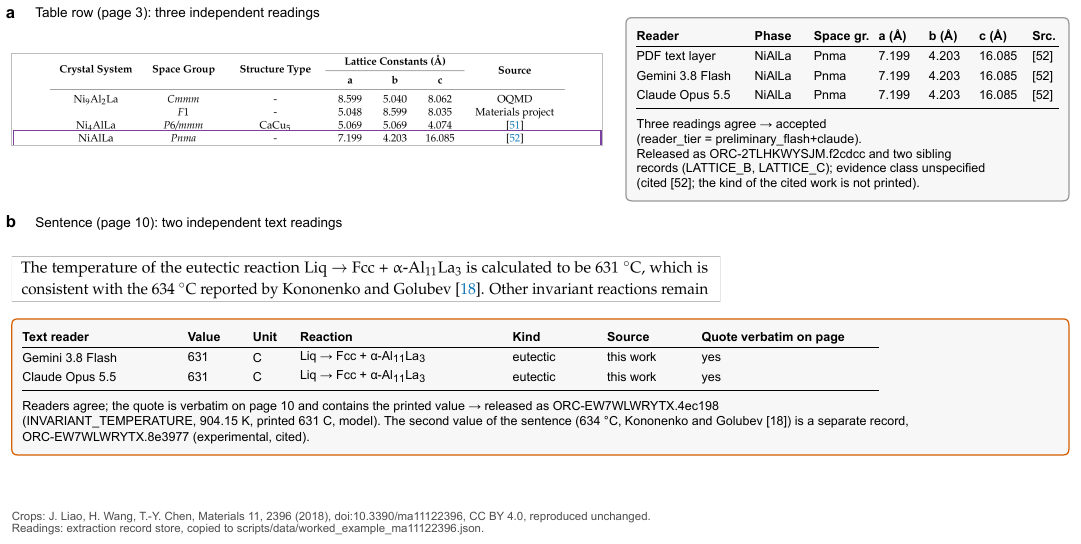}
\caption{\textbf{Two independent readers agree before a value is released.} \textbf{a}, One table row (outlined) as read independently by the PDF text layer and two vision--language models (Gemini~3.8~Flash and Claude~Opus~5.5); because the three readings agree, the row is accepted and released as three records, one for each lattice constant ($a$, $b$, $c$). \textbf{b}, One sentence as read by two text readers, both of which return the value, its unit, the reaction and the verbatim sentence; the sentence is found verbatim on the cited page with the printed value inside it, so the value is released, as class \texttt{model} because the paper states that it was calculated, while the second value in the sentence (634~$^\circ$C, cited from Kononenko and Golubev) becomes a separate record of class \texttt{experimental}. Crops are reproduced unchanged from Liao, Wang and Chen\cite{liao2018} (CC~BY~4.0).}
\label{sfig:dualreader}
\end{widefigure}

\begin{widefigure}[tbp]
\nfig{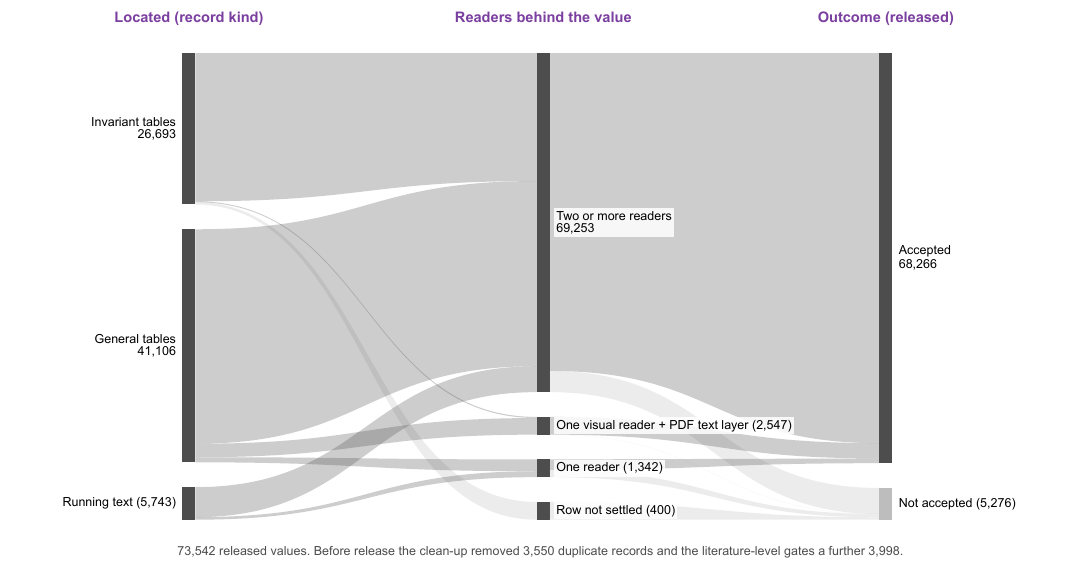}
\caption{\textbf{Nearly all released values rest on two or more independent readings.} Flows of the \nValues{} released values from the kind of record they were located in (left), through the number of independent readers behind them (centre, field \texttt{reader\_agreement}), to whether they are accepted (right). Two or more readers: \nRAMulti{} values (\pctRAMulti\%). One visual reader whose numbers are all found in the PDF text layer: \nRASingleText. One reader: \nRASingle, including running-text values read by one text reader, which are released only with a verbatim quote verified on the page. Row not settled: \nRANotClassified{} invariant-table rows still awaiting a second reader, none of them accepted. Values removed before release (failed quotes, fields outside their window, the \nDupRemoved{} duplicates removed in the clean-up and the \nGateDupMerged{} merged by the gates) are not drawn. Node heights are proportional to counts, except that small nodes are drawn at a minimum height for legibility.}
\label{sfig:sankey}
\end{widefigure}

\begin{widefigure}[tbp]
\nfig{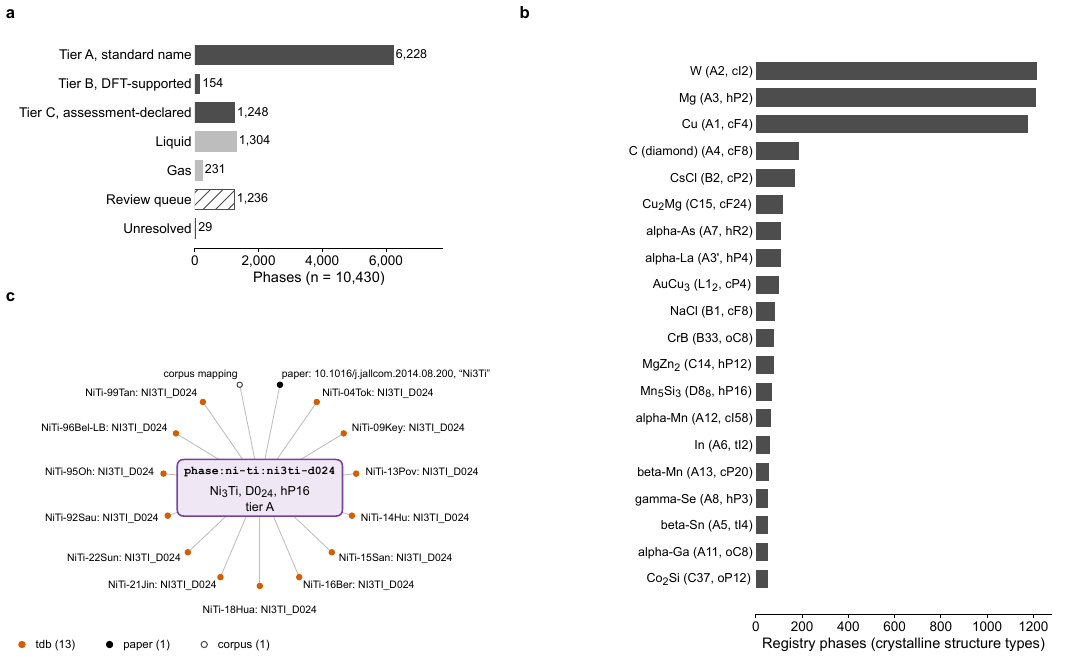}
\caption{\textbf{Each phase receives one identity that gathers the names used for it across assessments and papers.} \textbf{a}, Structure assignment of the \nRegPhases{} phases by tier (Supplementary Table~S4). \textbf{b}, The 20 most frequent crystalline structure types, labelled by prototype, Strukturbericht designation and Pearson symbol. \textbf{c}, The names collected for one identity, Ni$_3$Ti (D0$_{24}$): the same TDB phase name in \nNiThreeTiTdb{} SGTE assessments of Ni--Ti, the printed label in \nNiThreeTiPaper{} \nNiThreeTiPaperWord, and the mapping used in extraction.}
\label{sfig:registry}
\end{widefigure}

\begin{widefigure}[tbp]
\nfig{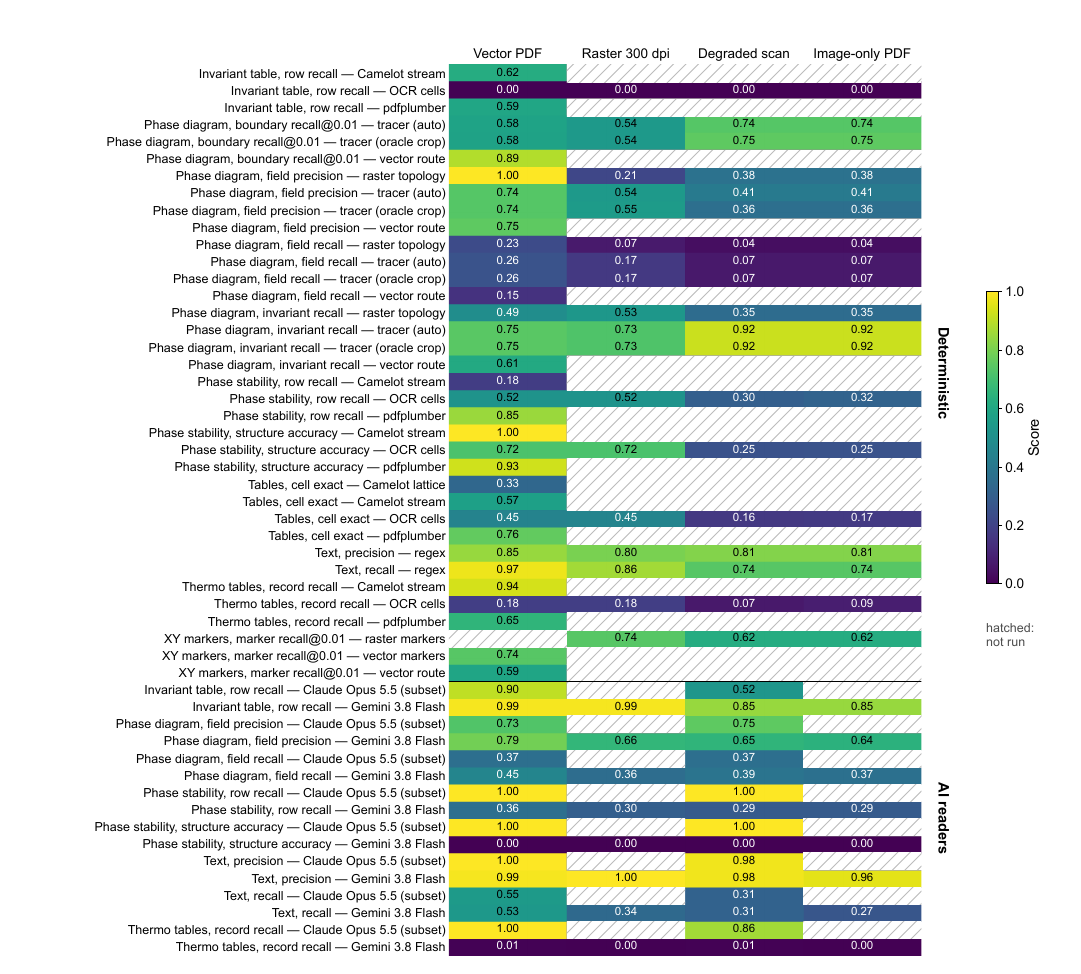}
\caption{\textbf{On synthetic papers whose truth is known, the AI readers recover most tables and text, whereas deterministic routes fail on many tasks.} Scores of deterministic routes (top) and AI readers (bottom) on the \nBenchDocs{} synthetic papers of the extraction benchmark in four renderings; hatched cells were not run for that rendering. The Gemini reader used the production prompts, and its zero scores for structure labels and property records come from the prompt version that returned placeholder rows (Supplementary Note~1). Claude Opus~5.5 was run on a subset of the documents in two renderings. Headline values are given in Supplementary Table~S6.}
\label{sfig:bench}
\end{widefigure}

\begin{widefigure}[tbp]
\nfigOr{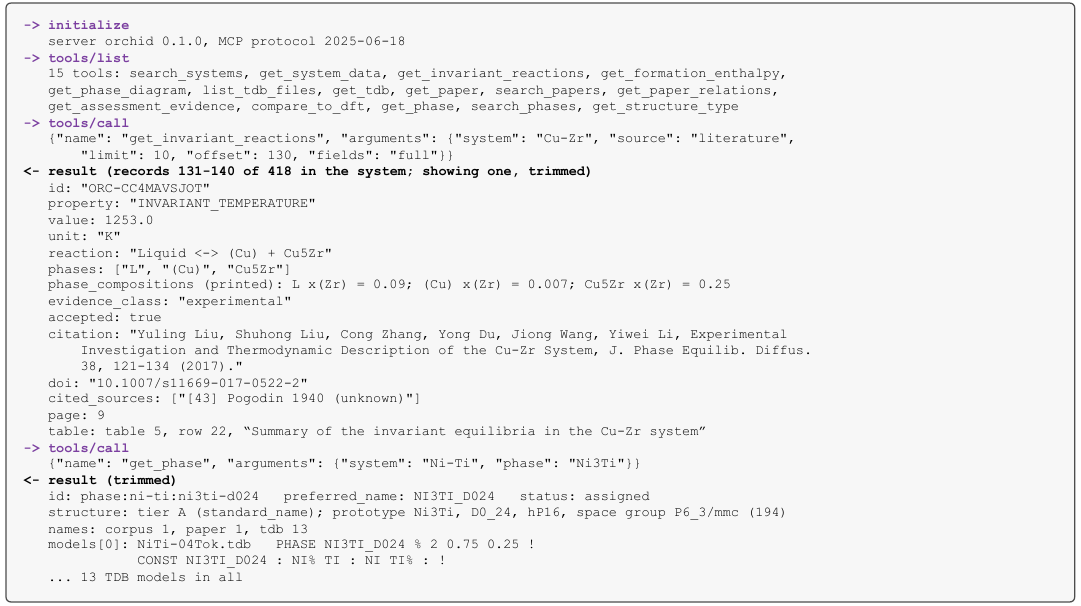}{Figure pending: figures-DD/figS\_mcp.pdf}

\caption{\textbf{An AI agent receives each value with its full state and its source.} The \texttt{orchid-mcp} server was run locally against the released files, and a client listed its tools and called two of them; requests are shown as sent and responses are trimmed for print. The invariant reaction arrives with its phases, temperature, evidence class, citation, page and the verbatim sentence it was read from, and the phase query returns the registry identity with the names used for it in the SGTE assessments and in the literature. The full exchange is released with the examples repository (\texttt{orchid-examples}, \texttt{quickstart/mcp\_exchange.json}).}
\label{sfig:mcp}
\end{widefigure}

\begin{widefigure}[tbp]
\centering
\IfFileExists{figures-DD/figS_cuzr.pdf}{\includegraphics{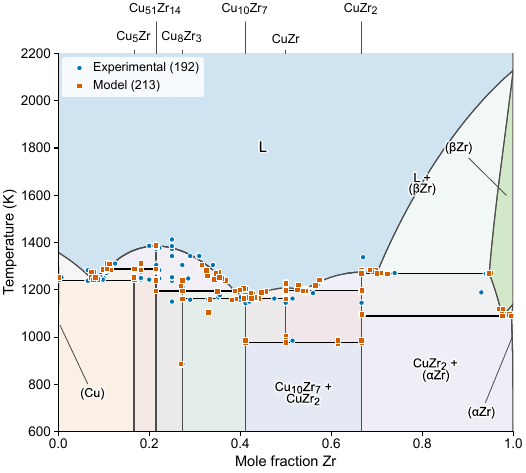}}{\tbd{Figure pending: figures-DD/figS\_cuzr.pdf}}
\caption{\textbf{Most invariant compositions printed in the literature fall on the invariant lines of an assessed Cu--Zr diagram, and the largest exceptions are reactions that this assessment places differently.} Phase fields are coloured by assemblage as in \mainFigRef{fig:espei}{the Cu--Zr figure of the main text}{b} (single-phase fields in the colour of the phase, two-phase fields as light tints), with dark-grey phase boundaries, vertical lines for the stoichiometric compounds and black invariant equilibria, calculated with pycalphad 0.11.2 from the Cu--Zr assessment of Zeng \emph{et al.}\ (1994; \texttt{CuZr-94Zen-LB}), one of the \nCuZrAssessments{} Cu--Zr assessments in the SGTE binary collection and the reference assessment of the Cu--Zr example in the main text. Symbols: printed phase compositions of accepted literature invariant reactions, by evidence class (number of points in the legend); compositions filled from stoichiometry are not shown. The points near 1,375--1,410~K at $x_\mathrm{Zr}\approx0.22$ record the congruent melting of Cu$_{51}$Zr$_{14}$, printed above the calculated liquidus, and those near 885~K record the decomposition of Cu$_8$Zr$_3$, which the calculated diagram does not show.}
\label{sfig:cuzr}
\end{widefigure}

\clearpage
\addcontentsline{toc}{section}{Supplementary Tables}

\begin{widetable}[htbp]
\raggedright{\normalfont\normalsize\bfseries Supplementary Tables\par}\vspace{0.8ex}\centering
\centering
\footnotesize
\caption{\textbf{Property vocabulary: the 20 most frequent codes.} Unit: the most frequent unit of the values of each code after normalization, where \texttt{kJ/mol (basis unstated)} marks values whose per-atom or per-formula basis is not printed and which are therefore not converted. Lattice lengths are in \AA\ where a unit was printed, and the \nLatticeNoUnit{} lengths printed without a unit are left unconverted; other units are as read. A coefficient of a printed expression in $T$, such as the $A$ or $B$ of $A + BT$ or a term of a polynomial for $C_p$, is stored under a code that ends in \texttt{\_EXPRESSION\_TERM} (\nExprTermRecords{} records, \nExprTermGibbs{} of them for the Gibbs energy), together with the printed expression, the term and its range of temperature, and is not exported to ESPEI. ESPEI output: the names of the ESPEI datasets that carry values of the code (\texttt{ZPF}: phase-equilibrium data); --: not exported. All \nPropertyCodes{} codes are listed in \texttt{literature/SCHEMA.md}.}
\label{stab:properties}
\renewcommand{\arraystretch}{1.15}
\begin{tabularx}{\linewidth}{lYP{3.6cm}lS[table-format=5.0]}
\toprule
Code & Quantity & Unit & ESPEI output & {Values} \\
\midrule
\tabPropBody
\bottomrule
\end{tabularx}
\end{widetable}

\begin{widetable}[htbp]
\centering
\footnotesize
\caption{\textbf{Main fields of a literature value.} Parquet column names (the JSON nests some of them); the release has \nParquetColumns{} columns in all, documented in \texttt{literature/SCHEMA.md}.}
\label{stab:fields}
\renewcommand{\arraystretch}{1.0}
\begin{tabularx}{\linewidth}{P{5.6cm}Y}
\toprule
Field & Meaning \\
\midrule
\texttt{id}, \texttt{pid} & stable ORCHID id; persistent locator (paper/page/table or passage/row) \\
\texttt{system}, \texttt{elements} & chemical system \\
\texttt{source\_system} & system of the source table or paper, which differs from \texttt{system} when the state belongs to a lower-order subsystem \\
\texttt{record\_kind} & \texttt{invariant\_table}, \texttt{general\_table} or \texttt{running\_text} \\
\texttt{property}, \texttt{property\_group} & controlled quantity code (e.g.\ \texttt{INVARIANT\_TEMPERATURE}, \texttt{HM\_FORM}, \texttt{CPM}, \texttt{LATTICE\_A}) and its group \\
\texttt{value}, \texttt{unit}, \texttt{value\_printed}, \texttt{unit\_printed} & normalized and printed value and unit \\
\texttt{value\_min}, \texttt{value\_max}, \texttt{value\_expression} & printed range; printed expression (e.g.\ model parameters) \\
\texttt{uncertainty\_value}, \texttt{uncertainty\_basis} & source-reported uncertainty in the record unit (\texttt{printed}, \texttt{printed\_range}), or an explicitly identified placeholder (\texttt{stated\_default}, \texttt{reading\_resolution}); scatter and composition-offset estimates are stored separately in the budget \\
\texttt{sigma\_printed}, \texttt{sigma\_reading}, \texttt{sigma\_method}, \texttt{sigma\_composition}, \texttt{sigma\_scatter}, \texttt{sigma\_conversion}, \texttt{sigma\_effective}, \texttt{sigma\_method\_basis}, \texttt{uncertainty\_budget} & uncertainty budget: reported and estimated components kept separate; reported coverage and its conversion to a $\sigma$ convention are documented; \texttt{sigma\_effective} determines the fitting weight (Methods, uncertainty budget) \\
\texttt{uncertainty\_budget}, \texttt{sigma\_x\_*}, \texttt{sigma\_T\_*} & uncertainty budget with its components kept separate; uncertainties of composition and temperature \\
\texttt{temperature\_K}, \texttt{temperature\_basis}, \texttt{pressure\_Pa}, \texttt{pressure\_basis} & state variables and how they were established \\
\texttt{phases}, \texttt{model\_phases} & printed phase labels; mapped CALPHAD model phases \\
\texttt{compositions} & list of \{phase, element, value, basis, source\}; mole fractions \\
\texttt{overall\_composition} & composition of the alloy or sample, where given (mole fraction) \\
\texttt{reaction}, \texttt{reaction\_type}, \texttt{transition\_kind} & reaction and its controlled class; \texttt{transition\_kind\_printed} keeps the label \\
\texttt{reference\_state\_json} & reference state of the quantity, where printed \\
\texttt{structure\_*} & Pearson symbol, prototype, Strukturbericht designation and space group, where printed \\
\texttt{flag\_state\_fully\_specified}, \texttt{specification\_gaps} & whether the record states every variable that its equilibrium state requires, and which are missing \\
\texttt{evidence\_class}, \texttt{origin}, \texttt{origin\_measured\_in} & evidence class, the rule that decided it, and the work in which the value was obtained \\
\texttt{accepted}, \texttt{reader\_tier}, \texttt{readers}, \texttt{reader\_agreement} & acceptance, which readings agreed, reader identities, and the number of independent visual readers (tables) \\
\texttt{flag\_*}, \texttt{review\_flags} & quality flags (SI~2, section ``Choosing what to trust'') \\
\texttt{method}, \texttt{conditions\_kind} & measurement method when printed with the value; where the conditions apply \\
\texttt{paper\_id}, \texttt{paper\_doi}, \texttt{paper\_citation}, \texttt{cited\_dois} & source paper and cited works \\
\texttt{quote}, \texttt{quote\_page}, \texttt{artifact\_caption}, \texttt{table\_row} & verbatim sentence (running text) or table locator \\
\texttt{model\_tdb}, \texttt{model\_tdb\_sha256} & linked TDB file of a model value (path, SHA-256 and catalog entry), only where the link is unambiguous (\nTdbLinked{} values) \\
\texttt{LATTICE\_*} units & lengths in \AA\ for \nLatticeAngstrom{} of the \nLatticeABC{} lattice constants; the \nLatticeNoUnit{} without a printed unit are unconverted \\
\texttt{unit} = \texttt{kJ/mol (basis unstated)} & formation values whose per-atom basis could not be established (\nBasisUnstatedFormation{} values); \texttt{flag\_basis\_inferred\_from\_magnitude} marks the \nBasisInferred{} taken as per atom from their magnitude alone \\
\texttt{plausibility\_*}, \texttt{calphad\_check} & comparison with a calculated phase diagram, where one exists \\
\texttt{flag\_espei\_ready}, \texttt{flag\_espei\_certified} & usable in ESPEI; usable and reproduced by the CALPHAD calculation \\
\texttt{gate\_failures}, \texttt{corrections\_json} & literature-level gates: demotions and flags with reasons; corrections with the stored value \\
\texttt{tool\_release}, \texttt{split}, \texttt{fold\_lso}, \texttt{fold\_lpo} & extraction tool release; recommended machine-learning split (by paper); five folds by chemical system and by paper \\
\bottomrule
\end{tabularx}
\end{widetable}

\begin{widetable}[htbp]
\centering
\footnotesize
\caption{\textbf{Quality gates.} Top: the gates that a literature value passes before release, with what each checks and its effect on the values, in the order applied. Bottom: the window rule, which keeps a field read by the first reader only when it is printed in the quote or in an allowed window around it; its counts are from the release fix-up of 22 October 2026 and are recorded in the extraction store, not in a released file.}
\label{stab:gates}
\renewcommand{\arraystretch}{1.02}
\begin{tabularx}{\linewidth}{P{3.0cm}YP{4.6cm}r}
\toprule
Gate & What it checks & Effect & Values \\
\midrule
reader agreement (tables) & two or more independent readings of a row agree; disagreements go to a judge model or stay not accepted & accepted values & \nAccepted \\
quote verification (text) & the reader's sentence is found verbatim on one page and contains the printed number & every running-text value passes & \nText \\
window rule & every published field is printed in its quote or an allowed window & fields outside removed; values flagged & \nDroppedFields \\
page-image sign check & suspect formation-enthalpy signs read on the page image & sign verified / flagged unsigned & \nSignVerified{} / \nUnsigned \\
clean-up (Version~1) & duplicates, vocabularies, per-formula units, garbled fields & duplicates removed & \nDupRemoved \\
 & & per-formula values converted (basis read on the page) & \nPerFormulaFixed \\
 & & DFT outliers flagged, unchanged & \nOutlierUnexplained \\
literature-level gates (G1--G12, page checks) & temperatures, units, signs, systems, phase lists, property semantics, evidence class, compositions, duplicates, pressure; changes of unit, sign, scale or system confirmed on the page (Supplementary Note~3) & records corrected & \nGateCorrected \\
 & & records demoted (not accepted) & \nGateDemoted \\
 & & further duplicates merged & \nGateDupMerged \\
states without phases (R3) & an invariant, transition, phase-composition or formation value needs the phases of its state & phases recovered from the quote or page / missing-phase findings at R3 & \nGateStatePhasesRecovered{} / \nGateStatePhasesMissing \\
complete invariant reactions (R3) & phases, class and phase count against the printed reaction; same-page duplicates & phases added / classes filled & \nGateInvPhasesFromReaction{} / \nGateInvClassFilled \\
 & & flagged: class conflict / phase count / duplicate / temperature & \nGateInvClassConflict{} / \nGateInvPhaseRule{} / \nGateInvDuplicate{} / \nGateInvTempCheck \\
unit tokens (R3) & unit prefix of a caption (``k J''); formation values below 0.1~\kjma & values restored / small values flagged & \nGateUnitPrefix{} / \nGateSmallFormation \\
system assignment (R3) & system bounded by the printed formulas; composition of a compound from its formula & compositions recomputed / flagged & \nGateCompFromFormula{} / \nGateCompConflictsLabel \\
ESPEI publication gate & value published, accepted, ESPEI-ready, of the folder's class and equal to the published numbers & values cut, listed with the reason & \texttt{validation.json} \\
ESPEI checks & \texttt{check\_dataset} and \texttt{load\_datasets} of ESPEI~\espeiVersion; likelihood finite against the named TDB & datasets excluded whole & \nEspeiExcluded{} + \nEspeiEvalExcluded \\
\bottomrule
\end{tabularx}

\vspace{0.8em}
\begin{tabular}{ll*{3}{S[table-format=4.0]}}
\toprule
Field & Accepted windows & {Kept} & {Dropped} & {Total} \\
\midrule
transition kind & W0 & 4 & 1,327 & 1,331 \\
citation & W0, W1-clean & 383 & 642 & 1025 \\
phase & W0, W1 & 346 & 396 & 742 \\
method & W0, W1 & 274 & 351 & 625 \\
reaction & W0, W1 & 83 & 102 & 185 \\
temperature & W0 & 0 & 27 & 27 \\
\bottomrule
\end{tabular}

\parbox{\linewidth}{\footnotesize Counts are of field instances, one per observation, before the clean-up and the literature-level gates. W0: inside the quote; W1: the preceding sentence of the same paragraph; W1-clean: W1 with no intervening citation. In the release, \nDroppedFields{} values carry fields removed by this rule.}
\end{widetable}

\begin{widetable}[htbp]
\centering
\footnotesize
\caption{\textbf{Structure assignment in the phase registry.} Each phase is given a crystal structure only when a rule establishes it, and the tiers are listed in decreasing order of certainty; phases for which no rule is decisive are kept in a review queue with their candidates.}
\label{stab:registry}
\renewcommand{\arraystretch}{1.15}
\begin{tabularx}{\linewidth}{P{4.4cm}Yr}
\toprule
Status & Rule & Phases \\
\midrule
tier A, standard name & the TDB name encodes one structure; stoichiometry, sublattices and DFT do not contradict it, including \nRegTierAInterstitial{} \texttt{FCC\_A1} descriptions with an interstitial sublattice, (M)(X,Va), that model B1 or C1 compounds & \nRegTierA \\
tier B, DFT-supported & unique on-hull match in both Materials Project and OQMD, same space group & \nRegTierB \\
tier C, assessment-declared & one consistent structure in the SGTE crystallographic table for this name and system or, in one case (NiTi, B19$'$), in a comment of the TDB file & \nRegTierC \\
liquid; gas & by definition & \nRegLiquid; \nRegGas \\
review queue & candidates listed, none certain & \nRegReview \\
unresolved & no candidate & \nRegUnresolved \\
\midrule
total & & \nRegPhases \\
\bottomrule
\end{tabularx}
\end{widetable}

\begin{widetable}[htbp]
\centering
\footnotesize
\caption{\textbf{DFT calibration by stratum.} Residuals are DFT minus experiment in \kjma\ over the default subset, with literature DFT values compared with the median experimental value of the same compound. $\sigma$: robust $\sigma$ (1.4826 times the median absolute deviation), with its 95\% interval from 1,000 bootstrap resamples of whole chemical systems. Match: the hull status of the paired DFT entry, on the hull and the only entry at its composition (on hull, single), one of several entries at that composition (polymorphs), or above the hull. Bias: median residual, recorded but not applied. Weight: the ESPEI weight, 500~J\,mol$^{-1}$/$\sigma$, for the full stratum; the released datasets use the leave-system-out $\sigma$. Source: \texttt{espei/dft/CALIBRATION.json}.}
\label{stab:calibration}
\begin{tabular}{lllS[table-format=3.0]S[table-format=2.1]lS[table-format=-1.1]S[table-format=1.3]}
\toprule
Source & Chemistry & Match & {$n$} & {$\sigma$} & 95\% CI & {Bias} & {Weight} \\
\midrule
\tabCalBody
\bottomrule
\end{tabular}
\end{widetable}

\begin{widetable}[htbp]
\centering
\footnotesize
\caption{\textbf{Extraction benchmark: baseline scores by rendering.} det.: deterministic route; AI reader: Gemini~3.8~Flash with the production prompts, as a single reader; --: not run for that rendering. $^{a}$Scores from the prompt version that returned placeholder rows (Supplementary Note~1).}
\label{stab:bench}
\begin{tabular}{llrrrr}
\toprule
Metric & Extractor & Vector PDF & Raster 300 dpi & Degraded scan & Image-only PDF \\
\midrule
\tabBenchBody
\bottomrule
\end{tabular}
\end{widetable}

\begin{widetable}[htbp]
\centering
\footnotesize
\caption{\textbf{Programmatic access.} The \nMcpTools{} tools, 4 resources and 1 prompt of the \texttt{orchid-mcp} server (tool names as returned by \texttt{tools/list}; Supplementary Fig.~S6), the static file endpoints described by its OpenAPI specification, and the other routes to the data. Every answer carries the citation, DOI, page, quote or table locator, evidence class and flags of each value, with the source information needed to report a discrepancy. Direct links to pre-filled issues are part of the contribution workflow under development in SI~2.}
\label{stab:access}
\renewcommand{\arraystretch}{1.15}
\begin{tabularx}{\linewidth}{P{5.0cm}Y}
\toprule
Route & Purpose \\
\midrule
\multicolumn{2}{l}{\emph{MCP tools}} \\
\texttt{search\_systems} & find chemical systems by elements or text; counts by evidence class \\
\texttt{get\_system\_data} & literature values of a system, filtered by property group, code or evidence class \\
\texttt{get\_invariant\_reactions} & literature and calculated invariant reactions of a system, side by side \\
\texttt{get\_formation\_enthalpy} & formation enthalpies of a compound: literature, Materials Project, OQMD \\
\texttt{get\_phase\_diagram} & calculated binary diagrams: TDB file, SHA-256, citation, phases, invariants \\
\texttt{compare\_to\_dft} & literature formation enthalpies against both DFT sources, with residuals \\
\texttt{list\_tdb\_files}, \texttt{get\_tdb} & TDB catalog entries and files \\
\texttt{get\_paper} & bibliographic record of a paper, its values and experimental context \\
\texttt{search\_papers} & search \dsBibliography{} by text, system, role or year \\
\texttt{get\_paper\_relations} & typed relations of a paper: what it cites, which assessments and TDB files used it \\
\texttt{get\_assessment\_evidence} & the data papers behind the assessments of a system \\
\texttt{get\_phase}, \texttt{search\_phases} & phase-registry identities, names, models, structure and stability \\
\texttt{get\_structure\_type} & structure types by prototype, Strukturbericht, Pearson symbol or space group \\
\midrule
\multicolumn{2}{l}{\emph{Static endpoints (OpenAPI, \texttt{openapi/openapi.yaml})}} \\
\texttt{/literature/\ldots} & index, per-system files, papers, contexts, DFT parity, full dumps (JSONL, CSV) \\
\texttt{/phase-diagrams/\ldots} & index and one file per calculated diagram \\
\texttt{/hulls/\{layer\}/\ldots} & catalogs and per-system hulls (MP, OQMD, SGTE at 298.15~K), entries tiers \\
\texttt{/phases/\ldots} & registry (JSON, JSON-LD, CSV), per-system records, structure types \\
\texttt{/catalog.json}, \texttt{/databases/\ldots} & TDB catalog and bundled TDB files \\
\texttt{/croissant.json} & Croissant metadata \\
\midrule
\multicolumn{2}{l}{\emph{Other routes}} \\
Parquet, Hugging Face & \texttt{parquet/*.parquet}; dataset card with \nHfConfigs{} configurations \\
web pages & per-system and per-phase pages at \url{https://data.orchid-project.org}; \texttt{llms.txt} \\
Zenodo & \nZenodoDataRecords{} records, one for each dataset except \dsBench{} and one for the ESPEI datasets, each with its own DOI, and an umbrella record \\
examples & \texttt{orchid-examples}: runnable workflows (Cu--Zr with PhaseForge+ and with ESPEI, the one-shot score, the MLIP benchmark, a quick start) \\
contributions & issue forms and validated files for corrections and new data (\texttt{CONTRIBUTING.md}) \\
\bottomrule
\end{tabularx}
\end{widetable}

\begin{widetable}[htbp]
\centering
\footnotesize
\caption{\textbf{Literature-level gates: actions by gate.} Corrected: field changed with the stored value kept; demoted: not accepted; flagged: flag or not ESPEI-ready only; merged: duplicates merged into one record. A record can receive several actions. Source: \texttt{literature/quality\_fixes.json} (\texttt{gates}).}
\label{stab:gatecounts}
\begin{tabular}{llrrrr}
\toprule
Gate & Checks & Corrected & Demoted & Flagged & Merged \\
\midrule
\tabGateBody
\bottomrule
\end{tabular}
\end{widetable}

\clearpage
\section{Methods}
\label{sdetail:methods}

Turning the literature into a record of this kind requires answering, for every value, a sequence of questions: where it is printed, what exactly it says, which equilibrium state it describes, how it was obtained, and whether it is consistent with what is already known about the system.
The first of these questions are answered while a value is being read, through the corpus from which it was taken, the way it was located on the page and the agreement required between independent readers.
The remaining ones, in contrast, can only be answered by placing the value in context, against calculated phase diagrams, first-principles energies and the rules of phase equilibria, and it is in the ESPEI datasets that all of these answers finally come together.

\begin{widefigure}[tbp]
\nfig{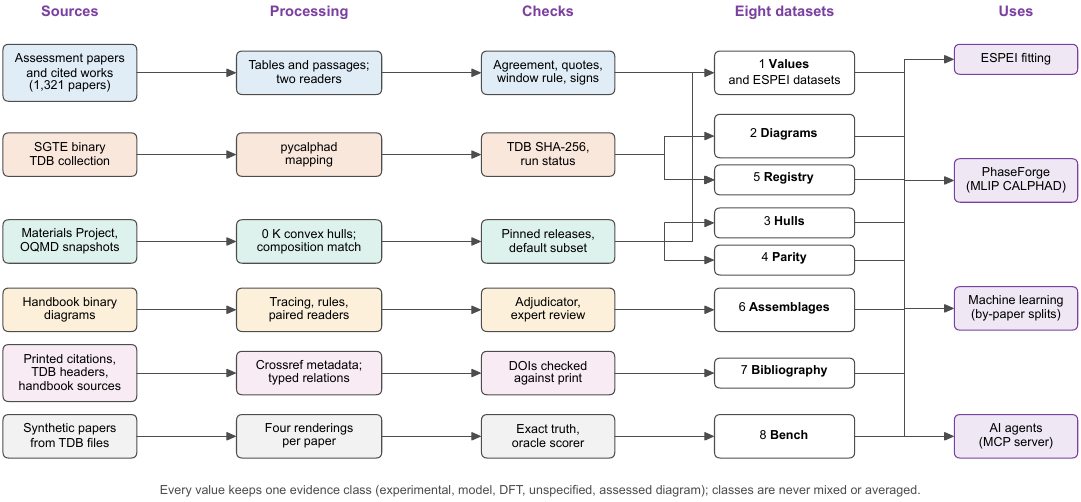}
\caption{\textbf{From sources to datasets.} Six sources pass through processing and checks into \nDatasetsWord{} datasets that serve four uses. Literature tables and running-text passages are read by independent readers; values pass reader agreement, verbatim-quote verification, the window rule for field placement, page-image sign checks, the Version~1 clean-up and the literature-level gates. The SGTE binary TDB files give calculated phase diagrams and, with the literature labels and the Materials Project and OQMD entries (arrow from lane 2), the phase registry. The bibliography (\dsBibliography) links the papers of the corpus, the cited works, the TDB files and the handbook diagrams. Handbook diagrams are traced and named by rules, paired readers, an adjudicator and an expert; only derived data are released. Synthetic papers generated from TDB files test the readers against exact truth. Rows are coloured by source. All datasets are served as files, web pages and through an MCP server.}
\label{fig:pipeline}
\end{widefigure}

\subsection{Corpus}
A corpus for CALPHAD data is most naturally built around the assessments themselves, since each assessment already identifies, through its references, the experimental work on which its model rests.
The corpus used here is therefore anchored in the assessments behind the SGTE collection of binary databases\cite{hallstedt2025}, and it extends to the primary sources these assessments cite and to further assessments of the same systems.
This extension was carried out in three steps, starting from the source paper named in the header of each SGTE binary TDB file, continuing with the works cited in these papers, and completing the set with further assessments of each system found through bounded title queries of the Crossref database.
The papers themselves were read from lawfully held copies, and, since their copyright remains with the publishers, they are not redistributed, although every value points to the published version through its DOI.
The values themselves were read from \nPapers{} papers published between \yearMin{} and \yearMax.
Because many of these values are reprints, each is attributed both to the paper in which it was read and to the primary source in which it first appeared, so that \dsValues{} refers to \nBiblio{} works in all.
The bibliography, \dsBibliography{}, in turn, places these works within a wider map of \nBibPapers{} papers, among them \nPrimarySources{} works cited as the original source of extracted values, and assessments that were identified but not yet read.

\subsection{Locating values}
Within each paper, values appear in two forms, in tables, where they are organized but often abbreviated, and in running text, where they are embedded in sentences that also state their context.
Tables were therefore located first and classified as invariant-reaction tables or as general tables for thermochemical, phase-composition and crystallographic data, and the running text was then read passage by passage.
Values shown only in figures were not read in Version~1, since recovering them requires the digitization of plots rather than the reading of printed numbers, a different task that is left to a later release.
Whatever its form, every value that was read keeps a locator, namely the paper, page, table label and row, or the passage and sentence in which it is printed, so that it can always be found again on the page.

\subsection{Dual-reader extraction}
Finding a value on the page, however, is not the same as reading it correctly.
To obtain independent readings of a table, the PDF text layer, where available, was compared with the outputs of two vision-language models reading the page image.
Gemini~3.8~Flash was the first reader and Claude~Opus~5.5 the second, with Gemini Pro used to confirm a reading or resolve disagreement (\suppfig{2}).
Both models received instructions encoding the rules of phase equilibria, including the number of phases required by the phase rule for an invariant reaction.
Agreement between these independent readings is the basis for accepting a value, since two readers that work separately are unlikely to make the same mistake.
A row was therefore accepted only when the readers agreed and the deterministic checks described below were passed, while disagreements were either settled by a judge model (\nTierJudge{} values) or left flagged as not accepted.
Whichever path a value followed is recorded in the field \texttt{reader\_tier} (Table~\ref{tab:readers}), so that a user can select values by the strength of the agreement behind them.
The running text called for a stricter form of the same principle, since a value in a sentence has no row or column to anchor it.
Two readers therefore read each passage, and each returned every value together with the verbatim sentence that states it and the page on which it appears, so that the claim could be checked against the printed text itself.
In both tables and text, finally, the phase labels used by the authors were mapped to the phases of a CALPHAD model only when the mapping was unambiguous, and otherwise kept as printed.

\begin{widetable}[tbp]
\centering
\footnotesize
\caption{\textbf{Reader agreement tiers of the literature values.}}
\label{tab:readers}
\renewcommand{\arraystretch}{1.2}
\begin{tabularx}{\linewidth}{YS[table-format=5.0]}
\toprule
\texttt{reader\_tier} & {values} \\
\midrule
Flash and Claude agree\newline(\texttt{preliminary\_flash+claude}) & \nTierFlashClaude \\
confirmed by Gemini Pro\newline(\texttt{confirmed\_pro}) & \nTierPro \\
first reader, second reader's verdict recorded\newline(\texttt{preliminary\_flash}) & \nTierFlash \\
judge decision\newline(\texttt{preliminary\_flash+claude+fable\_judge}) & \nTierJudge \\
\midrule
total & \nValues \\
\bottomrule
\end{tabularx}
\end{widetable}

The principle of agreement rests on an assumption that is easy to overlook, namely that each reader returns what it sees, so that an empty answer means an empty table.
During the preparation of this release, this assumption failed for one reader under one version of its instructions, which returned a single empty row for tables that the other reader transcribed in full.
The synthetic benchmark described in the Validation exposed this behaviour, and an audit of the production records showed that, because such empty readings had been counted as votes, a number of rows were in effect supported by a single reader.
To correct this failure, empty readings were excluded from agreement and affected rows were sent to an independent second reader.
The resulting field \texttt{reader\_agreement} records the number of independent readers supporting a table value.
Values supported by a single reader remain in the release, visibly flagged (\suppnote{1}, \suppfig{3}).
At list prices, the language-model calls behind this release would cost about US\$\costTotalUSD{}, of which about US\$\costPipelineUSD{} went to reading the literature and the handbook diagrams, roughly \costCentsPerAccepted{} US cents for each accepted value; this is a lower bound and a counterfactual one, since some sessions left no usage record and part of the work ran under subscriptions. Supervising and checking the agents took more than \costAuthorHours{} hours of one author's active time.

\subsection{Verification against the printed page}
Agreement between readers shows that they saw the same thing, but not that it is printed where they claim.
For this reason, a value read from running text is kept only when its quoted sentence is found verbatim on a single page of the PDF with the printed number inside it, and any further detail that a reader attached to the value, such as a phase, a method or a citation, is kept only when it is printed in the same sentence or immediately around it, the exact rules being listed in \supptab{3}.
Details that fail this test are removed rather than guessed, and the removal is recorded with each value.

\subsection{Evidence classes}
A value that has passed these tests still says nothing, by itself, about how it was obtained, and yet the difference between a measurement, a calculation and an expert's evaluation is precisely what a user needs to know before including the value in a fit.
Each value is therefore assigned exactly one evidence class, namely experimental, model (calculated with a CALPHAD or other thermodynamic model), first-principles (DFT) or unspecified, while the assemblages taken from expert-drawn phase diagrams form a class of their own in \dsAssemblages.
The class is decided from what the paper prints, in the column or label of a table, in a caption or in the text, and, for a value reprinted from another work, from that work itself, whose title, abstract or text usually states whether it reports a measurement or a calculation.
Where neither source settles the question, a cited work published before 1980 is taken to be experimental and any other value remains unspecified, and in every case the field \texttt{origin} records which of these rules made the decision, so that the classification itself can be checked (Table~\ref{tab:evidence}).

\begin{widetable}[tbp]
\footnotesize
\caption{\textbf{Evidence classes.} Every value has exactly one class; classes are never mixed or averaged. Counts: literature values (\dsValues) and assessed points (\dsAssemblages).}
\label{tab:evidence}
\renewcommand{\arraystretch}{1.2}
\begin{tabularx}{\linewidth}{P{3.6cm}>{\hsize=0.7\hsize}Y>{\hsize=1.3\hsize}YS[table-format=5.0]}
\toprule
Class & Definition & Rule (recorded in \texttt{origin}) & {Values} \\
\midrule
\texttt{experimental} & measured & the paper's column, label, caption or text names a measurement; for a cited value, the cited work's title, abstract or full text does; cited works published before 1980 are taken as experimental & \nExp \\
\texttt{model} & calculated from a CALPHAD or other thermodynamic model & the printed source names a calculation or an assessment; linked to a TDB file only when unambiguous (\nTdbLinked{} values) & \nModel \\
\texttt{dft} & first-principles calculation & printed as DFT or \emph{ab initio}; never compared with DFT & \nDFT \\
\texttt{unspecified} & origin not settled & no printed or resolvable kind of source & \nUnspec \\
\texttt{assessed\_phase\_diagram} & read from an expert-drawn handbook diagram & \dsAssemblages{} only; the original evaluation is cited first, the handbook second & \nAsmPoints \\
\bottomrule
\end{tabularx}
\end{widetable}

\subsection{Printed-page sign checks}
Reader agreement and quote verification can both miss errors already present in the PDF text layer.
In this text layer, a minus sign may be dropped, separated by a line break or interpreted as a digit.
To check for these errors, formation enthalpies with suspect signs were examined on images of the printed pages.
The record retains either the sign confirmed on the page or a flag indicating that no sign is printed, so that a sign is never supplied without printed evidence (\suppnote{1}).

\subsection{Units and conditions}
The same principle governs the conversion of values to common units, which is necessary if values from different papers are to be compared, but which must not obscure what the authors actually printed.
Temperatures are therefore converted to kelvin and compositions to mole fractions, while the printed string and the original basis are kept with each value.
Formation and mixing quantities are expressed per mole of atoms only when that basis is printed or can be established from the paper, and are otherwise left unconverted, with their basis marked as unstated.
This distinction also applies to pressure, which is recorded as an assumed one atmosphere for all but \nPressureNotOneAtm{} values because it is almost never printed.
The basis field identifies the assumption, and it is withdrawn when the page specifies another pressure.
Lattice lengths printed in nm or pm are converted to \AA, with the stored value retained.
Other quantities retain their printed units, including \nGibbsCoefT{} Gibbs energies that are coefficients of $T$ in expressions of the form $A + BT$ and therefore have units of J\,mol$^{-1}$\,K$^{-1}$.

Uncertainties are treated in the same way.
Only \nSigmaPrinted{} of \nValues{} values have a printed uncertainty.
Where present, the reported uncertainty is retained in the primary uncertainty field, with documented conversions to the record unit and preservation of the original information.
Estimates of scatter between studies and of uncertainty due to composition differences are stored separately in the uncertainty budget, together with their derivation.
The distinction allows the reported precision to be separated from the effective uncertainty used to weight ESPEI datasets.
To retain the experimental context, samples, purity, heat treatment and measurement methods are stored as quoted items with page references.
These items are attached only to values from the paper's own experiments, preventing the transfer of one study's conditions to another study's data.

\subsection{Calculated phase diagrams}
After extraction of a value and its conditions, an assessed CALPHAD description provides a reference for interpreting the result within its binary system.
For this purpose, the release includes \nDiagrams{} phase diagrams calculated with pycalphad\cite{otis2017} from the SGTE collection of binary assessments\cite{hallstedt2025}.
The diagrams were mapped using \texttt{BinaryStrategy} in pycalphad 0.11.2 at 101,325~Pa, with composition steps of 0.02 and temperature steps of 10~K.
The requested temperature range was 300--2,500~K, extended to 300--4,000~K when a refractory element was present.
This mapping provides phase boundaries as zero-phase-fraction lines together with the invariant equilibria.
A grid of stable phases, calculated by direct equilibrium calculations at 0.01 in composition and 10~K, was added so that the stable phase assembly at any point of the diagram can be looked up rather than recomputed.
The SHA-256 checksum associates a diagram with its TDB file, and the accompanying citation identifies the assessment.
The calculated invariants are therefore model values whose reliability is limited by that assessment and the resolution of the calculation grid.

\subsection{Agreement with the CALPHAD model}
The predictions of an assessment provide a second description of a reaction that can be compared with the literature record.
Although this comparison is useful for detecting inconsistencies, model values do not constitute a standard for establishing that literature values are correct.
When a value read from a paper and the corresponding value calculated from an assessment agree closely, the agreement supports both, whereas a large discrepancy points to an error in the reading, in the paper or in the model, or to a reaction that the model does not reproduce.
The comparison was performed for three-phase invariants whose phases could be mapped to a linked TDB file and for pure-element transitions using the SGTE unary data.
Agreement required a temperature difference of at most 25~K and, where printed compositions could be mapped, phase-composition differences of at most 0.05 in mole fraction.
This comparison does not yet include congruent and polymorphic transitions of compounds.
Across these comparisons with \nPlausTdb{} TDB files, \nPlausChecked{} values were evaluated, of which \nPlausPass{} agree and \nPlausFail{} disagree.
For these evaluated values, the \texttt{plausibility} field records the model and the temperature and composition differences, while the extracted value remains unchanged.
Agreement is reported separately for experimental values, of which \nPlausPassExp{} of \nPlausCheckedExp{} agree, since a model value checked against the TDB file of its own paper tests the self-consistency of the record rather than its agreement with independent evidence.

\subsection{Convex hulls}
A second, independent reference is provided by first-principles energies, from which binary convex hulls at 0~K were built using pinned snapshots of the Materials Project (release 2026.04.13)\cite{jain2013} and OQMD (v1.8, February 2026 release)\cite{saal2013,kirklin2015}.
These energies are calculated without reference to the extracted literature values and thus provide an independent comparison with those values and the assessments.
However, both databases use experimental references in their corrections or fitting, through the fitted Materials Project corrections (MP2020)\cite{wang2021mp2020} and the OQMD fitted elemental reference energies (the standard, or fit-partial, set of Kirklin \emph{et al.}\cite{kirklin2015}), which are used as stored by each database.
To make the stability information accessible, a hull is stored as a sequence of composition intervals from which stable phases and relative chemical potentials can be read directly.
In this representation, polymorphs on or very near the hull are retained and flagged.
The two databases, however, use different energy references and corrections, and they are therefore never averaged or compared with one another directly.
Their stability information is also treated differently, since the Materials Project hulls reproduce the stability reported by the database itself for all but one system, whereas the stability of OQMD entries, which could not be reproduced from its stored values, is derived from the hulls.

\subsection{DFT comparison and calibrated uncertainty}
Each source can, however, be compared with experiment, since many of the compounds on the hulls also appear in the literature with measured formation enthalpies.
To pair a literature value with a DFT entry, their reduced compositions must be identical.
The formula is taken from the printed phase label, and records naming more than one formula are set aside.
For a given source, the match uses the lowest entry in the space group identified by the record through a Strukturbericht designator or prototype.
If no structure is identified, the lowest-energy entry at that composition is used.
Restricting the comparison to accepted experimental values whose sign and magnitude raise no doubt then gives, for each database, a set of pairs whose differences measure how far DFT formation energies can be trusted.
The differences are analysed separately by database and by groups of chemically similar compounds.
This analysis gives robust uncertainty estimates of \sigMP~\kjma\ for the Materials Project and \sigOQMD~\kjma\ for OQMD, together with a small systematic bias that is recorded but not applied.
The ESPEI datasets use the calibrated uncertainty to weight first-principles data (Fig.~\ref{fig:dftsigma}).
Source-verified repeated extractions or later reports of the same DFT result contribute once to the calibration and fitting inputs, while the literature records retain both identifiers and the reported values through an explicit canonical evidence link.
The same identity links alternate unit representations of one observation. For experimental values reported alongside DFT calculations, the evidence class follows the description of the measurement; when the original measurement conditions remain unresolved, both representations are retained in the literature release and excluded from calibration and fitting.

\subsection{Data clean-up}
The two references also serve a more practical purpose, since a literature value that disagrees with both of them by a wide margin often reveals an error of reading or of reporting rather than a genuine discrepancy.
The clean-up of Version~1 used this and simpler criteria to address four kinds of defect: duplicated records, inconsistent labels for reactions and properties, formation enthalpies printed per formula unit but labelled per atom, and fields garbled during reading.
In every case, the clean-up merges, relabels, converts or flags a record but never invents a value, the original is kept wherever a value changed, and each change is listed with its evidence in the release (\texttt{QUALITY\_FIXES.md}; counts in \suppnote{4}).

\subsection{Literature-level consistency gates}
The clean-up treats records one at a time, but many errors only become visible when a value is compared with its table, its paper, the physics of its quantity and the calculated diagrams of its system, and for this reason a second set of gates was applied to the whole release (\texttt{tools/literature\_gates.py}).
These gates correct a record only when the correction is deterministic and evidenced, by a printed header, caption or heading or by the internal consistency of a whole table, and for changes of unit, sign, scale or chemical system the correction was in addition confirmed on the rendered page of the PDF.
Of the \nGateRecordsIn{} records that entered the gates, \nGateCorrected{} received a correction.
Most corrections concerned the conditions of a value, namely \nGateTempCorrected{} temperatures (\nGateTableCelsius{} in tables printed in degrees Celsius but stored as kelvin, and \nGateTempStdRef{} stored at 298~K that denote 298.15~K) and \nGateSystemReassigned{} assignments to a chemical system, while others concerned its meaning, namely \nGatePhaseListFixes{} phase lists of invariant reactions and \nGateMixToForm{} formation quantities that had been recorded as mixing quantities, and the remaining \nGatePageValueFixes{} values and \nGatePageUnitFixes{} units were read again on the page.
A record that could not be corrected in this way was demoted or flagged instead, with the reason in \texttt{gate\_failures}, and the gates merged a further \nGateDupMerged{} duplicates that differ only in the spelling of a phase label, leaving \nGateRecordsOut{} records, of which \nGateAcceptedAfter{} were accepted at that stage (\nGateAcceptedBefore{} before the gates). The current release contains \nAccepted{} accepted records.
The confirmations on the page were made by AI agents that inspected the rendered page, not by a person, and each of them is recorded in a ledger of \nGateVisualEntries{} entries (\texttt{tools/literature\_visual\_checks.json}) that gives the page, the hash of the page image, the verdict and a short note of what is printed, without any page image or text.
No printed value or quote is altered, and the stored value is kept beside every correction.
Three rules introduced in the third round determine which states can be released as accepted. A value of an invariant, a transition, a phase composition or a formation quantity requires the phases of its state, recovered from the verbatim quote (\nGateStatePhasesQuote{} records) or read on the page (\nGateStatePhasesPage); labels such as ``unknown'', ``other'' and ``?'' do not satisfy this requirement. The R3 gate recorded \nGateStatePhasesMissing{} missing-phase findings at that stage; values without identified phases were not accepted. A later page-checked recovery restored printed phases for 1,267 records demoted only by the placeholder rule and accepted them again, while retaining their exclusion from ESPEI output.

For invariant reactions, the phases are checked against the printed reaction, a missing class is filled only for binaries written in a deciding form and a printed class is never overwritten. The count of phases is also checked against the phase rule, and a second record of the same invariant on one page is flagged (\nGateInvPhaseRule{} and \nGateInvDuplicate{} records). In turn, every element in the printed formulas of the phases provides a lower bound on the system of the state, so that compound data cannot be narrowed to a unary system. Finally, a formation value below 0.1~\kjma, or one whose unit prefix is printed separately from the unit, is accepted only after a check on the page.

\subsection{Phase registry}
Combining values from different sources raises the question, already noted in the introduction, of whether two of them refer to the same phase.
The registry answers it by giving each phase of a binary system a single identity, built from the phases declared in the TDB files of the SGTE collection, the phase mappings made during extraction and the labels printed with the literature values, and by collecting under that identity every name the phase carries, each with the source in which it appears (Fig.~\ref{fig:registry}).
Structure types (prototype, Strukturbericht designation, Pearson symbol and space group) were assigned with the same caution, through three kinds of evidence of decreasing directness.
The most direct is a standard phase name that encodes a single structure, such as \texttt{L12\_FCC}, accepted when the model's stoichiometry, sublattices and DFT data do not contradict it.
Where the name is not informative, a stoichiometric compound can take its structure from DFT, but only when it matches exactly one entry on the hull of both the Materials Project and OQMD, with the same space group and no other phase at that composition.
The crystallographic table that accompanies the SGTE collection\cite{hallstedt2025} provides the third kind of evidence, accepted when all of its rows for a phase give the same structure and nothing in the first two contradicts it.
A structure type is thus assigned only when the evidence makes it certain (the first two kinds) or when it is stated by the assessors themselves (the third), and every other phase is placed in a review queue together with its candidate structures and the reasons why none was assigned.
For each phase, the registry also keeps the TDB models that describe it, with their definitions copied verbatim, its stability range in the calculated diagrams, and examples of the papers that model it or report values for it.

\begin{widefigure}[tbp]
\centering
\nfig{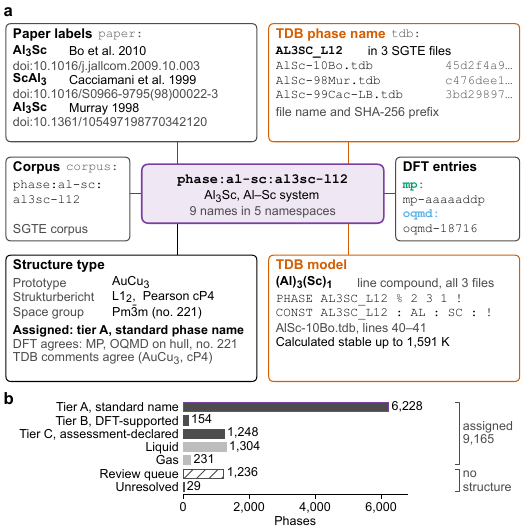}
\caption{\textbf{The phase registry (\dsRegistry).} \textbf{a}, The identity \texttt{phase:al-sc:al3sc-l12} (Al$_3$Sc, L1$_2$), read from the released file \texttt{phases/systems/AL-SC.json}: its names in five namespaces (paper labels with their DOIs, the TDB phase name with its files, the corpus identifier, and the Materials Project and OQMD entries), its structure type with the kind of evidence that assigned it, and its TDB model with the verbatim \texttt{PHASE} and \texttt{CONSTITUENT} statements. \textbf{b}, Phases by the way their structure type was assigned: standard name (Tier~A, \nRegTierA), DFT-supported (Tier~B, \nRegTierB), assessment-declared (Tier~C, \nRegTierC), liquid (\nRegLiquid), gas (\nRegGas), review queue (\nRegReview) and unresolved (\nRegUnresolved), of \nRegPhases{} phases.}
\label{fig:registry}
\end{widefigure}

\subsection{Assessed phase assemblages}
Not all of the evidence about a system takes the form of printed values, since much of it has been condensed by experts into evaluated phase diagrams, drawn from the experimental literature without an underlying thermodynamic model.
\dsAssemblages{} draws on the binary diagrams of the Okamoto \emph{Desk Handbook: Phase Diagrams for Binary Alloys} (2nd ed., ASM International, 2010), and because such a diagram is the outcome of an evaluation rather than a measurement or a calculation, the assemblages read from it form an evidence class of their own.
Reading such diagrams relies, above all, on the rules of phase equilibria.
A deterministic pass first traces the drawn lines, calibrates the axes and identifies the fields, and printed labels and the law of adjacent phase regions then name as many fields as the diagram allows, a name being accepted only when a single assemblage is consistent with the whole diagram (main-paper Fig.~\ref{fig:annotation}).
Machine readers are consulted only for the fields that the rules leave open, and even then they choose among the assemblages that the rules permit, while an adjudicator, and finally an expert, settle the cases that the readers cannot (\suppnote{2}).

\subsection{Uncertainty budget}
To weight evidence in a fit, the uncertainty must describe the reliability of the value rather than only the statistical precision usually reported for a measurement.
For example, Knudsen-effusion mass spectrometry values for Cu$_{51}$Zr$_{14}$ and Cu$_{10}$Zr$_{7}$ report uncertainties of \cuzrKemsSigmaLo{} to \cuzrKemsSigmaHi~\kjma, yet differ from calorimetric values by \cuzrKemsDiffLo{} to \cuzrKemsDiffHi~\kjma.
The comparisons below show that differences of this magnitude between independent studies are common.
An uncertainty budget is therefore retained for all values, preserving the reported uncertainty alongside additional contributions.
The reported interval and its stated coverage remain distinct from the effective uncertainty used in fitting, so that a calculated estimate does not replace the uncertainty attributed to the source.

These are the resolution of the printed number, a systematic contribution that depends on the class of method used, and, where they apply, terms for a measurement made at a historical formula, for the scatter of values found on the page to be mutually inconsistent, and for a conversion to another reference state.
To calibrate the systematic contribution, we compared independent measurements of the same state obtained by the same class of method in different primary studies.
As in the DFT calibration, the scatter across these measurements determines the floor, rather than the precision claimed within an individual study.
The scatter was estimated robustly as 1.4826 times the median absolute difference divided by $\sqrt{2}$, with a bootstrap over states providing the 95\% interval.
Classes with fewer than \nFloorMinPairs{} pairs or fewer than 10 states received a documented literature prior\cite{kubaschewski1993,drowart2005}.
If such a class still contained at least \nFloorMinPairsCleaned{} independent pairs from \nFloorMinStatesCleaned{} states, the larger of this prior and its own robust scatter was used (\nFloorCleanedClasses{} classes).
In applying these rules, no floor was permitted below the default ESPEI uncertainty.
For formation enthalpies, the calibrated floors are \floorHmCal~\kjma\ for solution and reaction calorimetry (\nPairsHmCal{} pairs; 95\% interval \ciHmCal), \floorHmEmf{} for EMF measurements and \floorHmNotPrinted{} when the method is not printed.
For invariant temperatures and phase compositions, the floors are \floorTinv~K and \floorXphase{} in mole fraction (Table~\ref{tab:sigma}).
These floors are larger than the agreement usually claimed for good calorimetry, because they also absorb the errors of extraction and of reference state that remain in Version~1, and they are therefore conservative.

The effective uncertainty combines the larger of the printed uncertainty and the floor with the other contributions.
This value determines the default ESPEI weights, while the retained components allow users to adopt a different combination rule.

\begin{widetable}[tbp]
\footnotesize
\caption{\textbf{Method floors of the uncertainty budget calibrated from the release.} Floor: robust scatter between independent measurements of the same state made with the same class of method in different primary studies (1.4826 times the median absolute difference divided by $\sqrt{2}$, rounded); pairs (states) and the bootstrap 95\% interval of the scatter. Bottom: model-calculated values against experimental values of the same state. The other \nFloorPriorClasses{} classes, with fewer than \nFloorMinPairs{} pairs or fewer than 10 states, take a documented prior\cite{kubaschewski1993,drowart2005}. Every value carries the budget; \nSigmaMethodCalibrated{} values take a calibrated floor and \nSigmaMethodPrior{} a prior. Source: \texttt{literature/UNCERTAINTY\_CALIBRATION.json}.}
\label{tab:sigma}
\renewcommand{\arraystretch}{1.2}
\begin{tabularx}{\linewidth}{>{\hsize=0.8\hsize}Y>{\hsize=1.2\hsize}Yllc}
\toprule
Quantity & Method class & Floor & Pairs (states) & 95\% interval \\
\midrule
\tabFloorBody
\bottomrule
\end{tabularx}
\end{widetable}

\subsection{ESPEI datasets}
To convert the extracted evidence into ESPEI datasets, the preceding checks must establish the state, its mapping to the phases of a model, and the consistency and uncertainty of the value.
These conditions are recorded for every value in two fields.
The first, \texttt{espei\_ready}, marks a value that can be used in a fit, because its state is complete, its phases can be mapped unambiguously to those of a CALPHAD model and are physically possible at its temperature, and it is consistent with the rules of phase equilibria.
The field \texttt{espei\_certified} further identifies ready values whose states are reproduced by the pycalphad equilibrium of the corresponding assessment within 25~K and 0.05 in composition.
Ready values that disagree with this calculation, or lack a corresponding calculation, remain available with the appropriate flag.

The influence of these values on a fit depends on how ESPEI represents uncertainty.
ESPEI expresses the uncertainty used in a fit through weights relative to default experimental uncertainties.
These default uncertainties in ESPEI~0.9.1 are 500~J~mol$^{-1}$ for enthalpies and 0.2~J~K$^{-1}$~mol$^{-1}$ for entropies and heat capacities\cite{bocklund2019}.
The weight of an observation is therefore the experimental reference divided by its effective uncertainty.
When a literature dataset contains values with unequal effective uncertainties, ORCHID exports them as separate fitting records so that a common weight does not replace those differences; values with equal uncertainties can remain grouped.
To allow this weight to be recomputed under another policy, the uncertainty budgets are retained in physical units, together with the basis of their components.
These bases include uncertainties printed by the authors, method floors calibrated against independent measurements, and comparisons with experiment for DFT and model values.
The records retain their links to the original dataset and source study because separation for weighting does not establish statistical independence, and the likelihood used here does not represent covariance between observations from a shared source.

The uncertainty budgets also allow the choice of evidence to be reconsidered.
First-principles formation energies are included only when the compound can be assigned to a single endmember of a specific TDB phase model, and receive weights based on the calibrated DFT uncertainty.
Model-calculated values instead use a floor obtained by comparison with experimental values of the same state (\floorModelHm~\kjma\ for formation enthalpies).
The resulting weights are choices that users are expected to revisit.
Every dataset is tagged with its evidence class and its quality flags, so that what enters a fit, and with what weight, is decided by the user rather than fixed by the release (Fig.~\ref{fig:espeirecord}).

\begin{widefigure}[tbp]
\nfig{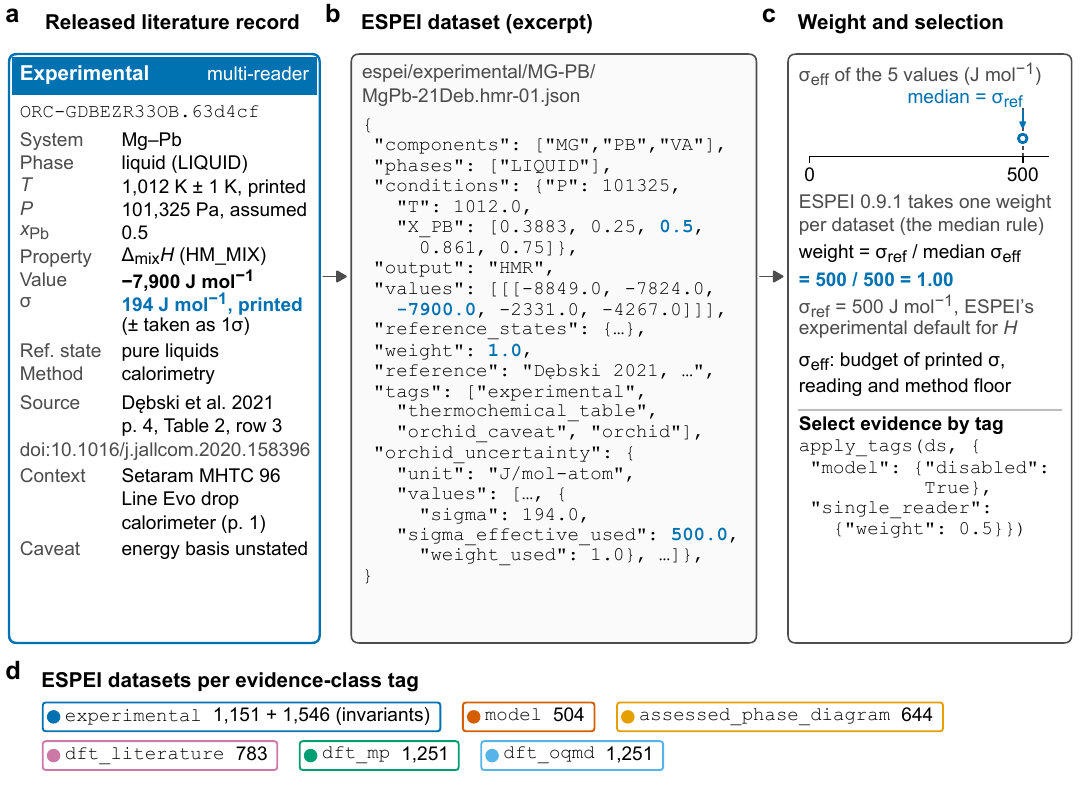}
\caption{\textbf{From a literature value to an ESPEI dataset.} One released ESPEI-ready value followed through the release. \textbf{a}, The literature record (\texttt{literature/systems/MG-PB.json}, record \texttt{ORC-GDBEZR33OB.63d4cf}): system, phase, conditions with their bases, property, value, printed uncertainty, reference state, method, reader agreement and source; the colour of the card is its evidence class. \textbf{b}, The ESPEI dataset that contains it, trimmed, with the value highlighted and the \texttt{orchid\_uncertainty} block that records the budget of each value and the weight used. \textbf{c}, The weight: ESPEI~0.9.1 takes one weight per dataset, here the experimental reference $\sigma$ divided by the median effective uncertainty of the values of the dataset, and data are selected by tag with \texttt{apply\_tags}. \textbf{d}, ESPEI datasets per evidence-class tag (experimental \nEspeiExpDatasets{} plus \nEspeiInvDatasets{} invariant datasets, model \nEspeiModelDatasets, assessed phase diagram \nEspeiAssessedDatasets, and the three DFT sources). Every line of panel b is rendered from the released file.}
\label{fig:espeirecord}
\end{widefigure}

\section{Validation}
\label{sdetail:validation}

A user who is to decide what enters a fit needs, above all, an estimate of how far each kind of value can be trusted.
The validation described in this section provides such estimates in three steps, moving from the reading itself, through the consistency of the extracted values with independent references, to the derived datasets built from them.

\subsection{Quote verification}
The most direct test of a reading, that the value is printed where the record says it is, was applied to every value taken from running text, and all \nText{} of them carry a quoted passage found verbatim on the cited page with the printed number inside it.
This test establishes where a value comes from, but not whether it was interpreted correctly, and the remaining checks in this section address that second question.

\subsection{Reader comparison pilot}
A first indication of how often a reading goes wrong came from a pilot in which the two text readers answered the same \pilotRequests{} requests.
Claude Opus found more values than Gemini Flash and quoted them verbatim far more reliably (\pilotVerbatimOpus{} of \pilotRowsOpus{} rows, against \pilotVerbatimFlash{} of \pilotRowsFlash), and the errors of both readers concentrated in passages whose PDF text layer is corrupted by control characters, the same passages in which Flash also lost minus signs.
These differences are the reason the release relies on two readers together with verification against the page rather than on either reader alone, although a comparison between readers is not, in itself, an estimate of precision (\suppnote{1}).

\subsection{Extraction benchmark}
A controlled estimate of extraction precision can be obtained from a reference with known generating values, which is the purpose of the synthetic papers in \dsBench{}.
Because every table, diagram and sentence in them was generated from an SGTE database before being rendered as a born-digital PDF, a clean image or a degraded scan, any extraction tool can be scored against the truth, and in this release the benchmark served both to compare the readers and to expose the failure described in the Methods, where one reader returned empty tables under one prompt version.
Its scores, summarized in \supptab{6} and \suppfig{5}, show that the AI readers recover invariant-reaction tables almost completely from born-digital PDFs and largely from degraded scans, while deterministic tools recover far fewer invariant rows even from born-digital PDFs and fail on most degraded diagrams.

\subsection{Scope of the numerical human-review sample}
\label{sdetail:precision}
Synthetic papers cannot reproduce every difficulty of the published literature, from corrupted text layers to tables that continue across pages.
A stratified random sample of accepted values was frozen before the corrections described below for independent reviewers to check against the printed pages field by field, with overlapping items intended to measure reviewer agreement.
The sample holds \nPrecisionSample{} values from the export of \precisionExportDate, of which \nPrecisionSurvive{} remain in release v2026.10 (\nPrecisionSurviveAccepted{} of them accepted), the other \nPrecisionGone{} having been merged with duplicates.
This sampling record is retained as part of the audit history, but the review was not completed and no aggregate human precision estimate is reported for the numerical collection.
The preliminary responses described below do not constitute a completed evaluation of the sample, and subsequent source adjudication was AI-assisted.
Because the sample predates the corrections, it also does not establish the precision of later releases.

\subsection{Consistency with independent references}
Verification of a reading establishes fidelity to the source, while comparison with thermodynamic references can reveal a source error or an incorrect assignment of the system or state.
We therefore compare the eligible records across the release with first-principles formation energies, invariant reactions calculated from the SGTE assessments and an independent compilation of experimental formation enthalpies.
The independence of these references varies, since an assessment may have used the same measurements and a compilation may reproduce them, as detailed below.

Formation enthalpies offer the broadest of these comparisons, since many of the compounds reported in the literature also appear in the Materials Project and OQMD.
Over the default subset of accepted experimental values, DFT and experiment agree with a mean absolute error of \maeMP~\kjma\ for the Materials Project and \maeOQMD~\kjma\ for OQMD, and with Spearman rank correlations of \spearmanMP{} and \spearmanOQMD{} (Fig.~\ref{fig:parity}, Table~\ref{tab:parity}), values consistent with earlier assessments of DFT accuracy\cite{kirklin2015}.
The residuals are, however, far from Gaussian, with a robust spread of about \sigOQMD--\sigMP~\kjma\ and a tail of large disagreements, of which \nOutliersExamined{} were examined, \nPerFormulaFixed{} being per-formula values labelled per atom and \nUnsignedFixed{} unsigned magnitudes, which were explained and corrected or flagged, while \nOutlierUnexplained{} have no deterministic explanation and are flagged \texttt{dft\_outlier\_unexplained}, \nOutlierUnexplainedDefault{} of them remaining in the default subset, with the flag, and the others being excluded by its point rules (\suppnote{4}).
The comparison is not entirely independent of experiment, since the Materials Project corrects its energies, and OQMD fits its elemental references, against measured enthalpies, but it is independent of the literature values and of the reading.

\begin{widetable}[tbp]
\footnotesize
\caption{\textbf{Literature formation enthalpies against DFT (default subset unless stated).} MAE, median $|$residual$|$ and RMSE in \kjma; band: fraction within $\pm$10~\kjma.}
\label{tab:parity}
\begin{tabular}{llS[table-format=4.0]*{3}{S[table-format=2.1]}*{3}{S[table-format=1.2]}}
\toprule
Source & Subset & {$n$} & {MAE} & {Median} & {RMSE} & {Spearman} & {Pearson} & {Band} \\
\midrule
\tabParityBody
\bottomrule
\end{tabular}
\end{widetable}

Phase equilibria provide a second and more demanding comparison, since an invariant reaction fixes a temperature, a set of phases and their compositions at once.
When the experimental invariant temperatures in the release are matched, by system, reaction type and phase set, to the invariants calculated from the SGTE assessments, the two agree with a median offset of \invValExpMedian~K and a robust spread of \invValExpSigma~K, \invValExpWithinTen\% of them within 10~K (main-paper Fig.~\ref{fig:invariants}).
Since the assessments were themselves fitted to many of these measurements, close agreement mainly confirms that the values were read and assigned correctly, while the small number of large disagreements points either to assessments that depart from the data or to values that require a second look.

Of the three references, the most direct is an independent compilation of experimental formation enthalpies\cite{kim2017}, because its values were gathered by other authors, often from the same original sources, so that a disagreement points to the reading or to the compilation rather than to any model.
For the \nKimOverlap{} compounds present in both, the two agree with a median difference of \medKim~\kjma\ and a mean absolute difference of \maeKim~\kjma, with every sign in agreement, while the \nKimDiffer{} compounds that differ by more than 10~\kjma\ are listed in the release so that the source of each discrepancy can be traced.

\subsection{Independent audits and corrections}
To determine whether individual discrepancies reflected systematic errors, separate audits searched the whole dataset for recurring classes of error.
These audits found several, among them tables in which temperatures printed in degrees Celsius had been stored as kelvin, values assigned to the wrong chemical system, formation quantities recorded as mixing quantities, and an export step that labelled thermochemical data as liquid by default.
Each class was then either corrected, when the printed page settled the value unambiguously, which was verified by inspecting the page itself, or the affected values were withdrawn from the fit-ready subset and flagged, and a second audit confirmed the result (\suppnote{3}).
The release records the checks passed and corrections received by each value so that users can select evidence according to the requirements of their analysis.

\subsection{Derived datasets}
Checks on derived datasets address both the source values and the transformations used to prepare them for their intended use.
For the ESPEI datasets, the relevant test is whether ESPEI can use them, and each was therefore evaluated against the TDB file it names, with every one of the \nEspeiEvaluated{} datasets in the release yielding a finite likelihood, while the \nEspeiEvalExcluded{} that did not were excluded with the reason recorded.

For the phase registry, the corresponding test is whether the identities it asserts hold up when different sources are compared, and building the registry was, in effect, such a test, since every disagreement between sources was recorded rather than resolved silently.
The recorded conflicts (\texttt{CONFLICTS.md}) identify uncertainty in phase identity, including names used for different structures in different systems, disagreement between Materials Project and OQMD ground states, names unsupported by DFT, and families of polymorphs.
In all such cases no structure type was assigned, while the verbatim TDB definitions were checked against the files from which they were copied.
Structure types assigned from the assessors' own tables (the third kind of evidence) rest on their statement rather than on an independent check, and are marked accordingly.

The phase assemblages read from handbook diagrams required the most extensive validation, because their names were assigned partly by machine readers.
To assess the naming procedure, a blind audit of \asmAuditFields{} fields used independent model instances that saw the assigned name but not its basis.
Name precision ranged from \asmAuditPrecLo{} to \asmAuditPrecHi{} across naming routes, and was about \asmAuditPrecWeighted{} after weighting by the number of fields.
Auditors from the same model family as some readers may, however, judge those readers too favourably.
The assessed assemblages agree with SGTE equilibria in \nAsmSgteAgree{} of \nAsmSgteComparable{} comparable point--assessment pairs.
Considering agreement at the point level, \nAsmSgtePointsAgreeAny{} of \nAsmSgtePointsComparable{} comparable points agree with at least one assessment.
The disagreements identify differences between handbook evaluations and assessments, and also exposed two systematic annotation faults corrected before release.
The expert, finally, reviewed the \asmExpertReviewed{} most difficult fields with the adjudicator's reasoning in view and accepted \asmExpertAccepted{} of the \asmExpertProposals{} named proposals, a figure that measures verification rather than independent agreement.
An independent estimate comes from a blind sample of \nBlindExpertN{} fields, one per diagram and drawn from diagrams the expert had not reviewed, in which the expert named each field without seeing the machine's answer.
In this blind sample, the expert and machine names agreed for \nBlindExpertAgree{} of \nBlindExpertN{} fields (\BlindExpertAgreeRawPct\%, 95\% interval \BlindExpertWilsonRaw).
Agreement increased to \nBlindExpertAgreeEquiv{} fields (\BlindExpertAgreeEquivPct\%, \BlindExpertWilsonEquiv) when two notation differences were recognized, namely repetition of the same phase and a prefix unresolved by the drawing.
Within this sample, all \nBlindRuleAgree{} fields named using printed labels and phase-equilibrium rules agreed.
The one substantive disagreement concerns a field bounded by three closely spaced, separately labelled line compounds, where the drawing itself leaves the assignment ambiguous, and the expert's answer has been adopted.

\subsection{Adversarial audit and native error rates}
\label{sdetail:audit}
The adversarial audits searched for error classes beyond those anticipated by the existing checks.
For this reason, before the release, we also searched for errors deliberately, using \nAuditsWord{} independent auditors, language-model agents that could read the release, the code and the printed pages but could not change them, each instructed to find errors rather than to confirm quality within a single scope.
Four of these scopes concerned the data themselves (elements and compositions; phases and their mapping to CALPHAD models; units, conditions and uncertainties; and DFT energies with their pairing to experiment), while the other two concerned how the data are presented, tracing every figure and every number and claim in this paper back to the release.
Each auditor combined targeted scans for whole classes of error with a random sample of records drawn with a recorded seed, and every suspected error was checked against the printed page of its source before it was counted.
The sampled fractions therefore estimate the native error rate of the release, that is, the rate of errors that remained after all the gates and corrections described above (Fig.~\ref{fig:audit}, Table~\ref{tab:audit}).
For elements and compositions, phases, and the pairing of experimental and DFT energies, the native rates of wrong values or states lay between \auditCoreRateLo{} and \auditCoreRateHi{} per cent.
The rate was highest, \auditValueRateUnits{} per cent (\auditValueCIUnits{} per cent), among thermochemical records, because thermochemical tables mix bases, reference states and property types more often than tables of phase equilibria do, and it is these tables that the uncertainty budget described above treats most conservatively.
Random samples of this size cannot, however, reveal errors that are rare among records but systematic in their cause, and it was the targeted scans that found these, among them compositions divided by one hundred twice when a header announced atomic per cent over values that were already fractions, polymorphs matched to the wrong DFT structure, intermediate phases labelled $\varepsilon$ mapped to the bcc solution, absolute entropies recorded as entropies of formation, and lower-order data filed under the system of the paper rather than that of the state.
Each of these classes was corrected from the printed pages and incorporated into a regression test that subsequent releases must pass.
The native rates document the audited snapshot before correction, providing a baseline for comparison with subsequent audits.
When the same protocol was applied to new random samples of the corrected release, the rate of wrong values or states was \auditPostRateElem{} per cent for elements and compositions, \auditPostRatePhase{} per cent for phases, \auditPostRateUnits{} per cent for units, conditions and uncertainties, \auditPostRateDft{} per cent for the pairing of DFT with experiment, \auditPostRateFig{} per cent for figures and \auditPostRateClaims{} per cent for the claims of this paper (Table~\ref{tab:audit}), and every rate is equal to or lower than in the first round, although for most scopes the intervals overlap, because the samples are small. The incomplete human-review record (Section~\ref{sdetail:precision}) contains 240 submitted responses covering 210 distinct records in the frozen 300-record sample. These preliminary responses do not establish a sampling-adjusted estimate of population accuracy, and the subsequent source adjudication was AI-assisted.

The re-audit itself found new classes of error, and some of them had been introduced by the fixes: the rule that moved lower-order states out of higher-order systems had trusted a composition that was already corrupt and narrowed about 290 compound records to a unary or smaller system, the floor of the uncertainty had been scaled by the number of atoms for formulas that were already normalised, and the rule that set temperatures printed as 0~K to 298.15~K had been applied to the DFT rows of papers whose own values were labelled experimental. The post-fix rates are therefore not those of a release free of error, and for units, conditions and uncertainties the rate of all in-scope errors (\auditPostAllRateUnits{} per cent) cannot be compared with that of the first round (\auditAllRateUnits{} per cent), because the re-audit also counted labels that had lost their basis and temperatures that were missing.

We corrected these classes in a third round, and because the re-audit had shown that a fix can introduce errors, we sampled \nRThreeResample{} records again in \nRThreeScopes{} pools of records that the round had changed, rendering and viewing every item. This caught \nRThreeFixIntroduced{} errors that the first build of the round had introduced, among them \rThreeFirstWrongFive{} of \rThreeNFive{} sampled invariant reactions, where derived classes had overwritten printed ones, and they were corrected before release; the counts for each pool, with the residuals that predate the round, are in \suppnote{3}.

A final AI-assisted re-audit of the third-round release used new seeded samples checked against the printed pages. It found wrong values or states in 3 of 110 literature values (2.7\%, Wilson 95\% interval 0.9--7.7\%), 1 of 300 phase-registry records (0.33\%), 5 of 110 records checked for thermochemical units, temperatures and reference states (4.5\%), and 4 of 110 experimental--DFT parity pairs (3.6\%); no error was found in 200 figure checks, whereas 7 of 100 record-level claims in the paper failed the stricter claim protocol (7.0\%). Separate samples found errors in 2 of 50 invariant records and 2 of 30 recovered-phase records. Although none of the 30 sampled C4 system reassignments had the wrong system, 7 contained a wrong value or state, concentrated in three tables. Corrections made on 27--28 October followed this audit. In a subsequent AI-assisted review of the fit-facing release, 189 of 200 ESPEI-ready records were fit-safe (94.5\%, Wilson 95\% interval 90.4--96.9\%).
The Y11 review used a frozen, stratified sample from the data snapshot \texttt{7ac9ca9f18}, with rare evidence categories deliberately oversampled. The value, composition, phase, property and evidence class passed in all 200 records. Two temperatures conflicted with the source pages, and nine other records lacked a source-supported molar basis: six DFT mixing-enthalpy records retained an unstated basis, while three experimental formation enthalpies used a normalization inferred from magnitude. These nine cases were precautionary fitting exclusions, not demonstrated errors in the printed numerical values. The strict composite criterion required all seven fields, including temperature and fitting-compatible unit normalization, to pass. The 94.5\% proportion describes the reviewed sample; applying the population stratum weights gives a 97.59\% point estimate, for which no confidence interval was reported. The audit proposed two temperature corrections and nine fitting-exclusion flags for subsequent integration; its score characterizes the frozen pre-correction snapshot.

\begin{widetable}[tbp]
\footnotesize
\caption{\textbf{Adversarial audits of release v2026.10: samples, native error rates and systematic classes.} Wrong value or state and all in-scope errors (including metadata): records with an error / records sampled (rate; Wilson 95\% interval in per cent). Systematic classes: number found by the targeted scans and three examples. Post-fix: the same protocol applied to the corrected release (\texttt{orchid-data} 1e5f4ce7d), as the rate of wrong values or states and, below it, of all in-scope errors (Section~\ref{sec:audit}).}
\label{tab:audit}
\renewcommand{\arraystretch}{1.15}
\begin{tabularx}{\linewidth}{P{2.3cm}>{\hsize=0.9\hsize}Y>{\hsize=1.0\hsize}Y>{\hsize=0.8\hsize}Y>{\hsize=1.3\hsize}YP{1.2cm}}
\toprule
Audit scope & Sample; seed & Wrong value or state & All errors & Systematic classes & Post-fix \\
\midrule
\tabAuditBody
\bottomrule
\end{tabularx}
\end{widetable}

\clearpage
\section{Matched MLIP benchmark and one-shot construction}
\label{sdetail:mlip}
The energy benchmark uses the identical 1,138 records, 455 compounds and 185 binary systems for all eight MLIPs and both DFT references. The primary variant relaxes Materials Project structures and applies the specified MP2020 correction convention. Robust residual spreads use $1.4826$ times the median absolute deviation, with 1,000 cluster-bootstrap replicates resampling whole chemical systems. The matched-set results are distinct from the source-level calibration of all available DFT pairs.

The one-shot Cu--Zr models use a common PhaseForge construction for GRACE-2L-OMAT, MACE-MPA-0, MACE-OMAT-0, ORB-v3 MPA, ORB-v3 OMat, SevenNet-Omni, MatterSim 5M and CHGNet 0.3.0. Each model completed 21 structure relaxations and five liquid runs. The liquid schedules comprised 2,000, 3,000 and 10,000 steps; all 40 liquid runs passed the recorded diffusion check. Some special quasirandom structures did not meet the force convergence threshold within 1,000 FIRE steps and entered the construction at their last energies; relaxed-away structures were flagged rather than dropped. These choices are part of the tested workflow and must accompany interpretation of the phase-stability results.

The figure and reported field scores use a pinned evidence snapshot, with 28 assessed field points and 287 boundary points. A later corrected release contains 276 boundary points, while the field points and field scores are unchanged. Field confidence intervals use 2,000 point-bootstrap replicates with seed zero. The overlapping intervals do not resolve the ranking of the eight models; a paired significance test against SGTE was not performed. The final PhaseForge+ field score is in sample and is kept distinct from one-shot predictions and held-out activity errors.

\begin{widetable}[htbp]
\centering\small
\caption{One-shot Cu--Zr field scores on the pinned evidence. Intervals are 95\% point-bootstrap intervals; precision and F1 use the corresponding phase-prevalence weights.}
\begin{tabular}{lccc}\toprule
Model & Recall [95\% interval] & Precision & F1\\\midrule
MACE-MPA-0 & 0.56 [0.42, 0.69] & 0.53 & 0.51\\
ORB-v3 MPA & 0.54 [0.40, 0.67] & 0.50 & 0.42\\
ORB-v3 OMat & 0.48 [0.36, 0.60] & 0.45 & 0.41\\
MACE-OMAT-0 & 0.46 [0.32, 0.60] & 0.60 & 0.46\\
SevenNet-Omni & 0.46 [0.34, 0.59] & 0.41 & 0.40\\
MatterSim 5M & 0.44 [0.31, 0.58] & 0.41 & 0.36\\
GRACE-2L-OMAT & 0.43 [0.30, 0.55] & 0.41 & 0.33\\
CHGNet 0.3.0 & 0.35 [0.22, 0.49] & 0.44 & 0.34\\\midrule
SGTE assessment & 0.72 [0.59, 0.84] & 0.65 & 0.66\\
PhaseForge+ (in sample) & 0.76 [0.64, 0.87] & 0.65 & 0.69\\\bottomrule
\end{tabular}
\end{widetable}

\subsection{Runtime of the final Cu--Zr refinement}
The final run, \texttt{runs/final2/F10b\_s3}, records 1,618.98~s (26.98~min) and 409 objective evaluations on a Mac Studio (Mac13,1, Apple M1 Max, macOS 14.3). The optimizer script \texttt{fit\_pr268.py} uses analytical ESPEI gradients, first with conjugate gradients and then with scaled L-BFGS-B. This run used starting seed 3, the prepared \texttt{CuZr-PhaseForge-MLIP-B2b.tdb} description and the \texttt{data/a4d\_F10} fitting datasets together with the MLIP anchors. The timer starts after loading the datasets and constructing the residual context and covers optimization and final run-output preparation. It excludes literature extraction and review, MLIP relaxation and molecular dynamics, construction of the starting description, preparation of the fitting datasets, and exploration of alternative starts. The reported time therefore measures the final refinement run, rather than the development of the complete demonstration. The earlier \texttt{tuning.wall\_s} entry in the aggregate results file refers to a different fit. The timing provenance and source checksum are retained in \texttt{figures-DD/data/cuzr\_final\_runtime.json}.

\section{Full construction and refinement figure notes}
The following legends retain the detailed methodological qualifications of the source manuscript. The figures themselves appear in the main paper.

\paragraph{Figure~\ref{fig:annotation}.} \textbf{Annotating an assessed diagram, shown on a synthetic benchmark diagram.} The Ni--Ti diagram of the synthetic paper \texttt{syn-NiTi-04Tok-v0} (\dsBench), calculated from the SGTE assessment of Tokunaga \emph{et al.} (2004) and drawn as in a paper; no handbook drawing is reproduced. Field names are exactly those of the benchmark's truth key, which uses conventional labels for the TDB phases: (Ni) = \texttt{FCC\_A1}, ($\alpha$Ti) = \texttt{HCP\_A3}, ($\beta$Ti) = \texttt{BCC\_A2}, NiTi = \texttt{NITI\_B2}, Ni$_3$Ti = \texttt{NI3TI\_D024}, NiTi$_2$ = \texttt{NITI2}, L = \texttt{LIQUID}. \textbf{a}, Lines traced automatically from the vector PDF (black) over the drawn lines (grey); closed traces smaller than 0.03 in mole fraction and 60~K, which are plot markers, are removed. \textbf{b}, Fields enclosed by the drawn lines (truth field map); a field name can label more than one enclosed region, so that regions and names are counted separately. Temperatures are shown up to 2,100~K; fields are clipped to this range before their points are placed. \textbf{c}, Field names; names of fields too narrow for a label are placed outside with a leader line, and names that would overlap are not printed. \textbf{d}, One point per field: the area centroid when it lies at least 0.03 (axis fraction) from every line and the frame, otherwise the interior point farthest from all field boundaries (the pole of inaccessibility, i.e.\ the centre of the largest inscribed circle), since the centroid of a curved or crescent-shaped field can lie outside the field or on a boundary; fields too thin for half that margin give no point. Dashed: the tie-line on the isotherm through the point of the largest two-phase field, between the two boundaries it meets; it is drawing-based and a weak constraint. Point coordinates and margins: \texttt{scripts/data/fig\_annotation\_points.json}.

\paragraph{Figure~\ref{fig:mlip}.} \textbf{MLIPs: saturated on energies, discriminating on phase diagrams.}
\textbf{a}, Robust standard deviation $\sigma$ ($1.4826\times$ the median absolute deviation) of the formation enthalpy
minus experiment for the two DFT databases and 8 universal machine-learning interatomic potentials (MLIPs), all
evaluated on the identical set of 1,138 records (455 compounds, 185 binary systems); MLIP energies are
relaxations of the MP structures with the MP2020 correction (Methods). Bars are 95\% confidence intervals from a
cluster bootstrap that resamples chemical systems with all their records (1,000 replicates). Violet shades separate
OMat24-trained models from models with an MP-compatible head. The green band is the interval of Materials Project; the
blue band on the left is the experimental noise floor, the calibrated between-study $\sigma$ of solution and reaction
calorimetry for formation enthalpies (3.2~kJ\,(mol of atoms)$^{-1}$, from 80 pairs of independent
measurements of 47 states; \suppinfo{}). The MLIPs span $\sigma=8.6$--$9.7$~kJ\,(mol of atoms)$^{-1}$
against 8.8 for MP, about 2.8 times the noise floor (PBE ceiling). The column gives
$\Delta\sigma=\sigma_\mathrm{MLIP}-\sigma_\mathrm{MP}$ with its cluster-bootstrap interval; every interval includes zero.
\textbf{b}, Residual of GRACE-2L-OMAT against the residual of MP for the same records (hexagonal bins, logarithmic density;
dashed line $y=x$; 25 of 1,138 pairs lie beyond $\pm40$~kJ\,(mol of atoms)$^{-1}$ and are not drawn).
The strip lists the Pearson correlation of each MLIP's residuals with those of MP ($r=0.961$--$0.997$).
\textbf{c}, Prevalence-weighted recall of the phase assemblages at the 28 assessed field points of Cu--Zr
(the fraction of the phases present at those points that a description also predicts, each phase weighted by its
prevalence; Methods and \suppinfo{}) for one-shot descriptions built by PhaseForge from each MLIP alone, with no fit to
experiment, against the $\sigma$ of panel a (same axis). Vertical bars are point-bootstrap 95\% intervals. GRACE-2L-OMAT reaches 0.43.
The grey band spans the 9 later CALPHAD assessments in the collection (0.70--0.98), with a tick per
assessment, and the dashed line is the SGTE assessment of Zeng et al.\ (1994; 0.72). The star is the PhaseForge+
description tuned with ORCHID data (0.76), drawn at the $\sigma$ of its starting MLIP; it is in-sample, because the
tuning used Cu--Zr data that include these points.

\paragraph{Figure~\ref{fig:espei}.} \textbf{Constructing a Cu--Zr description from MLIP energies and ORCHID data.}
\textbf{a}, Workflow. PhaseForge\cite{zhu2025machinelearningpotentials,zhu2025accelerating,sariturk2025phaseforge}
builds a starting description (TDB, MLIP only) from GRACE-2L-OMAT energies\cite{bochkarev2024grace} of the six
compounds, of special quasirandom structures (SQS) generated with m\_scraps\cite{arroyave2026mscraps,singh2021scraps}
instead of ATAT, and of the liquid from molecular dynamics, on SGTE unary data. ORCHID supplies phase equilibria, liquid
activities and experimental formation enthalpies, each with its uncertainty, and its DFT-versus-experiment calibration
gives the MLIP energies an uncertainty of \sigmaMlipFit~kJ\,(mol of atoms)$^{-1}$, so that experiment governs where it exists
(bars drawn to scale against a typical experimental uncertainty); the fit used the data and the calibration of release
\cuzrFitRelease{} of \texttt{orchid-data}, whose calibration has since changed slightly (release v2026.10 gives
\sigmaMlipNow~kJ\,(mol of atoms)$^{-1}$). PhaseForge+\cite{kunselman2025construction,sariturk2025phaseforgeplus}
then tunes the interaction parameters of the liquid and the solid solutions (constant and linear in $T$), the compound
energies and an entropy term for CuZr by gradient-based maximization of the likelihood, with analytic gradients
(Jansson derivatives for the phase equilibria) from the gradient-capable ESPEI of Kunselman et al.\cite{kunselman2025gradient,kunselman2024espeibranch}
(pull request \#268, not yet merged). The badge counts the \cuzrInvTotal{} invariant temperatures within twice their uncertainty (2$\sigma$).
\textbf{b}, Phase diagram tuned with ORCHID data (PhaseForge+, violet lines, fields coloured by phase), compared with the
same model before tuning (PhaseForge from MLIP energies alone, dashed) and with the SGTE assessment of Zeng et al.\ (1994;
\texttt{CuZr-94Zen-LB}, one of the \nCuZrAssessments{} Cu--Zr assessments of the SGTE collection; grey).
Symbols are ORCHID phase-equilibrium points (filled, fitted; open, held out) and the points of \dsAssemblages{} placed in the
fields of the assessed diagram (orange; the \nCuZrAsmFit{} of the \nCuZrAsmPoints{} released Cu--Zr points whose phases
map to this assessment). Tuning brings \cuzrInvFinal{} of the \cuzrInvTotal{} invariant temperatures within twice their uncertainty (MLIP alone \cuzrInvMlip,
SGTE \cuzrInvSgte), lowers the liquid-activity error from \cuzrActHoldMlip{} to \cuzrActHoldFinal~kJ\,mol$^{-1}$ on held-out isotherms (SGTE \cuzrActHoldSgte) and the
held-out phase-equilibrium error from \cuzrZpfHoldMlip~kJ\,mol$^{-1}$ to \cuzrZpfHoldFinal{} (SGTE \cuzrZpfHoldSgte). It still places the congruent melting of
CuZr \cuzrCongruentHigh~K too high and the eutectoid decomposition of CuZr \cuzrEutectoidLow~K too low, and shows a narrow spurious ($\beta$Zr) field near
$x_\mathrm{Zr}=0.1$ between 1220 and 1290~K. Besides the points drawn, the fit used \nCuZrDemoDense{} further points
inside the fields of the same assessed diagram, which are released with ORCHID-Assemblages as a dense grid; the model is post hoc relative to the pre-registered fit (\cuzrInvPrereg{} of \cuzrInvTotal{} invariants), which is reported in the
demonstration repository.

\clearpage
\section{Supporting records, audits and DFT calibration}
\label{sdetail:supporting}
The figures and calibration below support the main-paper evaluation and MLIP comparison (Section~\ref{sec:benchmark} of the main paper). The final six-scope re-audit, the measured precision of the ESPEI-ready subset and the preliminary human review are reported in the Validation section; none of them is used as a precision estimate for the whole numerical collection. The DFT calibration covers the default experimental subset and is distinct from the matched set used to compare eight MLIPs.

\subsection{Representation of the source evidence}
Figure~\ref{fig:anatomy} illustrates how a printed table row and sentence are represented as records, retaining the distinction between the reported values, their evidence classes and assumed conditions.
\begin{widefigure}[!htbp]
\nfig{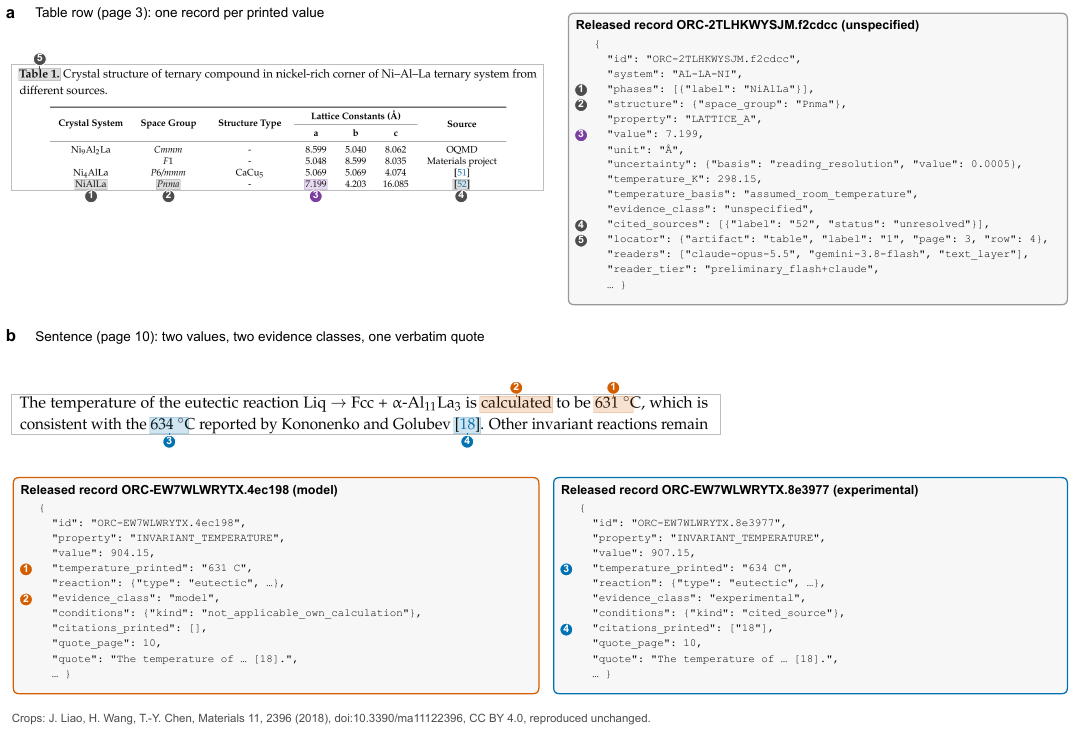}
\caption{\textbf{Anatomy of a record.} \textbf{a}, A table row of an open-access paper and the released record of its first lattice constant; numbered marks pair printed items with the fields that record them (3, the value). Temperature and pressure are not printed and are recorded with their bases (assumed); the row's other lattice constants are two sibling records. The kind of the cited work [52] is not printed, so the evidence class is \texttt{unspecified}. \textbf{b}, One sentence gives two records: the calculated eutectic temperature (631~$^\circ$C, \texttt{model}) and the measured value the paper cites (634~$^\circ$C, \texttt{experimental}); both carry the same verbatim quote and page. JSON excerpts are trimmed (\ldots). Crops are reproduced unchanged from Liao, Wang and Chen\cite{liao2018}, an open-access article distributed under a CC~BY~4.0 licence.}
\label{fig:anatomy}
\end{widefigure}
\FloatBarrier
\subsection{Audit snapshots and correction history}
The audit rates refer to the specific snapshots identified in the caption and in the Validation section above. They distinguish wrong values or states from other metadata errors and are not independent human precision estimates.
\begin{widefigure}[!htbp]
\nfig{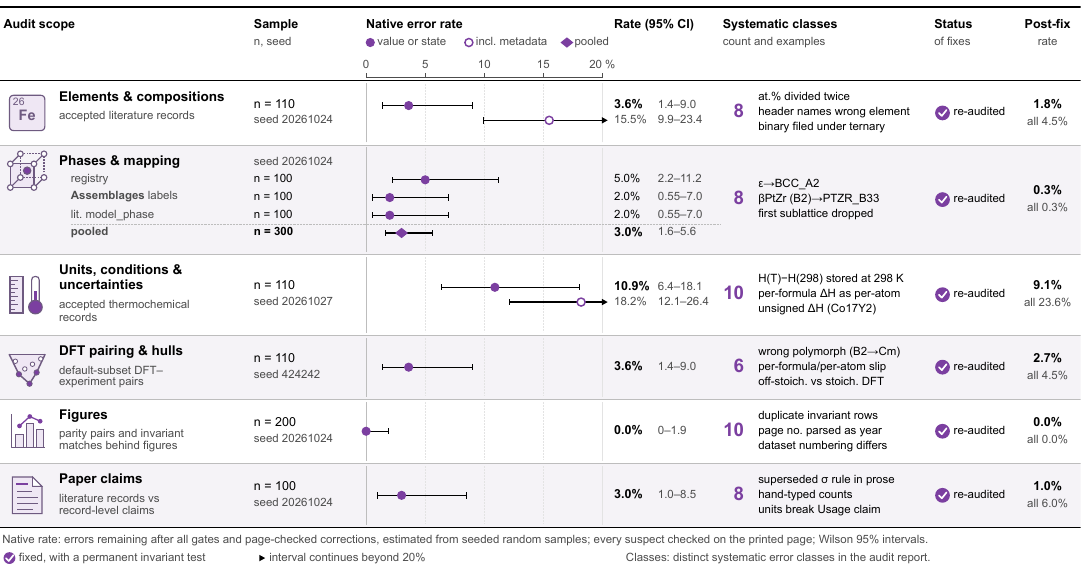}
\caption{\textbf{Errors identified by adversarial audits.} Seeded samples across six scopes distinguish wrong values or states from all in-scope errors, with Wilson 95\% intervals. Initial rates refer to the audited release snapshot \texttt{0d9f7c4c3}; post-fix rates refer to \texttt{1e5f4ce7d}. Targeted scans identified additional systematic classes. These are agent-assisted audits of specified snapshots, not human precision estimates or a final error rate for every later revision. Full denominators, correction history and subsequent resampling are retained in SI~1.}
\label{fig:audit}
\end{widefigure}
\FloatBarrier
\subsection{Experimental calibration of formation energies}
For the default subset of accepted experimental observations, the comparison with DFT gives mean absolute errors of \maeMP~\kjma\ for the Materials Project and \maeOQMD~\kjma\ for OQMD, with Spearman correlations of \spearmanMP{} and \spearmanOQMD{}, respectively (Fig.~\ref{fig:parity}).
The residual distributions have robust spreads of \sigMP{} and \sigOQMD~\kjma\ and tails containing both possible source errors and disagreements that remain unexplained after inspection.
The comparison is independent of the extraction, but not wholly independent of experiment, since the databases use measured information in their energy corrections or elemental references\cite{wang2021mp2020,kirklin2015}.
The resulting uncertainty therefore describes agreement under the stated matching and reference conventions, not an intrinsic error bound applicable to every DFT calculation.

\begin{widefigure}[!htbp]
\nfig{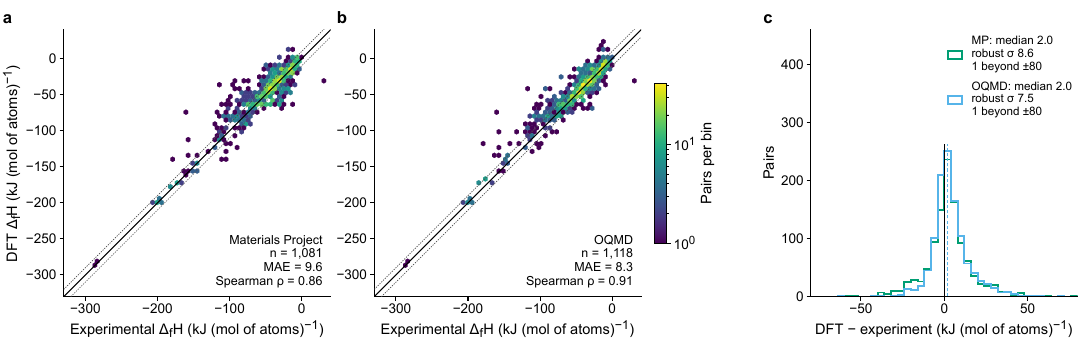}
\caption{\textbf{DFT against experiment for formation enthalpies.} Default subset: accepted experimental values, signed in print, without uncorrected sign loss or magnitude flags, each paired with the DFT entry of the structure that the paper names (the lowest-energy entry of that space group), or with the lowest-energy entry at the same composition when no structure is printed. Pairs are left out when the named structure is absent from the DFT database, when the measured composition is more than 0.02 from that of the compound, when the reference state is not the solid standard reference of 0~K DFT, and when the literature value has a gate failure. \textbf{a}, Materials Project (n = \nMPDefault; MAE \maeMP~\kjma). \textbf{b}, OQMD (n = \nOQMDDefault; MAE \maeOQMD~\kjma). Hexagonal bins on a log colour scale; solid line, $y = x$; dotted lines, $\pm$10~\kjma. \textbf{c}, Residuals (DFT minus experiment) of both sources; dashed lines, medians (bias \biasMP{} and \biasOQMD~\kjma, recorded but not applied); robust $\sigma$ (1.4826 times the median absolute deviation) \sigMP{} and \sigOQMD~\kjma; \nResBeyondMP{} (MP) and \nResBeyondOQMD{} (OQMD) residuals beyond $\pm$80~\kjma\ are not shown.}
\label{fig:parity}
\end{widefigure}

The residual distributions provide a basis for placing computed energies and experimental evidence on a common uncertainty scale for fitting.
The spread depends on the chemical stratum and the quality of the structure match, which are accounted for in the calibration.
Within this calibration, the system under consideration is excluded when estimating the uncertainty assigned to its first-principles observations, thereby avoiding the use of the same residuals for calibration and subsequent weighting.
Figure~\ref{fig:dftsigma} connects these spreads to the resulting ESPEI weights, showing how a computed energy can remain useful without receiving the precision of a well-characterized experimental observation.
The small systematic biases are recorded but not applied, and strata with insufficient data fall back to the source-level estimate.
The calibrated weights thus express a documented comparison with experiment while leaving the choice of evidence and weighting policy open to the assessor.

\begin{widefigure}[!htbp]
\nfig{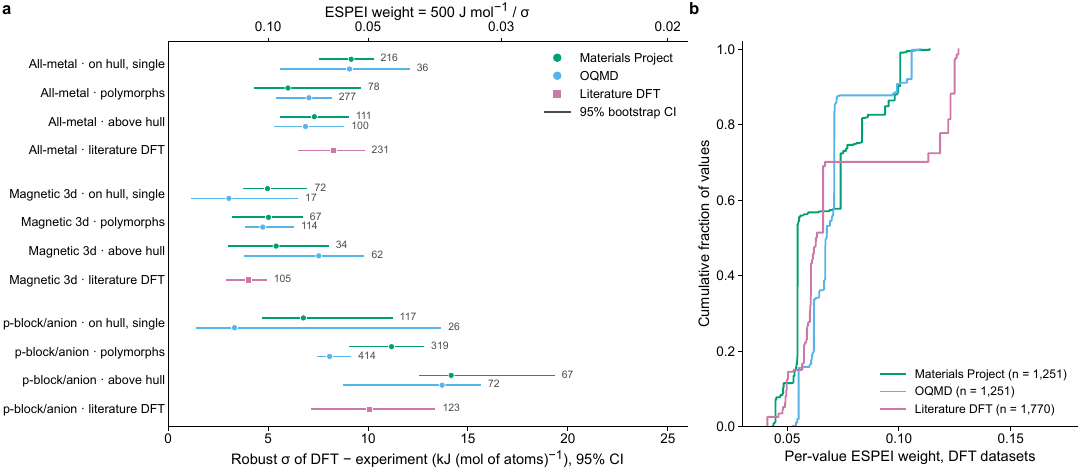}
\caption{\textbf{Calibrated DFT uncertainty and the resulting ESPEI weights.} \textbf{a}, Robust $\sigma$ of DFT minus experiment with 95\% intervals from 1,000 bootstrap resamples of whole chemical systems for each stratum (chemistry and DFT match quality) of the Materials Project and OQMD, and for literature DFT values by chemistry; numbers give the pairs per stratum. The match-quality stratum of a compound is assigned separately in each database, according to its own polymorphs and convex hull, so that the same compound can fall in different strata and the numbers of pairs per stratum differ between the databases. Top axis: the ESPEI weight, 500~J\,mol$^{-1}$ divided by $\sigma$. Strata with fewer than 30 pairs fall back to the source-level $\sigma$ in the datasets. \textbf{b}, Cumulative distribution of the per-value weights written to the DFT datasets (leave-system-out $\sigma$). Stratum values: \supptab{5}.}
\label{fig:dftsigma}
\end{widefigure}

\section{Additional coverage and state-resolved benchmark results}
\label{sdetail:science-r4}
The following figures document the source history, coverage and collection-wide comparison with calculated invariant temperatures.
\begin{widefigure}[tbp]
\nfig{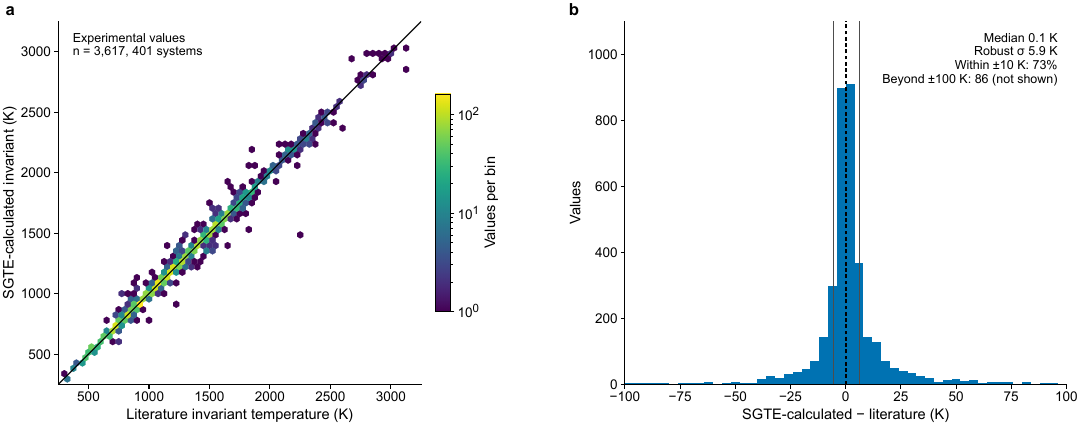}
\caption{\textbf{Literature invariant temperatures against SGTE-calculated invariants.} Accepted experimental temperatures of binary eutectic, peritectic, eutectoid and peritectoid reactions, matched by system, reaction class and phase-registry identities (n = \nInvValExp, \nInvValExpSystems{} systems). Eutectic, peritectic, eutectoid and peritectoid values number \nInvValEutectic, \nInvValPeritectic, \nInvValEutectoid{} and \nInvValPeritectoid. \textbf{a}, Calculated against literature temperature; hexagonal bins on a log colour scale; line, $y = x$; grey band, $\pm$10~K. \textbf{b}, Residuals (calculated minus literature); dashed line, median (\invValExpMedian~K); solid lines, median $\pm$ robust $\sigma$ (\invValExpSigma~K); \nInvValBeyond{} residuals beyond $\pm$100~K are not shown. Where several assessments of a system match, the median over assessments is used, each distinct invariant counted once. The comparison is in sample: \nInvValExpInSample{} of the \nInvValExp{} values come from a paper that is the source of one of the SGTE assessments of its system.}
\label{fig:invariants}
\end{widefigure}
\begin{widefigure}[tbp]
\nfig{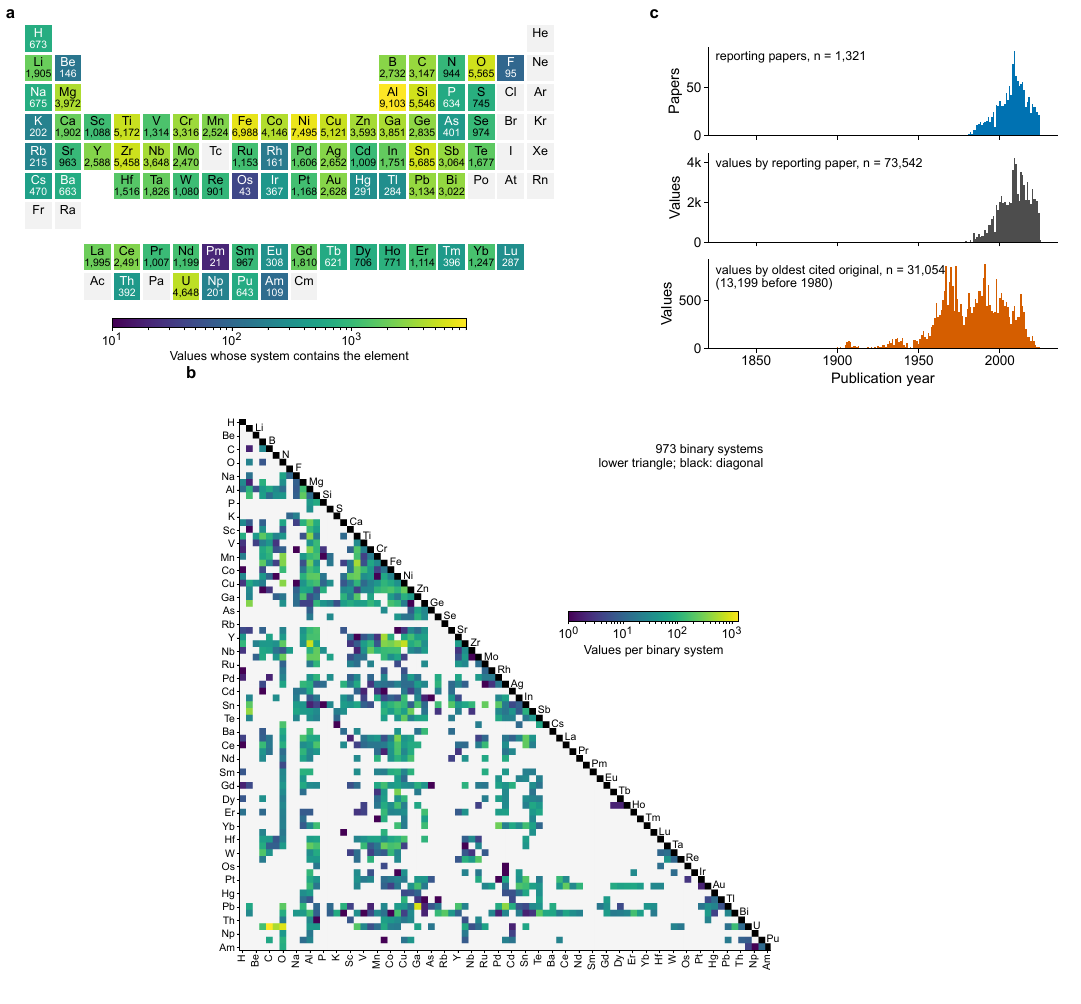}
\caption{\textbf{Breadth and history of the literature evidence.} \textbf{a}, Values by constituent element (logarithmic scale; grey indicates none). \textbf{b}, Values for the \nSystemsBinary{} binary systems, with elements ordered by atomic number. \textbf{c}, Publication years of the reporting papers, their extracted values and the oldest cited original for values with such a reference. Counts describe the released collection and its source history, not an exhaustive survey of all published thermodynamic measurements.}
\label{sfig:coverage-history}
\end{widefigure}
\subsection{Temperature-coverage classification}
The temperature panels use the same 73,542-record Parquet collection as the manuscript inventory. An explicit scalar record has a finite positive temperature with a printed, row-temperature, stated-room-temperature or caption-ambient basis, or represents a scalar invariant or transition temperature. The latter temperatures are themselves reported properties and may have no separate temperature-basis field. Ranges, bounds, expression terms and assumed conditions are excluded from the scalar set. This gives 55,516 explicit scalar records, 4,856 assumed-temperature records, 1,582 range, bound or expression records, and 11,588 other or unresolved records. The temperature histogram uses 150~K bins; 23 scalar values above 4,500~K are omitted from the displayed range but retained in the counts. This partition describes metadata availability, not independent measurement counts, acceptance status or fitting eligibility. The complete predicate, grouped counts and input hash are retained in the figure-generation script and its JSON sidecar.

\subsection{Field-wise and phase-wise failures}
The failure maps use the cached results and pinned evidence described in Section~\ref{sdetail:mlip}. No new equilibrium calculations or changes to the evidence snapshot enter these maps. For an assessed phase set $O$ and predicted set $P$, the two field maps show $|O\setminus P|$ and $|P\setminus O|$, respectively. The phase map gives the number of false negatives divided by the number of assessed field points containing that phase. Supports, in the plotted phase order, are 2, 3, 4, 5, 5, 4, 7, 3, 1, 11, 2, 2 and 5. They are overlapping phase-presence counts across 28 points, not independent sample sizes. All three unrepresented classes, Cu$_{24}$Zr$_{13}$, Cu$_2$Zr and Cu$_5$Zr$_8$, are retained in the benchmark, with hatching distinguishing the missing model vocabulary from errors among represented phases. C15b combines the Cu$_5$Zr and Cu$_9$Zr$_2$ labels by structural identity under the benchmark's physical mapping; CuZr$_2$ groups the specified polymorph family. The existing strict mapping remains available in the benchmark outputs. Point coordinates below follow the temperature-then-composition order of the main figure; full-precision coordinates and observation identifiers are retained in \texttt{figures-DD/data/mlip\_field\_coordinates.csv}.

\begin{widefigure}[tbp]
\nfig{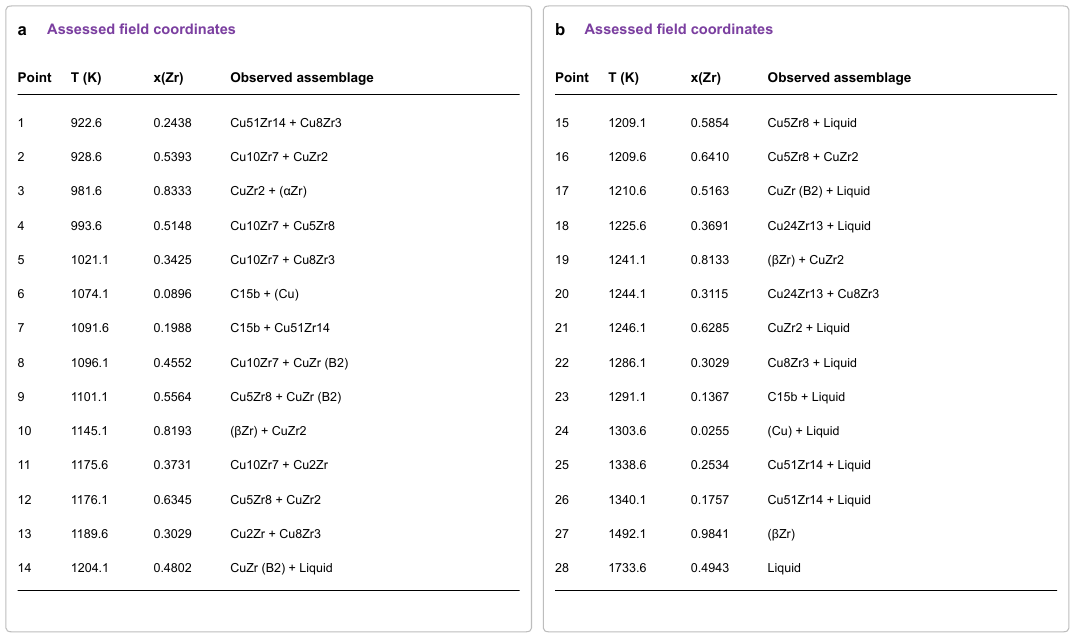}
\caption{\textbf{Coordinates and assessed assemblages of the field benchmark.} Point indices match the main-paper failure maps. Temperatures and compositions are rounded only for this display; evaluation uses the full-precision coordinates. Phase labels follow the physical-phase mapping specified above.}
\label{sfig:field-coordinates}
\end{widefigure}

\subsection{Predictions at withheld activity observations}
The activity panels evaluate the original MLIP-only and final refined TDB files against all 12 activity observations in the demonstration's holdout directory. The calculation uses the same ESPEI equilibrium residual implementation as the reported benchmark, including the allowed phase set, recorded conditions and pure-liquid reference states. The predicted activity is recovered from the residual through $a_\mathrm{pred}=a_\mathrm{obs}\exp[\Delta\mu/(RT)]$. It therefore represents the equilibrium prediction used in the reported score, rather than imposing a metastable liquid where the model predicts another equilibrium. Re-evaluation reproduces the cached RMSE values of 5.09299 and 1.70697~kJ\,mol$^{-1}$ for the initial and final descriptions. No printed activity uncertainties are available, and no measurement error bars or model confidence intervals are inferred from the residual spread. The split withholds observations within the same system and can share a primary study with fitting data; it is not an independent source- or system-level validation. Input file hashes, observation identifiers and individual predictions accompany the generated figure.

\FloatBarrier
\clearpage
\section{Automated source-verification pilot}
\subsection{Protocol for automated source verification of a frozen literature snapshot}
\label{sdetail:automated-source-audit}
The preceding audits identify defects within their stated scopes, while a source review of complete records must also check whether a correctly transcribed value has been associated with the correct thermodynamic state. To address this distinction, we froze a separate working-tree export on 27 October 2026 and specified an automated audit of the accepted literature records. This snapshot contains 73,542 records in total, of which 65,310 are accepted records in the three source kinds covered by the audit. These counts describe the audit snapshot and are distinct from the manuscript's earlier release population. The source files, sampling code and record identifiers are retained with their hashes so that the audit can be repeated against the same evidence. The sampling protocol was fixed before source judgments were obtained.

Within this population, we drew 300 records without replacement using seed 20261027, stratifying by evidence class and source kind. The 11 occupied strata received a proportional allocation with a minimum of 15 records, capped at the size of a stratum, which includes all three records in the smallest stratum. The resulting sample contains records from 229 papers. We retain the inclusion probability and inverse-probability weight of every selected record, and inaccessible or ambiguous source pages remain unresolved rather than being replaced. Records from the same paper can describe related observations, so the number of selected records does not represent a count of independent experiments.

For source verification, the protocol assigns an initial reader to reconstruct the observation from its original page and surrounding context without seeing the stored extraction. A separate verifier checks the frozen record against the source before either judgment is disclosed to the other reader. Adjudication then examines the original evidence and both judgments, including the association of the property and value with its units, composition, phases, temperature, pressure and provenance. It also distinguishes printed conditions from assumptions and measurement conditions from processing history. Deterministic thermodynamic checks provide additional evidence about consistency, while agreement among readers alone is insufficient to establish that an extraction is supported. The model identities and original judgments are retained, and any use of the same model family must accompany the reported outcomes.

For the sampled records, a demonstrated defect in a required field constitutes failure, full source support of the applicable fields constitutes support, and the remaining cases are unresolved. For a completed audit, population fractions would be estimated with the sampling weights, retaining unresolved records in the denominator. If the weighted supported and unresolved fractions are denoted by $P$ and $U$, respectively, the corresponding bounds on correctness are $[P,P+U]$, conditional on the validity of the audit judgments. Finite-population uncertainty is calculated within the sampling strata, with conservative intervals for sparse outcomes. This uncertainty does not account for shared mistakes by the auditors. Separate known-answer challenges are specified to assess error detection and false alarms, and a future page-level completeness study would be required to estimate missed observations. The automated audit is therefore reported separately from the preliminary human review and from any subsequent corrections to the release.

\subsection{Pilot outcomes and limitations of automated source verification}
\label{sdetail:automated-source-pilot}
The protocol above provides a means of checking the association of a numerical value with its source context, although its use also requires an assessment of the auditors themselves. Verified source documents were recovered for all 300 sampled records. The initial reconstruction used GPT-5.5 (\texttt{gpt-5.5-2026-04-23}), while a separate GPT-5.4 call (\texttt{gpt-5.4-2026-03-05}) checked the stored record before GPT-5.5 adjudicated the two outputs against the original pages. These are separate review passes within one model family. The intended second provider could not be used because its account lacked credit, and the subsequent run stopped at the OpenAI project's enforced spending limit. Thus, only 46 records completed all three stages, with initial labels of 30 supported, 14 erroneous and two unresolved; 254 records remain pending. Because completion is neither random nor independent of record complexity, these outcomes do not provide an estimate of population accuracy.

\begin{figure}[htbp]
\centering
\makebox[\linewidth][c]{\includegraphics[width=183mm]{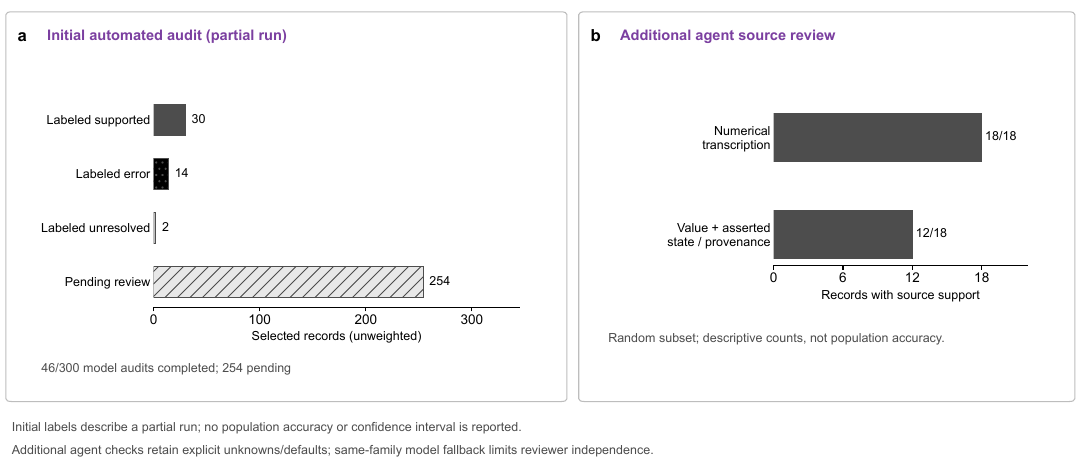}}
\caption{\textbf{Automated source audit and additional agent review.} The initial 300-record audit remains incomplete, so panel~a shows unweighted execution counts rather than extraction accuracy. Panel~b distinguishes numerical transcription from the asserted state and provenance in a separate random subset of 18 records. All 18 numerical values were supported, while six records contained a demonstrated state or provenance defect. Transparent unknowns and explicitly labeled assumptions were retained as limitations rather than counted as errors. These are AI-assisted source judgments, not independent human reference labels; no population precision is inferred from either panel.}
\label{fig:automated-source-pilot}
\end{figure}

To examine the distinction between numerical transcription and the thermodynamic state to which it belongs, we also drew 18 records from the full 300-record sample using seed 20261028 for an additional source review by three agents. The numerical values and their direct unit conversions were supported in all 18 records, but six contained errors in state association, phase labels or provenance (Fig.~\ref{fig:automated-source-pilot}). These included a sample identifier interpreted as a phase, the term ``congruent'' retained as a phase name, a binary observation assigned the chemistry of a broader table, a calculated invariant classified as DFT, and an earlier cited measurement marked as the reporting paper's own work. Another record assigned a composition reported at 878~K to an equilibrium state at 910.7~K after losing the qualifying footnote. The remaining 12 records were consistent with the source within their explicitly retained assumptions and unknowns. These descriptive counts do not establish the completeness of the states or the correctness of an inferred molar basis, and they are not a population precision estimate.

The additional source reviews also expose limitations of the automated judgments. In two targeted checks, an initial error label was overturned because the judge had selected a different row sharing the same temperature or had treated a valid lower-value-plus-upper-bound representation as a lost range. In a third check, the source confirmed that a Sc--Cr calculation had been assigned to a six-element system and incorrectly attributed to an experimental reference. Likewise, a synthetic challenge containing six correct records and six deliberate corruptions detected all six introduced defects, but one correct record was flagged because a general pure-solid reference was judged insufficiently specific for a Cu activity. This challenge is small and deliberately simple; its outcome cannot calibrate error rates on the historical literature. The original labels, subsequent reviews and supporting evidence are retained separately so that disagreements remain inspectable.

These results support the use of multiple agents to identify defects and organize source review, while also showing why neither agreement between readers nor a correct numerical value is sufficient to establish extraction reliability. The 18-record review was additional AI assessment rather than human validation; one reviewer had already seen the initial judgment for one of its records. This automated evaluation of the 300-record sample therefore remained incomplete; the preliminary human review of the same frozen sample and the later measurement of the fitting subset are reported separately in this section. The frozen export has SHA256 \texttt{8de3831eed304f1f845e368c4f1cb0d28050}\allowbreak\texttt{bb8b372aea79bc40552c811ace7f}. Its 65,310 accepted records must not be conflated with the earlier manuscript inventory of 68,266 or with later revisions of the live release. No corrections were applied to the frozen data during this audit.

\section{Diagram-extraction demonstration and phase-recovery review}
\label{sdetail:review-r6}
The synthetic Ni--Ti example in main-paper Fig.~\ref{fig:annotation} illustrates the extraction procedure. It uses a calculated diagram with known phase identities, so that tracing and phase assignment can be examined against a defined reference.

The additional phase-recovery review in main-paper Fig.~\ref{fig:reliability}b concerns a proposed correction batch rather than the complete numerical collection. A sample of 100 proposals was drawn with seed 20261027 from 1,268 phase-only proposals, spanning 36 page groups and one continuation page. A reviewer in a separate assistant context completed all page readings and froze the answers before opening the proposal key. The comparison gives 84 exact unordered-label matches, 15 notation-only differences and one unresolved phase-identity case. Wilson 95\% intervals are 75.58--89.90\% for exact agreement and 94.55--99.82\% for phase identity. The latter retains the unresolved record in the denominator.

The unresolved Ni--Mo row prints the same composition label for bcc-SQS16 and fcc-SQS16 prototypes. It was withheld because a phase-only correction must preserve that distinction. The remaining notation differences concern parentheses, solid-state suffixes and liquid abbreviations. The review establishes agreement for the proposed phase labels under these comparison rules, while temperature, composition, property interpretation and units require their own checks. It is AI-assisted review with shared-model and source dependence, and possible prior exposure during earlier source inspections. The intervals assume independent field or record outcomes and do not describe source-cluster uncertainty. No aggregate human precision estimate is inferred.

\FloatBarrier
\clearpage
\renewcommand{\suppfig}[1]{Fig.~S#1 in SI~1}\renewcommand{\supptab}[1]{Table~S#1 in SI~1}\renewcommand{\suppnote}[1]{Supplementary Note~#1 in SI~1}
\setcounter{figure}{0}\setcounter{table}{0}
\renewcommand{\thefigure}{R\arabic{figure}}
\renewcommand{\thetable}{R\arabic{table}}
\twocolumn[\begin{center}{\Large\bfseries\sffamily Supplementary Information 2: Repository organization and access\par}\end{center}\vspace{1em}]
\section*{Scope of this document}
This SI describes how the ORCHID evidence is organized, retrieved and selected for analysis. The dataset inventory is followed by access routes, evidence-selection rules and machine-learning splits, with licensing and release information at the end. SI~1 documents extraction, calibration and validation. Counts use the same snapshot as the main paper, and features awaiting release are identified explicitly.

\section{A guide to the ORCHID datasets}
\label{sec:repository}

The evidence assembled and checked through the procedures described in SI~1 is organized into \nDatasetsWord{} datasets, which are released together as version v2026.10.
Within this common release, the datasets are deposited separately so that users can retrieve and cite them independently.
An umbrella record (\doiUmbrella) links the eight individual data deposits on Zenodo, giving \nZenodoDataRecords{} records including the umbrella and \nZenodoRecords{} when the runnable examples deposit (\doiExamples) is also counted.
Of these deposits, seven contain the \dsValues{} (\doiLiterature), \dsDiagrams{} (\doiPhaseDiagrams), \dsHulls{} (\doiHulls), \dsParity{} (\doiDftBenchmark), \dsRegistry{} (\doiPhaseRegistry), \dsAssemblages{} (\doiAssessed) and \dsBibliography{} (\doiBibliography) datasets.
The eighth deposit contains the ESPEI datasets derived from \dsValues{} (\doiEspei), which are deposited separately so that users can retrieve and use them independently.
Within each deposit, the files are packaged in a single archive that preserves their repository paths and includes a manifest of SHA-256 checksums.
In addition to the archived deposits, these seven datasets and the ESPEI export are available at \url{https://data.orchid-project.org}, where they are served from the \texttt{orchid-data} repository\cite{orchid_data_2026}.
The eighth named dataset, \dsBench{}, contains the synthetic extraction benchmark and is maintained separately in \texttt{orchid-bench}, alongside its generator and scorer, with which it is archived in the repository's software release (\href{https://doi.org/10.5281/zenodo.23748414}{10.5281/zenodo.23748414}).
The deposits are connected through shared phase identities, source relations and descriptions of thermodynamic states. As summarized in the Appendix of the main paper (Fig.~\ref{fig:dataset-guide}), \dsValues{} and \dsAssemblages{} provide observations of thermodynamic states, while \dsDiagrams{}, \dsHulls{} and \dsParity{} supply calculated references against which those observations can be interpreted. The connection among these products is maintained through \dsRegistry{} and \dsBibliography{}, which preserve phase identities and source relations, and \dsBench{} provides documents with known answers for evaluating extraction. Their shared identifiers and state representation allow the same evidence to support CALPHAD assessment, finite-temperature benchmarks, machine learning and agent-based analysis. Table~\ref{tab:overview} gives the corresponding file-level inventory, while main-paper Fig.~\ref{fig:coverage} shows the coverage of the literature values on which the release is built.

Within these datasets, the JSON files provide the reference representation of the records, from which the Parquet tables are derived.
The Parquet tables add only a \texttt{split} column to this representation, with its value determined from the paper identifier according to the hash rule in \suppnote{4}.
To support the use of these files, the release includes checksum manifests, citation files and field descriptions (Table~\ref{tab:overview}), so that users can verify, cite and interpret the data without consulting the code that produced them.

\begin{widetable}[tbp]
\footnotesize
\caption{\textbf{File-level overview of the \nDatasetsWord{} ORCHID datasets (release v2026.10).} All datasets except \dsBench{} are in the \texttt{orchid-data} repository and on Zenodo (the ESPEI datasets of \dsValues{} as a record of their own); \dsBench{} is in the \texttt{orchid-bench} repository.}
\label{tab:overview}
\renewcommand{\arraystretch}{1.2}
\begin{tabularx}{\linewidth}{P{3.2cm}>{\hsize=1.1\hsize}Y>{\hsize=0.9\hsize}Y}
\toprule
Dataset & Files & Content \\
\midrule
1 \dsValuesFormal{} (\dsValues) & \texttt{literature/} (\texttt{index.json}, \texttt{systems/<SYS>.json}, \texttt{orchid-literature.jsonl.gz}, \texttt{.csv.gz}, \texttt{papers.json}, \texttt{contexts.json}, \texttt{SCHEMA.md}, \texttt{UNCERTAINTY\_CALIBRATION.*}); Parquet \texttt{literature\_values}, \texttt{papers}, \texttt{experimental\_context}; \texttt{espei/} (own record) & \nValues{} values; \nSystems{} systems; \nPapers{} reporting papers; \nBiblio{} papers and cited works; \nContextItems{} context items; \nEspeiDatasetsAll{} ESPEI datasets, of which \nEspeiDatasets{} hold literature and DFT values (\nEspeiValues{} values, \nEspeiSystems{} systems) \\
2 \dsDiagramsFormal{} (\dsDiagrams) & \texttt{phase-diagrams/index.json}, \texttt{phase-diagrams/sgte/*.json}; Parquet \texttt{phase\_diagram\_index}, \texttt{phase\_diagram\_invariants} & \nTdbFiles{} TDB files, \nDiagrams{} calculated (\nDiagramSystems{} systems); \nInvariants{} distinct invariants \\
3 \dsHullsFormal{} (\dsHulls) & \texttt{hulls/dft-mp/}, \texttt{hulls/dft-oqmd/}, \texttt{hulls/calphad-sgte/}; Parquet \texttt{hull\_vertices} & MP: \nHullMPSystems{} systems, \nHullMPVertices{} vertices; OQMD: \nHullOQMDSystems{} systems, \nHullOQMDVertices{} vertices; SGTE at 298.15~K: \nHullSGTEAssessments{} assessments \\
4 \dsParityFormal{} (\dsParity) & \texttt{literature/dft\_parity.json}, \texttt{kim2017\_crosscheck.*}; Parquet \texttt{dft\_parity} & \nParityPoints{} pairs (MP \nParityMP; OQMD \nParityOQMD) from \nParityRecords{} records \\
5 \dsRegistryFormal{} (\dsRegistry) & \texttt{phases/} (\texttt{registry.json}, \texttt{systems/<SYS>.json}, \texttt{registry.jsonld}, \texttt{context.jsonld}, \texttt{registry.csv}, \texttt{registry.parquet}, \texttt{structure\_types.*}, \texttt{review\_queue.json}, \texttt{CONFLICTS.md}) & \nRegPhases{} phases in \nRegSystems{} binary systems; \nRegNames{} names; \nRegModels{} TDB models; \nRegStructureTypes{} structure types \\
6 \dsAssemblagesFormal{} (\dsAssemblages) & \texttt{assessed-phase-assemblages/} (\texttt{records.jsonl.gz}, \texttt{.csv}, \texttt{.parquet}, \texttt{exclusions.*}, \texttt{espei-zpf/}, \texttt{espei-zpf-with-tie-lines/}, \texttt{summary.json}) & \nAsmPoints{} points from \nAsmDiagrams{} diagrams; \nAsmEspeiDatasets{} ESPEI ZPF datasets (\nAsmEspeiPoints{} points) \\
7 \dsBibliographyFormal{} (\dsBibliography) & \texttt{bibliography/} (\texttt{papers.jsonl}, \texttt{edges.jsonl}, \texttt{.csv}, \texttt{.parquet}, \texttt{orchid-bibliography.bib}, \texttt{index.json}, \texttt{SCHEMA.md}) & \nBibPapers{} papers (\nBibWithDOI{} with a DOI); \nBibEdges{} typed relations \\
8 \dsBenchFormal{} (\dsBench) & \texttt{orchid-bench}: \texttt{docs/<doc>/} (four renderings), \texttt{truth/<doc>.json}, \texttt{outputs/}, \texttt{results/} & \nBenchDocs{} synthetic papers, \nBenchPages{} pages, four renderings each \\
\bottomrule
\end{tabularx}
\end{widetable}

\begin{widefigure}[tbp]
\centering
\nfig{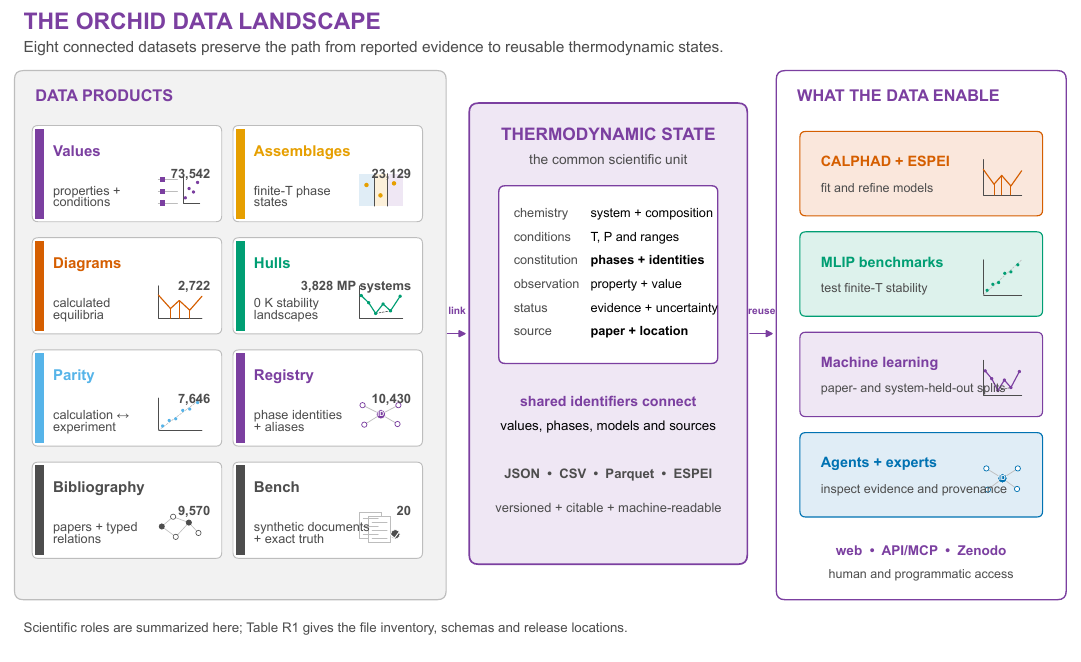}
\caption{\textbf{The ORCHID datasets are connected through a common representation of thermodynamic state.} The eight products retain different forms of evidence and calculated reference data, while shared record, phase and source identifiers connect them to the state that they describe. The same organization supports thermodynamic assessment, finite-temperature benchmarking, machine learning and inspection by people or agents; Table~\ref{tab:overview} gives the corresponding file-level inventory.}
\label{fig:repo-dataset-guide}
\end{widefigure}

\subsection{Literature values (\dsValues)}
The connection between the literature evidence and its representation in ORCHID can be seen in Fig.~\ref{fig:anatomy} in SI~1, where released records are shown alongside the printed table row and sentence from which they were extracted.
These records are grouped according to the chemical system to which they refer and stored in \texttt{systems/<SYS>.json}, where \texttt{<SYS>} lists the elements in alphabetical order, in upper-case letters and joined by hyphens, as in \texttt{AL-NI}.
Within this file, the records are retained together with the associated papers, contexts, tables or passages and links to TDB files.
To allow analyses across chemical systems, the complete collection is also provided in a single JSON dump (\texttt{orchid-literature.jsonl.gz}) and in CSV form, in which nested fields are encoded as JSON strings.

The literature dataset contains \nValues{} values, of which \nAccepted{} are accepted, and cover \nPropertyCodes{} property codes organized into \nPropertyGroups{} property groups (Table~\ref{tab:counts}, \suppfig{1}).
Most of these values concern binary systems, with \nValuesBinary{} values in \nSystemsBinary{} systems, although the collection also includes \nValuesTernary{} values in \nSystemsTernary{} ternary systems, \nValuesUnary{} in \nSystemsUnary{} unary systems and \nValuesHigher{} in \nSystemsHigher{} systems of higher order.
Across these systems, the collection covers \nElements{} elements (main-paper Fig.~\ref{fig:coverage}a,b).
The values collected across these systems were extracted from invariant-reaction tables, which contributed \nInvTable{} values, from general tables, which contributed \nGenTable{}, and from running text, which contributed \nText{}.
The fields used to represent these values are listed in \supptab{2} and the Parquet columns are documented in \texttt{SCHEMA.md}, while \supptab{1} gives the most frequent property codes together with their units and ESPEI outputs.

To interpret these values, it is also necessary to know how the conditions to which they refer were established.
The records therefore distinguish conditions printed in the source from those assumed or inferred, as in the assumption of a pressure of one atmosphere for \nPressureAssumed{} of the \nValues{} values and the inference of a per-atom basis solely from the magnitude of a formation value for \nBasisInferred{} values.
Alongside this information, the uncertainty is retained as a budget in which the contribution printed in the source is kept separate from the systematic contribution of the measurement method.
The contributions retained in this budget are combined according to the uncertainty model described in SI~1 to obtain the effective uncertainty.
Alongside the uncertainty budget, the records retain information on ESPEI readiness and the comparison with the CALPHAD model to support subsequent fitting and evaluation, together with the chemical system, source paper and fold assignments needed for machine learning.
The names and meanings of these fields are given in \supptab{2}, together with the units used for lattice lengths and the links connecting model values to the corresponding TDB files in the ORCHID catalog.

From these records, ESPEI datasets are derived and made available in the folder \texttt{espei/} (Table~\ref{tab:espei}).
These datasets follow the ESPEI 0.9.1 format and retain the source in the \texttt{reference} field, together with a \texttt{bibtex} key that links to its entry in \texttt{references.bib}.
Alongside this source information, the datasets retain the ORCHID observation identifiers of the values, the agreement between readers and the caveats, together with an \texttt{orchid\_uncertainty} block that records the uncertainty budget of each value and the weight used.
Before a dataset is written, the values it contains are checked against those in the published literature release.
For an observation to be retained, it must be published and accepted in the literature release and belong to the evidence class of the folder, while its numerical value, temperature and phase compositions must match those in the release.
Values that do not meet these requirements are excluded, and the reason for excluding them is recorded.
The results of these checks are recorded in \texttt{validation.json} and \texttt{VALIDATION.md}, which identify the dropped values and excluded datasets and report the outcome of \texttt{check\_dataset} and \texttt{load\_datasets} in ESPEI \espeiVersion, while the datasets themselves are listed in \texttt{index.json} and \texttt{index.csv}.
The basis of the uncertainty assigned to the different folders and outputs is documented in \texttt{UNCERTAINTY\_SUMMARY.md}, while \texttt{dft/CALIBRATION.json} and \texttt{.md} describe the DFT calibration and report the leave-system-out $\sigma$ for the individual systems.
The datasets also retain these qualifications through their tags, with \nEspeiSingleReader{} carrying \texttt{single\_reader} and \nEspeiNoPrintedSigma{} carrying \texttt{no\_printed\_uncertainty}.

\begin{widetable}[tbp]
\footnotesize
\caption{\textbf{ESPEI datasets in \texttt{espei/}.} Folders are never mixed: each holds one evidence class or one DFT source.}
\label{tab:espei}
\renewcommand{\arraystretch}{1.2}
\begin{tabularx}{\linewidth}{lYS[table-format=4.0]S[table-format=4.0]S[table-format=3.0]}
\toprule
Folder & Content & {Datasets} & {Values} & {Systems} \\
\midrule
\tabEspeiBody
\bottomrule
\end{tabularx}
\end{widetable}

\begin{widetable}[tbp]
\footnotesize
\caption{\textbf{Literature values by property group and evidence class (release v2026.10).}}
\label{tab:counts}
\begin{tabular}{l*{5}{S[table-format=5.0]}}
\toprule
Property group & {Experimental} & {Model} & {DFT} & {Unspecified} & {Total} \\
\midrule
\tabCountsBody
\bottomrule
\end{tabular}
\end{widetable}

\subsection{Calculated phase diagrams (\dsDiagrams)}
The reported evidence preserved in the literature records can be considered alongside the calculated phase diagrams, which describe the predictions of the assessments.
The association of these predictions with their underlying models is preserved by organizing the diagrams according to the TDB file used in the calculation, rather than the chemical system.
This association is recorded in \texttt{phase-diagrams/sgte/<TDB>.json}, where the TDB file is identified by its name and SHA-256 checksum, together with the components, composition variable (\texttt{x\_component}) and temperature range of the calculation.
Alongside this description of the calculation, the file records the phases and their boundaries (\texttt{regions}), together with the invariants and their temperatures and phase compositions, and the grid of stable phases.
To allow these diagram files to be located by chemical system, \texttt{phase-diagrams/index.json} lists them together with the citation and DOI of the corresponding assessment.
Phase diagrams were calculated from \nDiagrams{} of the \nTdbFiles{} TDB files, covering \nDiagramSystems{} chemical systems, while the remaining \nDiagramsFailed{} files did not yield a calculated diagram and their status is recorded.

In calculating these diagrams, the mapper can reach the same invariant from different starting points, so that the \nInvariantHits{} detections it returned were consolidated into \nInvariants{} distinct invariants.
In this consolidation, two detections are treated as the same invariant when they involve the same phases, their temperatures differ by no more than 0.1~K and their phase compositions agree to within 0.01.
The distinct invariants obtained through this consolidation are stored in the Parquet tables alongside rows for the diagrams, with \texttt{n\_mapper\_hits} recording how many times an invariant was detected.
The JSON files, in turn, retain all detections returned by the mapper, including those consolidated into the same invariant.
The coverage of these calculated diagrams overlaps with that of the literature values, with a calculated diagram available for \nBinWithDiagram{} of the \nSystemsBinary{} binary systems represented in the literature collection.
For these systems, the availability of calculated diagrams allows a direct comparison between the \nBinWithDiagramValues{} literature values and the predictions of the corresponding assessments.
An example of this comparison is shown for Cu--Zr in \suppfig{7}, where the literature values are considered alongside the predictions of the assessment of Zeng \emph{et al.} (1994)\cite{zeng1994cuzr}, one of the \nCuZrAssessments{} Cu--Zr assessments in the SGTE collection.

\subsection{Convex hulls (\dsHulls)}
The calculated phase diagrams are complemented by DFT convex hulls, which identify the compounds predicted to be stable at 0~K.
These hull data are organized by binary system, so that the data for a given system within a layer are stored in a separate JSON document.
Within this organization, the Materials Project layer covers \nHullMPSystems{} binary systems involving \nHullMPElements{} elements, of which \nHullMPWithCompound{} systems contain a stable intermediate compound.
For \nHullMPAgree{} of these systems, the stability flags in this layer agree with those reported by the Materials Project.
The corresponding layer derived from OQMD covers \nHullOQMDSystems{} binary systems involving \nHullOQMDElements{} elements, with a stable intermediate compound present in \nHullOQMDWithCompound{} of these systems.
To allow these DFT predictions to be compared with those of the assessments, a third layer was calculated from the SGTE models using pycalphad\cite{otis2017} at 298.15~K, covering \nHullSGTEAssessments{} assessments of \nHullSGTESystems{} systems.

To make these comparisons, it is necessary to retain the reference conventions used by the source databases.
The catalog for a given layer therefore records the source snapshot and tolerances, together with the elemental reference and the energy corrections applied by the provider.
The elemental reference recorded in this catalog is \texttt{dft\_elemental\_ground\_state\_0K} for the Materials Project and \texttt{dft\_fitted\_elemental\_reference\_0K} for OQMD.

The stable phases are identified by the vertices of these hulls, while nearby entries are also retained in the released layers to allow comparison.
These nearby entries are retained when they lie within 0.05~eV\,atom$^{-1}$ of the hull, together with the hull vertices themselves.
The hull vertices are collected in the Parquet table \texttt{hull\_vertices}, which contains \nHullVertices{} vertices, including elemental endpoints, of which \nHullMPVertices{} come from the Materials Project and \nHullOQMDVertices{} from OQMD.

\subsection{DFT versus experiment (\dsParity)}
To determine how much weight these DFT energies should receive in a fit, formation enthalpies reported in the literature are paired with entries for the same compound in the Materials Project and OQMD.
In making this match, when the paper identifies the crystal structure of the compound, only the DFT entry for that structure is considered.
For these matched measurements, a departure from the stoichiometry of the compound is accounted for through an additional contribution to the uncertainty associated with composition.
Among the \nHmFormRecords{} formation-enthalpy records considered for matching, \nParityRecords{} matched at least one DFT entry, comprising \nParityExp{} experimental, \nParityModel{} model and \nParityUnspec{} unspecified values.
For records that could not be matched, the principal reasons for exclusion were an unstated per-atom basis (\nExclBasis), a phase label without a multi-element formula (\nExclNoFormula), or a value that was itself obtained from DFT (\nExclDFT).

A match alone does not establish that a pair is suitable for calibrating the DFT uncertainty.
Calibration therefore uses a \emph{default subset} of accepted experimental values that are not unsigned in print and have no flag of a suspect magnitude.
For a pair to enter this subset, the measured composition must lie within 0.02 of that of the compound, and the structure identified in the record must be present in the DFT database.
The reference state must also be the solid standard reference of 0~K DFT, excluding pairs with a liquid or gas elemental reference or a reference element outside the compound.
Finally, the record must have no literature gate failure.
These restrictions leave \nMPDefault{} Materials Project and \nOQMDDefault{} OQMD pairs.

To account for differences within these populations, the pairs are stratified by DFT match quality and chemistry (all-metal, p-block or anion element, magnetic 3d metal).
Within a stratum, a median bias and robust $\sigma$ are calculated, with 95\% intervals obtained by resampling whole chemical systems.
The $\sigma$ applied to a system is recomputed without the pairs from that system, while strata with fewer than 30 pairs use the source-level value (\supptab{5}).
The points, matching rules, exclusions and statistics are recorded in \texttt{dft\_parity.json}, and the calibration in \texttt{espei/dft/CALIBRATION.json}.
For this comparison, energies are converted using 1~\evat\ = 96.485~\kjma, and residuals are defined as DFT minus experiment.

\subsection{Phase registry (\dsRegistry)}
Comparison across sources depends on recognizing when different names refer to the same phase.
The registry assigns the identity \texttt{phase:<elements>:<slug>} to a phase and associates it with the names used in TDB files, papers, the extraction corpus and the Materials Project and OQMD.
The resulting \nRegNames{} names comprise \nRegNamesTdb{} TDB names, \nRegNamesPaper{} paper labels, \nRegNamesCorpus{} corpus labels, \nRegNamesMP{} Materials Project names and \nRegNamesOQMD{} OQMD names.
Through this identity, the record links to the TDB models and their verbatim definitions, stability in calculated diagrams (\nRegWithStability{} phases), literature values (\nRegWithValues{} phases) and example references (\nRegWithRefs{} phases, \nRegDOIs{} distinct DOIs).
Structure types are assigned on the basis of a standard name (Tier~A), DFT support (Tier~B), or the declaration of an assessment (Tier~C).
When none of these applies, the record retains the candidate structures and the reasons for review (\supptab{4}).

The compact index \texttt{phases/registry.json} points to the full records in \texttt{phases/systems/<SYS>.json}.
To make these identities available as linked data, \texttt{registry.jsonld} and \texttt{context.jsonld} express the registry as a schema.org \texttt{DefinedTermSet} in the manner of SKOS.
The release also lists the \nRegStructureTypes{} available structure types (\nRegStructureTypesUsed{} used), phases awaiting a structure in \texttt{review\_queue.json}, and unresolved disagreements in \texttt{CONFLICTS.md}.

\subsection{Assessed phase assemblages (\dsAssemblages)}
An assessed phase assemblage represents the phases drawn by an expert at a point deep inside a named field of a handbook diagram.
To describe this point, the record specifies temperature, composition and their uncertainty budgets, together with the placement method, either the area centroid or, for a nonconvex field, its pole of inaccessibility.
To retain the distinction between drawn and estimated boundaries, the record states whether the bounding lines are solid, dashed or both.
There are \nAsmBoundarySolid, \nAsmBoundaryDashed{} and \nAsmBoundaryMixed{} points in these categories, including \nAsmDiagramsDashed{} diagrams with dashed lines.
The release contains \nAsmPoints{} points from \nAsmDiagrams{} diagrams, placed at \nAsmCentroid{} centroids and \nAsmPole{} poles of inaccessibility.
Of these points, \nAsmTwoPhase{} lie in two-phase fields and \nAsmTieLines{} have an optional tie-line block.
Their records retain the handbook label, canonical phase name and phase-registry identity where available, together with the basis of the naming decision (Table~\ref{tab:asm}).
The cross-check against SGTE assessments (main-paper Fig.~\ref{fig:asm}), fields for machine learning and source citations are retained as well, citing the original evaluation before the handbook.

To use these assemblages in thermodynamic fitting, the points are also distributed as \nAsmEspeiDatasets{} ESPEI ZPF datasets containing \nAsmEspeiPoints{} points.
The \nAsmEspeiExcluded{} points whose phases have no TDB name are listed separately.
A further \nEspeiAssessedDatasets{} datasets map the same points to the phases of the SGTE assessments of the system.
Among these mapped datasets are those from a denser grid inside the fields of the Cu--Zr diagram, used in the final fit of main-paper Fig.~\ref{fig:espei}.
The underlying assemblage records are available as JSON, CSV and Parquet, accompanied by exclusions and their reasons for fields and diagrams in which no point could be placed.
As in the other datasets, the release contains only derived data and includes no page image, crop or text from the handbook.

\begin{table}[tbp]
\centering
\footnotesize
\caption{\textbf{Assessed phase-assemblage points by naming basis.} Deterministic printed label: a phase label printed inside the field. Deterministic adjacency and deterministic phase rules: names propagated by the solver from neighbouring fields, using the rule that adjacent fields differ by one phase and the constraints of invariant and compound lines. Reader consensus: agreement of the paired vision-model readers. Adjudicator: name assigned by the adjudicating model with a recorded reason. Expert reviewed: name confirmed or corrected by the domain expert. DFT hull corroborated: low-temperature field named with the help of the Materials Project and OQMD hulls (Supplementary Note~2 of SI~1).}
\label{tab:asm}
\begin{tabular}{lr}
\toprule
Basis & Points \\
\midrule
\tabAsmBasisBody
\bottomrule
\end{tabular}
\end{table}

\subsection{Papers and their relations (\dsBibliography)}
The bibliography places the extracted values within the literature from which they were obtained and the literature that remains to be read.
Its relations allow a value to be traced back to the original report and forward to the assessments and diagrams that used it.
The file \texttt{bibliography/papers.jsonl} contains records for \nBibPapers{} papers, \nBibWithDOI{} of them with a DOI.
A record gives the Crossref metadata, the roles of the paper and the number of values it reports or for which it is cited as the source.
The file \texttt{edges.jsonl} connects these papers to TDB files and handbook diagrams through \nBibEdges{} typed relations, retaining the provenance of the relations.
The collection includes \nBibRead{} corpus papers and assessments that were read, \nBibCitedSources{} original sources of extracted values, \nBibTdbSources{} TDB sources, \nBibHandbookSources{} evaluations reproduced in the handbook and \nBibLeads{} papers identified but not yet read.
Although a paper can have several of these roles, the collection also includes works to which no extracted value is yet attributed, so it is larger than the \nBiblio{} papers and cited works associated with \dsValues{}.
The records are also distributed as CSV, Parquet, JSON-LD, BibTeX (\texttt{orchid-bibliography.bib}) and CSL-JSON.
These records can also be accessed by chemical system through the files in \texttt{systems/}.

\subsection{Synthetic extraction benchmark (\dsBench)}
To evaluate extraction against known answers, the synthetic benchmark groups its files by document.
A document is supplied as a vector PDF, 300-dpi page images, degraded 150-dpi scans and an image-only PDF, together with the truth key \texttt{truth/<doc>.json} (schema \texttt{orchid-synthetic-paper-truth/\allowbreak v2}).
The truth key identifies the source TDB by SHA-256 and records the page transforms of the renderings.
For figures, it specifies the axes, boundaries, invariant lines and plotted series.
For the tables in the same documents, it identifies the semantic field and exact value of the cells, and for running text it records the stated values and their positions.
The outputs to be compared with this truth are stored in \texttt{outputs/<extractor>/<rendering>/}.
The corresponding scores, ledger of reader calls and \texttt{RESULTS.md} are stored in \texttt{results/}.
Within this benchmark, the documents identify themselves as synthetic, and the three fictitious references cited in a document are explicitly identified as such.

\section{Access and reuse}
\subsection{Related resources}
ORCHID complements existing resources in three broad groups.
The first consists of compilations of experimental values for particular properties, such as the formation enthalpies of intermetallic and other inorganic compounds collected by Kim \emph{et al.}\cite{kim2017}, the magnetic transition temperatures extracted automatically from the literature by Court and Cole\cite{court2018} and extended by Kelly and Cole\cite{kelly2025}, and the properties and observed phases of multi-principal element alloys curated by Borg \emph{et al.}\cite{borg2020}.
The second consists of high-throughput DFT databases, such as the Materials Project\cite{jain2013} and OQMD\cite{saal2013,kirklin2015}, which provide formation energies at 0~K for very large numbers of compounds, and the third of the CALPHAD resources themselves, the TDB collections and fitting software discussed in the introduction.
What none of these provides is the phase-equilibrium and thermochemical evidence behind CALPHAD models, recorded as equilibrium states with their provenance, and Table~\ref{tab:compare} sets out how ORCHID differs from the most closely related of them in this respect.

\begin{widetable}[tbp]
\footnotesize
\caption{\textbf{ORCHID and related resources.} Open licence: terms that allow reuse and redistribution of the data. CALPHAD-ready conditions: each datum carries its thermodynamic state (phases, temperature, pressure, compositions) in a form usable for fitting. Per-value provenance: each value points to its source publication (ORCHID: DOI, page and quote or table row). ML splits: official training and test splits. Agent access: an official Model Context Protocol (MCP) server or, failing that, an API. --: not applicable (computed data). Cells were checked against each resource's public documentation on 23 October 2026\cite{thermoml_archive,nist_license,mpds_developer,mpds_open_ccby,hallstedt2025,asm_apd,mp_terms,mp_mcp,oqmd_download,oqmd_api,espei_datasets}. $^{\dagger}$Not fully verified: the reuse terms of publisher-contributed ThermoML files, the scope of the open MPDS subset, the granularity of the former ASM Alloy Phase Diagram Database (APD, now ASM MPDS) citations; the Materials Project and OQMD licence statements were read from the legacy terms page and an archived download page.}
\label{tab:compare}
\renewcommand{\arraystretch}{1.2}
\setlength{\tabcolsep}{4pt}
\begin{tabularx}{\linewidth}{>{\hsize=0.95\hsize}Y>{\hsize=1.05\hsize}Y>{\hsize=1.1\hsize}Y>{\hsize=0.95\hsize}YP{1.25cm}>{\hsize=0.95\hsize}Y}
\toprule
Resource & Open licence & CALPHAD-ready conditions & Per-value provenance & ML splits & Agent access \\
\midrule
ORCHID (this work) & yes (CC~BY~4.0) & yes (states; ESPEI datasets) & yes (DOI, page, quote or row) & yes (by paper; system and paper folds) & MCP server; OpenAPI \\
ThermoML archive (NIST TRC)\cite{frenkel2003,thermoml_archive} & partial$^{\dagger}$ (NIST open data; files from a cooperation between NIST TRC and five journals (about 2003--2019)) & partial (state variables per property; mainly fluids; not ESPEI format) & partial (one file per article) & no & no (bulk download) \\
MPDS / PAULING FILE (incl.\ former ASM APD)\cite{mpds_developer,asm_apd} & partial$^{\dagger}$ (subscription; open subset under CC~BY) & partial (redrawn diagrams with phase labels, phase lists; no ZPF data or digitized equilibria) & yes (per entry, to publication; per diagram for the former APD$^{\dagger}$) & no & API (no MCP) \\
SGTE binary collection\cite{hallstedt2025} & yes (CC~BY~4.0) & no (model parameters, not observations) & partial (each TDB cites its assessment) & no & no \\
Materials Project\cite{jain2013,mp_terms} & yes (CC~BY~4.0) & partial (0~K DFT energies) & -- (calculation ids) & no & MCP server\cite{mp_mcp}; API \\
OQMD\cite{saal2013,oqmd_download} & yes (CC~BY~4.0) & partial (0~K DFT energies) & -- (calculation ids) & no & API (REST, OPTIMADE)\cite{oqmd_api}; no MCP \\
ESPEI-datasets\cite{espei_datasets} & yes (CC~BY~4.0) & yes (ESPEI JSON) & partial (reference key per dataset) & no & no \\
\bottomrule
\end{tabularx}
\end{widetable}

\subsection{Obtaining the data}
The choice of an access route depends on whether the analysis requires complete records, a tabular view or queries for selected systems and phases.
The data deposits described above are archived on Zenodo and available with documentation on the project's data site; the synthetic benchmark is archived with its software release. JSON files preserve complete records, while CSV files provide a flat view for spreadsheets and inspection. Parquet tables include fixed training and test splits and are also available as configurations of a Hugging Face dataset.
Programs and AI agents can, in addition, query the data directly through a Model Context Protocol server\cite{mcp2024spec}, whose tools answer questions such as which values are known for a system or how a phase is named across sources, and every answer returns with the provenance of the values it contains (Table~\ref{tab:aiready}).

For those who work with files, temperatures are given in K, pressures in Pa, formation enthalpies in \kjma\ or, where the basis is not printed, in kJ\,mol$^{-1}$ with the basis unstated, and compositions as mole fractions, with the exceptions, such as DFT energies in \evat, lattice lengths in \AA, Gibbs energies that keep their printed units and \nFormUnitNotNormalized{} formation values whose unit could not be normalized, which keep the unit as read and are never ESPEI-ready, listed in \supptab{1}.

The Parquet tables can be loaded directly with pandas, DuckDB or the Hugging Face \texttt{datasets} library\cite{lhoest2021datasets} (\nHfConfigs{} configurations: \hfConfigList).
For those who prefer to browse, every chemical system, compound and binary phase has a static page at \url{https://data.orchid-project.org}, and \url{https://orchid-project.org} provides viewers for the data, the phase diagrams, the DFT comparison and the hulls.
For programs and AI agents, the \texttt{orchid-mcp} server exposes \nMcpTools{} tools (\mcpToolList; \supptab{7}, with a recorded exchange in \suppfig{6}), an OpenAPI specification describes the static file endpoints for programs that do not use MCP, and Croissant metadata\cite{akhtar2024croissant} and \texttt{llms.txt} files describe the data for machines.

These routes return the same versioned records, so the choice of interface can be separated from the later choice of evidence for a particular analysis (Fig.~\ref{fig:repo-access}).

\begin{widefigure}[tbp]
\centering
\nfig{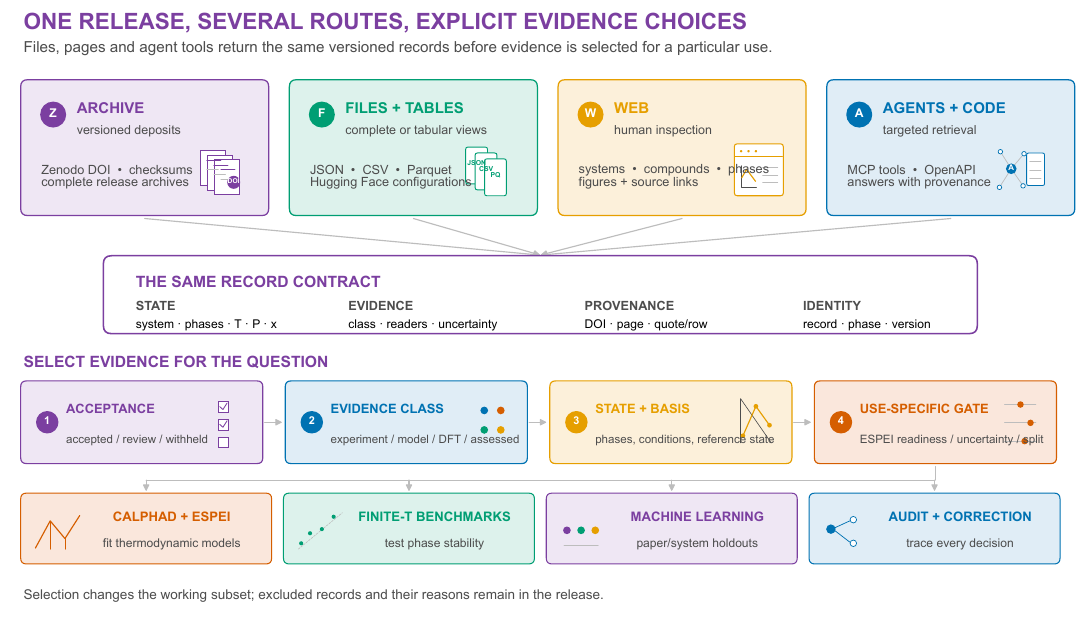}
\caption{\textbf{Access and evidence selection are separate operations.} Complete archives, tabular files, web pages and agent tools expose the same state, evidence, provenance and identity fields. A user can then select records according to acceptance, evidence class, completeness of the state and the requirements of the intended calculation, while the records left out of a working subset remain available with the reasons for their exclusion.}
\label{fig:repo-access}
\end{widefigure}

\begin{widetable}[tbp]
\footnotesize
\caption{\textbf{AI readiness of ORCHID.} Each aspect follows the ESIP AI-readiness checklist\cite{esip2022} and the FAIR principles\cite{wilkinson2016}. Paths are relative to the \texttt{orchid-data} repository unless another repository is named.}
\label{tab:aiready}
\renewcommand{\arraystretch}{1.2}
\begin{tabularx}{\linewidth}{P{2.5cm}YY}
\toprule
Aspect & How ORCHID meets it & Where (file / endpoint) \\
\midrule
Formats and schemas & documented JSON, CSV and Parquet; a schema per dataset; controlled vocabularies for property codes, evidence classes and reaction types & \texttt{parquet/}, \texttt{*/SCHEMA.md}, \texttt{literature/QUALITY\_FIXES.md} \\
Machine learning & Parquet with by-paper splits (training and test data never share a paper) and leave-system-out and leave-paper-out folds; Hugging Face dataset card with \nHfConfigs{} configurations & \texttt{parquet/literature\_values.parquet} (\texttt{split}, \texttt{fold\_lso}, \texttt{fold\_lpo}), \texttt{hf/README.md}, \texttt{DATASET\_CARD.md} \\
Metadata & Croissant\cite{akhtar2024croissant} and JSON-LD metadata; DataCite metadata through Zenodo; citation files; datasheet\cite{gebru2021} & \texttt{data/croissant.json} (\texttt{orchid-project.org}), \texttt{phases/context.jsonld}, \texttt{zenodo/*.metadata.json}, \texttt{CITATION.cff}, \texttt{DATASET\_CARD.md} \\
Discoverability & crawlable static pages with schema.org JSON-LD, sitemap, \texttt{llms.txt}; stable identifiers (\texttt{ORC-}, \texttt{phase:}) and DOIs & \texttt{systems/}, \texttt{phases/pages/}, \texttt{sitemap.xml}, \texttt{robots.txt}, \texttt{llms.txt}, \texttt{llms-full.txt}, \texttt{zenodo/dois.json} \\
Access for programs and agents & MCP server with \nMcpTools{} tools; OpenAPI specification; Python package installable from GitHub (\texttt{pip install git+https://github.com/orchid-calphad/orchid-mcp}; \doiCodeMcp{}) & \texttt{orchid-mcp}, \texttt{openapi/openapi.yaml} \\
Per-value provenance and trust & DOI, page, verbatim quote, cited original where resolved, readers and their agreement, evidence class, tool release, uncertainty budget & \texttt{literature/systems/<SYS>.json}, \texttt{paper\_doi}, \texttt{quote}, \texttt{readers}, \texttt{reader\_agreement}, \texttt{sigma\_effective} \\
Quality labels & flags, review states, adversarial audit rates, extraction benchmark & \texttt{flag\_*}, \texttt{review\_state}, \texttt{literature/QUALITY\_FIXES.md}, \texttt{orchid-bench/results/} \\
Reuse & ESPEI-validated datasets with $\sigma$, weights and tags; calibrated DFT $\sigma$ & \texttt{espei/}, \texttt{espei/dft/CALIBRATION.json}, \texttt{assessed-phase-assemblages/espei-zpf/} \\
Licensing & CC~BY~4.0 for ORCHID's data, which permits use for model training with attribution; quotes are short excerpts kept for verification under the publishers' copyright; third-party licences stated; restricted sources are never redistributed & \texttt{LICENSE}, \texttt{LICENSES.md}, \texttt{NOTICE.md} \\
Reproducibility & deterministic builders with \texttt{--check}; versioned DOIs; changelog; status ``Version~1, machine-extracted'' & \texttt{tools/*.py --check} (\texttt{orchid-project.org}), \texttt{CHANGELOG.md}, \texttt{zenodo/} \\
\bottomrule
\end{tabularx}
\end{widetable}

\subsection{Selecting evidence for an analysis}
Once the data have been obtained, their use depends on selecting evidence appropriate to the intended analysis.
The release makes this selection explicit by retaining the information needed to assess a value before it enters a fit.
The records specify the evidence class, number of agreeing independent readers, checks passed and corrections received.
Beyond these checks, the records distinguish readiness for ESPEI from certification by agreement with a CALPHAD calculation.
A user who needs the most conservative subset can therefore restrict a fit to certified experimental values, while one who wants the widest coverage can include single-reader or model values, knowing exactly what each choice admits.
The following coverage and evaluation details inform these selections.
The numerical literature values were extracted by machine, with page verification performed by AI agents; no independent human precision estimate is reported for the numerical collection. The completed expert evaluation of assessed-diagram fields reported in SI~1 concerns a separate population and task.
The coverage also excludes values reported only in figures, and pressure is almost never printed, so the one atmosphere recorded for all but \nPressureNotOneAtm{} values is an assumption.

To implement evidence selection, Table~\ref{tab:filters} gives the filters and their counts.
The accepted values provide the default subset for analysis. Within this subset, model values retain their status as calculations, while unspecified values retain an unresolved evidence class.
Assessed points should be kept separate from both, while the weight assigned to a value should follow its effective uncertainty.
Interpretation also requires considering how the sources were selected and read.
In particular, the corpus consists of the papers that could be read, rather than all papers on a system.
Within that corpus, an experimental value cited from older work retains the citing paper's interpretation of that work.
For tables covering several systems, values are assigned to the system identified by their row.

\begin{table}[tbp]
\footnotesize
\caption{\textbf{Recommended filters for the literature values.} Counts refer to release v2026.10.}
\label{tab:filters}
\renewcommand{\arraystretch}{1.15}
\begin{tabularx}{\linewidth}{P{3.6cm}rY}
\toprule
Field or flag & Values & Recommended use \\
\midrule
\texttt{accepted} & \nAccepted & restrict analyses to accepted values \\
\texttt{evidence\_class} model, unspecified & -- & do not treat as measurements \\
\texttt{assessed\_phase\_diagram} points & \nAsmPoints & keep apart from measured and model values \\
\texttt{reader\_agreement = single\_reader} & \nRASingle & exclude or down-weight \\
unit \texttt{kJ/mol (basis unstated)} & \nBasisUnstatedFormation & exclude from per-atom analyses \\
\texttt{flag\_basis\_inferred\_from\_magnitude} & \nBasisInferred & may be excluded from per-atom analyses \\
\texttt{flag\_unsigned\_in\_print} & \nUnsigned & exclude from signed comparisons \\
\texttt{flag\_source\_disputed} & \nSourceDisputed & check before use \\
\texttt{flag\_magnitude\_suspect} & \nMagnitudeSuspect & check before use \\
\texttt{flag\_system\_mismatch} & \nSystemMismatch & check before use \\
\texttt{flag\_dft\_outlier\_unexplained} & \nDftOutlier & check before use \\
\texttt{sigma\_effective} (\texttt{sigma\_method\_basis}) & all & weights; the basis states whether the method floor was calibrated or is a prior \\
\bottomrule
\end{tabularx}
\end{table}
Such errors can be reported through the issue trackers of the repositories, and identifiers (\texttt{ORC-...}, \texttt{phase:...}) are stable across releases.
\paragraph{Contribution workflow under development.}
A contribution branch that is not yet merged adds links from data pages and MCP responses to pre-filled issues in the \texttt{orchid-data} repository.
These links are intended to preserve the record identifier, source and page when a user or agent reports a discrepancy.
The same branch provides a pull-request workflow for corrections, new values and assessments, with automated release-rule checks followed by source-page review by a maintainer.
Under this proposed workflow, accepted contributions would enter a subsequent release with contributor credit and the original value retained in the record history.
These mechanisms remain prospective until the branch is merged and the corresponding release is verified.

\subsection{Community calibration and planned educational use}
The retained source information also provides a basis for community contributions to the coverage and calibration of the collection.
Using the released identifiers and provenance, a contributor can report a discrepancy in a specific record together with the source needed to examine it, while proposed additions can be evaluated against the existing requirements for state assignment and consistency.
An educational use is also being prepared in which students identify the phases at points of real diagrams, with known-answer questions used to calibrate their reliability before uncertain responses are considered for inclusion.
The first classroom use is planned for November 2026 and is therefore a prospective contribution mechanism, not a completed validation result.
Repository mechanisms that remain under development are documented separately rather than assumed to be operating in the released workflow.

\subsection{Machine learning}
The provenance retained for evidence selection also determines how the records can be divided into training and test sets for machine-learning models of thermodynamic properties. Such divisions must account for correlations among observations from the same chemical system and paper.
To support these evaluations, the records retain their chemical system and source paper together with two fixed assignments to five folds.
One assignment groups all values of a chemical system, while the other groups all values of a paper.
These groupings allow a model to be tested on chemistry or sources absent from its training data.
Within these folds, composition and structure can be used to learn formation enthalpies, invariant temperatures and phase stability.
For these learning tasks, the evidence classes allow training to be restricted to experiment, and the uncertainty budgets allow observations to be weighted according to their reliability.
For a single fixed division, the column \texttt{split} assigns all values of a paper to the training, validation or test set.
The resulting sets contain \nSplitTrain{}, \nSplitVal{} and \nSplitTest{} values from \nSplitTrainPapers{}, \nSplitValPapers{} and \nSplitTestPapers{} papers, respectively, using the hash rule in \suppnote{4}.
However, a measurement may be reprinted in several papers.
The strictest leakage control therefore groups values by the work in which they were first measured, recorded in \texttt{origin\_measured\_in}.
Alongside these property-prediction tasks, for the training and testing of extraction itself, the quotes and table locators of the records can be paired with their structured fields, and new extractors can be tested on \dsBench, whose answers are known exactly.

\subsection{Licence and citation}
All of these uses rest on the licence, and the derived data are released under CC~BY~4.0.
Within these terms, each \texttt{quote} is a single short excerpt, kept so that the value can be checked against its source, and it remains under the copyright of the original publisher.
Apart from such quotes, the source papers, page images and full texts are not redistributed.
For \dsAssemblages, only derived facts are released (phase names, assemblages, coordinates, phase metadata and citations), and the pages, images and text of the handbook are not.
Users are asked to cite this paper, the dataset DOI and the original source of a value (\texttt{paper\_doi}, \texttt{cited\_dois}, and, for \dsAssemblages, the original evaluation and then the handbook).
Third-party data keep their licences and citations, which applies to the SGTE binary collection\cite{hallstedt2025}, the Materials Project\cite{jain2013} and OQMD\cite{saal2013,kirklin2015}, all released under CC~BY~4.0, to the compilation of Kim \emph{et al.}\cite{kim2017}, whose files are released under the MIT licence, and to the bibliographic metadata from Crossref, which carry no restriction (CC0).
The dataset card, in turn, follows the datasheet questions of Gebru \emph{et al.}\cite{gebru2021}.

\section*{Data availability}
The data described in this paper are release v2026.10 of ORCHID, released under CC~BY~4.0 (third-party data under the licences stated in the licence section and in \texttt{LICENSES.md}).
The release is archived on Zenodo in \nZenodoDataRecords{} records, comprising an umbrella record (\doiUmbrella) and eight individual data deposits: literature values (\doiLiterature), ESPEI datasets (\doiEspei), calculated phase diagrams (\doiPhaseDiagrams), convex hulls (\doiHulls), DFT versus experiment (\doiDftBenchmark), phase registry (\doiPhaseRegistry), assessed phase assemblages (\doiAssessed) and bibliography (\doiBibliography).
The synthetic extraction benchmark, \dsBench{}, is archived with the software release of the \texttt{orchid-bench} repository (\href{https://doi.org/10.5281/zenodo.23748414}{10.5281/zenodo.23748414}).
The same files are served, with documentation, at \url{https://data.orchid-project.org} from the \texttt{orchid-data} repository at \url{https://github.com/orchid-calphad}, and the Parquet tables are mirrored as a Hugging Face dataset.
The GitHub repositories and the data site have been public since 28 October 2026.

\section*{Code availability}
The code is MIT-licensed at \url{https://github.com/orchid-calphad}.
The \texttt{orchid-project.org} repository holds the builders of the released files, namely \texttt{export\_literature\_data.py}, \texttt{literature\_quality.py}, \texttt{literature\_gates.py}, \texttt{phase\_rule.py}, \texttt{uncertainty\_budget.py} and \texttt{ml\_folds.py} (\dsValues, clean-up, gates, uncertainty budget and folds), \texttt{build\_espei\_bundle.py}, \texttt{espei\_constraints.py} and \texttt{espei\_evaluate.py} (ESPEI datasets, DFT calibration and likelihood evaluation), \texttt{compute\_binary\_diagrams.py} and \texttt{add\_phase\_grid.py} (\dsDiagrams), \texttt{build\_dft\_parity.py} and \texttt{kim2017\_crosscheck.py} (\dsParity), \texttt{build\_phase\_registry.py} (\dsRegistry), \texttt{build\_assessed\_assemblages.py} and \texttt{dashed\_boundaries.py} (\dsAssemblages), \texttt{build\_bibliography.py} (\dsBibliography), \texttt{calc\_invariants.py} and \texttt{build\_parquet.py}, \texttt{build\_static\_pages.py}, \texttt{build\_precision\_sample.py} and \texttt{score\_precision\_sample.py}.
The builders are deterministic, and each has a \texttt{--check} mode that verifies the released files.
The \texttt{orchid-data} repository holds the data, and \texttt{orchid-mcp} holds the MCP server and its OpenAPI specification.
Phase diagrams were calculated with pycalphad 0.11.2 under Python 3.13, and the ESPEI\cite{bocklund2019} datasets were validated with ESPEI 0.9.1.
The extraction pipeline, the hull builders and the phase-diagram annotator are in \texttt{orchid-pipeline}, and the synthetic extraction benchmark, with its generator, baselines and scorer, is in \texttt{orchid-bench}.
Each release is archived on Zenodo (\href{https://doi.org/10.5281/zenodo.23748335}{10.5281/zenodo.23748335}).
The counts and figures of this paper are generated from the released files by \texttt{scripts/make\_numbers.py} and \texttt{scripts/make\_figures.py}, which are distributed with the manuscript source.

\FloatBarrier
\clearpage
\bibliographystyle{rsc}
\bibliography{references_DD}
\end{document}